%% file: main_gromov.tex
\DocumentMetadata{
    lang = en,
    pdfversion = 1.7,
}
\documentclass[11pt,letterpaper]{article}

\newcommand{\mytitle}{Multi-Fidelity Particle Flow Filters based on Log Homotopy}

\input{arxiv/packages}

\input{arxiv/macros}

\title{\mytitle}
\date{\today}
\author[1]{Abhinab Bhattacharjee\thanks{\href{mailto:abhinab93@vt.edu}{abhinab93@vt.edu}}}
\author[2]{Andrey A. Popov\thanks{\href{mailto:apopov@hawaii.edu}{apopov@hawaii.edu}}}
\author[3]{Amit N. Subrahmanya\thanks{\href{mailto:famitnagesh@anl.gov}{famitnagesh@anl.gov}}}
\author[1]{Adrian Sandu\thanks{\href{mailto:asandu7@vt.edu}{asandu7@vt.edu}}}
\affil[1]{Computational Science Lab, Department of Computer Science, Virginia Tech, USA}
\affil[2]{Department of Information \& Computer Sciences, the University of Hawai'i at Manoa, Honolulu, Hawai'i, USA}
\affil[3]{Argonne National Laboratory, Lemont, Illinois, USA}

\usepackage[draft]{csl}
\csltitle{\mytitle}
\cslauthor{Abhinab Bhattacharjee, Andrey A. Popov, Amit N. Subrahmanya, Adrian Sandu}
\cslyear{2026}
\cslreportnumber{4}
\cslemail{abhinab93@vt.edu}

\begin{document}

    \csltitlepage

    \maketitle

\subfile{sections/abstract}

\subfile{sections/introduction}

\subfile{sections/background}

\subfile{sections/methodology}

\subfile{sections/numerical_experiments}

\subfile{sections/conclusion}

    \newpage
    \appendix
    \appendixpage
    \addappheadtotoc 
    \renewcommand{\thesection}{\Alph{section}}
    \counterwithin{figure}{section}
    \renewcommand{\thefigure}{\Alph{section}.\arabic{figure}}
    \subfile{sections/appendix}

    \newpage
    \bibliographystyle{plain}
    \bibliography{bib/reference,bib/sandu}


\end{document}

%% file: arxiv/packages.tex
\usepackage[inline]{enumitem}
\usepackage[dvipsnames]{xcolor}
\usepackage{setspace, enumitem, graphicx, amsmath, sectsty, pdfpages, amssymb, booktabs, float, array, tikz, pgfplots, colortbl, multirow, longtable, authblk, soul, bm, amsthm}
\usepackage{algorithm}
\usepackage{algpseudocode}
\usepackage{cite}
\usepackage{stmaryrd}
\usepackage[listings,skins,breakable]{tcolorbox}

\definecolor{DarkRed}{HTML}{800000}

\usepackage{hyperref}
\hypersetup{
	colorlinks = true, linkcolor = DarkRed, anchorcolor = blue,
	citecolor = blue,  filecolor = blue, urlcolor = blue}
\usepackage{cleveref}

\usepackage[justification=justified, font=footnotesize,labelfont=bf]{caption}
\usepackage{subcaption}

\usepackage{geometry}
\usepackage{todonotes}

\usepackage{tikz}
\usetikzlibrary{shapes,arrows,positioning,calc}

\usetikzlibrary{shapes, arrows.meta, positioning} 

\usetikzlibrary{decorations.text} 
\usetikzlibrary{fit, backgrounds, decorations.pathmorphing}

\usepackage{bbm}
\usepackage{authblk}
\usepackage[title,titletoc]{appendix}

\usepackage{lineno}
\usepackage{mwe}

\usepackage{subfiles}
\usepackage{mathtools}

\numberwithin{equation}{section}

\theoremstyle{remark}
\newtheorem{remark}{Remark}

\usepackage{soul}

%% file: arxiv/macros.tex
{%
\end{tcolorbox}\endgroup}

\newtcbox{\popovnotesmall}{breakable,enhanced jigsaw,nobeforeafter,tcbox raise base,boxrule=0.4pt,top=0mm,bottom=0mm,
  right=0mm,left=4mm,arc=1pt,boxsep=2pt,before upper={\vphantom{dlg}},
  colframe=Fuchsia!75!black,coltext=black,colback=Fuchsia!20,
  overlay={\begin{tcbclipinterior}\fill[Fuchsia!75!black] (frame.south west)
    rectangle node[text=white,font=\sffamily\bfseries\tiny,rotate=90] {AAP} ([xshift=4mm]frame.north west);\end{tcbclipinterior}}}
\MakeRobust\popovnotesmall
\ifdefined
\fi

\newtcbox{\arashnote}{breakable,enhanced jigsaw,nobeforeafter,tcbox raise base,boxrule=0.4pt,top=0mm,bottom=0mm,
  right=0mm,left=4mm,arc=1pt,boxsep=2pt,before upper={\vphantom{dlg}},
  colframe=Green!75!black,coltext=black,colback=Green!75,
  overlay={\begin{tcbclipinterior}\fill[Green!75!black] (frame.south west)
    rectangle node[text=white,font=\sffamily\bfseries\tiny,rotate=90] {ArSa} ([xshift=4mm]frame.north west);\end{tcbclipinterior}}}
\MakeRobust\arashnote
\ifdefined
\fi

{%
\end{tcolorbox}\endgroup}

\newcommand{\sandu}[1]{ {\color{blue}#1} }

\def\<#1>{\mathinner{\left\langle#1\right\rangle}}

\def\|#1|{\left|#1\right|}

\newcommand{\bfM}{\symbf{M}}
\newcommand{\bfH}{\symbf{H}}

\newcommand{\bfR}{\symbf{R}}
\newcommand{\bfQ}{\symbf{Q}}
\newcommand{\bfx}{\symbf{x}}
\newcommand{\bfy}{\symbf{y}}
\newcommand{\bfY}{\symbf{Y}}
\newcommand{\bfX}{\symbf{X}}

\newcommand{\bfXi}{\symbf{\Xi}}
\newcommand{\bfxi}{\symbf{\xi}}
\newcommand{\bfL}{\symbf{L}}
\newcommand{\bfN}{\symbf{N}}
\newcommand{\bfu}{\symbf{u}}

\newcommand{\bfA}{\symbf{A}}
\newcommand{\bfa}{\symbf{a}}
\newcommand{\bfb}{\symbf{b}}
\newcommand{\bfI}{\symbf{I}}
\newcommand{\bfP}{\symbf{P}}
\newcommand{\bfK}{\symbf{K}}
\newcommand{\bfD}{\symbf{D}}
\newcommand{\bfE}{\symbf{E}}
\newcommand{\bfU}{\symbf{U}}
\newcommand{\bfZ}{\symbf{Z}}
\newcommand{\bfz}{\symbf{z}}

\newcommand{\NLModel}{\mathcal{M}}
\newcommand{\NLObs}{\mathcal{H}}

\newcommand{\bfPhi}{\symbf{\Phi}}
\newcommand{\bfphi}{\symbf{\phi}}
\newcommand{\bfmu}{\symbf{\mu}}

\newcommand{\bfnu}{\symbf{\nu}}
\newcommand{\bfSigma}{\symbf{\Sigma}}
\newcommand{\bfone}{\symbf{1}}

\newcommand{\bfzero}{\symbf{0}}

\DeclareMathOperator{\sech}{sech}

\newcommand{\inner}[2]{\left\langle#1,#2\right\rangle}

\newcommand{\norm}[1]{\left\lVert#1\right\rVert}

\sectionfont{\fontsize{12}{15}\selectfont}
\subsectionfont{\fontsize{12}{15}\selectfont}

\emergencystretch=\maxdimen
\pgfplotsset{compat=1.17}
\providecommand{\keywords}[1]
{
  \small	
  \textbf{\textit{Keywords---}} #1
}

\algrenewtext{Procedure}[2]{%
  \algorithmicprocedure\ \textproc{#1}%
  \ifthenelse{\equal{#2}{}}{}{$\left(\mbox{#2}\right)$}}

%% file: sections/abstract.tex
\begin{abstract}

This work proposes a new Multi-Fidelity Particle Flow Filter (MFPFF) based on a linear control variate framework and multilevel Monte Carlo methods.
Conventional particle filters often suffer from weight collapse and particle degeneracy in high-dimensional spaces.
Particle flow filters address this by smoothly transporting prior particles toward the posterior distribution, thereby avoiding the need for resampling.
This work constructs multi-fidelity particle flow filters by augmenting a small number of expensive high-fidelity particles (full-order model runs) with a large number of inexpensive low-resolution particles (reduced-order surrogates constructed via intrusive Galerkin projection).
These surrogates act as control variates to enhance the statistical estimate of the full-order posterior, yielding enhanced analysis accuracy at comparable computational cost.
The work  focuses on two particle flows based on log homotopy, the stochastic Gromov flow and deterministic Exact flow. 
Numerical experiments with the Korteweg-De Vries (KdV) and Quasi-Geostrophic (QG) test problems demonstrate that the proposed MFPFF achieves accuracy comparable to high-fidelity simulations while significantly reducing computational expense.


\keywords{particle flow, bayesian inversion, data assimilation, reduced order models, multi-fidelity}
    
\end{abstract}

%% file: sections/introduction.tex
\newpage

\section{Introduction}
\label{sec:introduction}
Data assimilation (DA) aims to improve the estimation of states and parameters of nonlinear dynamical systems by fusing prior knowledge of the hidden states, imperfect numerical model predictions, and noisy measurements of observable states \cite{reich_2015_dabook, markasch_2016_dabook, vanleeuwen_2019_pf_highdimension, kovachki_kalman_inversion, smoothers_1, smoothers_2}. 
DA algorithms are broadly divided into variational, ensemble-based, and hybrid formulations.
Variational methodologies (e.g., 3D-Var and 4D-Var)  \cite{rabier_2000_ecmwf_4dvar, wang_2010_economical_4dvar} leverage an optimal control framework to identify the optimal states or parameters that minimize the mismatch between model predictions and observational data.
In contrast, ensemble-based frameworks (such as ensemble Kalman and particle filters \cite{evensen_1994_enkf, evensen_2003_enkf, leeuwen_pf_paper_1})  employ Monte Carlo simulations to approximate error statistics via a finite set of random state and/or parameter realizations.
Hybrid techniques, such as 4DENVar \cite{liu_2009_4DEnVar}, combines the advantages of both approaches; notably, they circumvent the mathematically rigorous and computationally intensive requirement of developing tangent linear and adjoint models.

Particle filters are a standard family of algorithms for sequential data assimilation in highly nonlinear, non-gaussian, multi-modal settings \cite{leeuwen_pf_paper_1, pf_vanLeeuwen2010, pf_vanLeeuwen2015}.
The bootstrap particle filter employs importance sampling technique where the particles are weighted by the observation likelihood.
When observations are highly informative and/or the dimension of the system is too high, 
the importance sampling weights get concentrated on a few particles (one particle in the worst case), leading to filter degeneracy of the\cite{Synder_ObstaclestoHighDimensionalParticleFiltering, daum_degeneracy_1}. 
In \cite{etpf_reich_2013}, the Ensemble Transport Particle Filter (ETPF) was proposed based on a discrete optimal transport problem; ETPF suffers from a similar degeneracy issue in higher dimensions. 
Several approaches have been considered to address the degeneracy problem \cite{highdimensions_pf_1, highdimensions_pf_2, highdimensions_pf_3, vanleeuwen_2019_pf_highdimension}.
Particle flow based filters  \cite{daum_log_homotopy, daum_exactflow, daum_2017_generalized_gromov_filter, gromov_pal_1, daum_generalized_pf_nonlinear_filter, zanetti_andrey_survey} have shown promise in alleviating particle filter degeneracy. 
A flow in phase space ``migrates'' the particles from prior to posterior distribution following a certain drift law (see \cref{subsec:particle_flow}). 
The weights remain constant throughout the evolution under the flow,  which removes the weight degeneracy issue. However, the flow can move particles to sample a distribution that is different from the sought-after posterior. 

Sampling based DA algorithms need a large number of model runs, which can quickly become computationally prohibitive when the underlying numerical model is complex.
For this reason, we consider multi-fidelity approaches where the information is systematically combined across a hierarchy of models with varying levels of complexity and computational costs.
The low fidelity models, which are computationally inexpensive but inaccurate and potentially biased, capture the bulk of the statistical information.
High-fidelity models, which are costly but accurate and potentially unbiased, are strategically used to correct the bias and reduce the variance of the estimator.
This coupling of models of varying complexity achieves the accuracy of a full high-fidelity simulation at a fraction of the total compute cost. 

In the context of multi-fidelity simulation of dynamical system, a Full Order Model (FOM) refers to high-dimensional discretization of the physical system that accurately simulates the governing equations (which are often partial differential equations).
In a sampling based DA methods, the evaluation of multiple FOMs constitutes the primary computational bottleneck.
Reduced Order Models (ROMs) \cite{rom_book_1, Brunton_Kutz_2019, rom_non_linear_nguyen} address this by projecting the FOM onto a low-dimensional subspace that retains the dominant dynamical behavior, thereby serving as the inexpensive low-fidelity surrogate in the multi-fidelity hierarchy.
Following \cite{Sandu_2021_MF-EnKF}, we construct this reduced subspace via Proper Orthogonal Decomposition (POD), which extracts the most energetically significant modes from snapshots of a FOM simulation.
ROMs have been used in inverse problems \cite{rom_inverse_problems_1, rom_inverse_problems_2, rom_inverse_problems_3},
for efficient and accurate numerical simulation of nonlinear problems \cite{park_mou_liu_sandu_illescu_2026_two_level_ROM}, and to accelerate data assimilation techniques in both variational \cite{4dvar_rom_1, 4dvar_rom_2, Sandu_2015_POD-inverse-problems} and ensemble based \cite{mlenkf,Sandu_2016_reducedSampling4DVar,Sandu_2026_model-forest-EnKF} approaches.

Multi-level Monte Carlo (MLMC) was first introduced by Giles \cite{giles2015multilevel} as a variance reduction technique for Monte Carlo simulations.
Rather than relying solely on single high fidelity simulation, MLMC exploits a hierarchy of models with varying levels of accuracy and computational cost. 
We leverage a large ensemble of inexpensive simulations (low cost, low accuracy) to capture the bulk of the statistical variation, complemented by a small number of expensive high-fidelity model runs to ensure accuracy and reduce bias.
This strategic combination of models of varying accuracy/complexity leads to better estimates (reduced variance) and faster convergence compared to standard Monte Carlo methods.
Inexpensive low-fidelity models can take several forms, including simplified physical approximations, reduced-order models, data-fit surrogates, or Machine Learning (ML) models.
While this framework generalizes naturally to a full hierarchical multi-level setting \cite{Sandu_2021_MF-EnKF}, we restrict our experimental results to the bi-fidelity case in this study.

Several works in the data assimilation community have been influenced by MLMC methodology. 
Hoel et al. \cite{mlenkf} proposed a multilevel Kalman filter in which multilevel statistical moments are estimated using a telescoping sum of empirical moments across a hierarchy of model fidelities (see eq. 2.12--2.19 in \cite{mlenkf}).
However, as indicated in \cite{mlenkf, Sandu_2021_MF-EnKF}, it can lead to nonpositive multilevel covariance leading to stability issues in calculation of Kalman Gain. 
Popov et al. \cite{Sandu_2021_MF-EnKF} improved upon that using control variate theory, where the multilevel empirical covariances are always nonnegative. 
Gregory et al. \cite{mletpf} proposed a multilevel particle filter based on optimal transport \cite{etpf_reich_2013}. 
An optimization based on rearrangement is used to couple the coarse- and fine-level ensembles while preserving their correlation, which is essential to achieve variance reduction in the multilevel estimator (see eq. 4.2--4.10 in \cite{mletpf}).
However, this approach assumes identical ensemble sizes at the fine and coarse levels. 
This assumption is not suitable for our setting, where full-order model (FOM) evaluations are significantly more expensive than their reduced-order counterparts.
We refer interested readers  to several follow-up studies in this domain \cite{popov2026learning, mlenkf_2,attia_smoothers_mf}.

A multilevel particle filtering approach based particle flows has not yet been unexplored to the best of our knowledge. 
This work proposes a novel Multi-Fidelity Particle Flow Filter (MFPFF), which integrates multi-fidelity data assimilation with a particular particle flow based on log homotopy, multivariate control variates \cite{multivariate_control_variate}, and ROM theory.
By formulating these particle flow algorithms within a linear control variate framework, we derive a bi-fidelity architecture where a limited ensemble of expensive FOM simulations is augmented by a significantly larger, computationally cheaper ensemble of ROM simulations with lower accuracy.
The ROMs act as the low-fidelity in this bi-fidelity setup.
The statistical uncertainty inherent in the coarser ROM level serves as a ``control variate'' for the finer FOM level, enabling a superior posterior estimate with minimal additional computational overhead.
While the proposed methodology is presented in a bi-fidelity context, the framework is extensible to a generalized multi-fidelity setting. 
We utilize ``optimal'' projection and interpolation operators to map across fidelities/resolutions.
Statistical analysis demonstrates that the proposed MFPFF achieves a significant reduction in estimator variance compared to standard single-fidelity filters.

This work brings the following novel contributions.
\begin{itemize}[topsep=3pt,itemsep=3pt,parsep=0pt,partopsep=0pt,leftmargin=4ex]
    \item A novel multi-fidelity particle flow filter (MFPFF)  based on the homotopy formulation of particle flows is proposed.
    It extends \cite{Sandu_2021_MF-EnKF} to particle flows.
    \item Two distinct classes of particle flows are investigated, namely the deterministic Exact Flow \cite{daum_exactflow} and the stochastic Gromov Flow \cite{daum_2017_generalized_gromov_filter, gromov_duam_2, gromov_pal_1}.
    \item Localization and nonlinear observations are explored in the multi-fidelity particle flow paradigm. 
\end{itemize}


The remainder of the paper is organized as follows. \Cref{sec:background} reviews basic data assimilation methods, particle flow filters, POD ROMs and control variate theory. 
\Cref{sec:methodology} derives the Gromov and Exact flow filter in a bi-fidelity setting and lists the algorithm.
We also dive into a statistical analysis of MFPFF in this section.
\Cref{sec:numerical_experiments} tests the newly proposed method in two test cases.
Localization and nonlinear observations are  explored in this study. 
\Cref{sec:conclusion} concludes the study with remaining questions and possible future research directions.

%% file: sections/background.tex
\section{Background}
\label{sec:background}

\renewcommand{\Re}{\symbb{R}}

\subsection{Data Assimilation}
\label{sec:DA}

    Data Assimilation involves uncertainty in both the model states and observations. 
    We consider the states to be random variables denoted by $\bfX \in \symbb{R}^{n}$, where $\bfx$ is a realization of $\bfX$.
    Similarly, we consider observations as random variables denoted by $ \bfY \in \symbb{R}^{m} $, where $\bfy$ is a realization of $\bfY$ .
    With a slight abuse of notation, we will use the realization and the random variable interchangeably.
    The nonlinear model operator $\NLModel_{k-1,k} : \symbb{R}^{n} \to \symbb{R}^{n} $ propagates the state (as a Markov process) at discrete time ``$k-1$'' to a future discrete time ``$k$'' as:
    \begin{equation}
    \label{eq:model_prop}
        \bfX_{k} = \mathcal{M}_{k-1,k} \left(\bfX_{k-1} \right) + \tau_{k-1},
    \end{equation}
    where $\tau_{k-1} \sim N(\mathbf{0}, \mathcal{Q}_{k-1})$ is the  model noise. 
    In this study, we consider a completely deterministic model (i.e., $ \mathcal{Q}_{k-1} = 0 $), often referred to as a strong constraint in variational settings \cite{evensen2022_book} .
    The observation at time $k$ is based around the latent ``true'' state and is denoted as:
    \begin{equation}\label{eq:observation}
        \bfY_{k} = \mathcal{H}_{k} (\bfX_{k}^{tr}) + \eta_{k},  \quad \eta_{k} \sim \mathcal{N} (\mathbf{0}, \bfR_{k}),
    \end{equation}
    where $\eta_{k}$ is the observation error at time $k$ , $\bfX_{k}^{tr} \in \mathbb{R}^{n}$ is the latent true state, and $\NLObs_{k}: \symbb{R}^{n} \to \symbb{R}^{m} $ is the nonlinear observation operator that maps from the state space to the observation space.
    In many practical applications, $\bfX_{}^{tr}$ is a latent variable which is unavailable; however, we have access to noisy sensor data which is equivalent to $\bfY$ in \cref{eq:observation}.
    The linearized observation operator is :
    \[
        \bfH_{} \coloneqq \frac{d \NLObs(\bfx)}{d \bfx} \bigg|_{\bfx = \bfx_{\text{current}}} ~ ,
    \]
    where $\bfx_{\text{current}}$ is the current state. 
    The observation errors are assumed to be independent and identically distributed (i.i.d.) over time and drawn from a Gaussian distribution, although the Gaussian assumption is not required in general.
    We also consider $\bfR \in \Re^{m \times m}$ to be a diagonal matrix; however, this assumption is not necessary.
    In most operational setups, observations are sparse, i.e. $m \ll n$.
    From here the observation likelihood at time $k$ under the assumption that observations at time $k$ depends only on the state at time $k$ is denoted as:     
    %
    \begin{equation}\label{eq:likelihood}
       \mathscr{P}(\bfY_{k} \mid \bfX_{k}) = \mathcal{N} \left(\mathcal{H}_{k} (\bfX_{k}),\ \bfR_{k} \right),
    \end{equation}
    where $\mathcal{H}_{k} (\bfX_{k})$ is the model predicted observables and $\bfY_{k}$ is given by \cref{eq:observation}.
    The prior probability (often referred to as background or forecast) at time $k$ is: 
    \begin{equation}\label{eq:prior}
        \mathcal{P}(\bfX_{k} \mid \bfY_{1:k-1}) = \int_{\Re^n} \mathcal{P}(\bfX_{k} \mid \bfX_{k-1}) \mathcal{P}(\bfX_{k-1} \mid \bfY_{1:k-1}) ~~ d\bfX_{k-1}, 
    \end{equation}
    where $\bfY_{1:k-1} = \{\bfY_{1}, \dots, \bfY_{k-1}\}$ are the sequence of observations/measurements from time $t_{1}$ up to, and including, time $t_{k-1}$.
    The distribution $\mathcal{P}(\bfX_{k-1} \mid \bfY_{1:k-1})$ denotes the probability of the current state given all previous observations, and $\mathcal{P}(\bfX_{k} \mid \bfX_{k-1})$ is the model transition probability density \cref{eq:model_prop}.
    The posterior probability (often called analysis) at time $k$ is: 
    \begin{equation}\label{eq:assimilation}
        \mathcal{P}(\bfX_{k} \mid \bfY_{1:k}) = \mathcal{P}(\bfY_{k} \mid \bfX_{k}) \int_{\Re^n} \mathcal{P}(\bfX_{k} \mid \bfX_{k-1})\, \mathcal{P}(\bfX_{k-1} \mid \bfY_{1:k-1}) \; d\bfX_{k-1},
    \end{equation}
    \Cref{eq:assimilation} is a direct application of Bayes' rule, which finds the posterior distribution by combining the prior and likelihood as follows:
    \begin{equation}\label{eq:bayes_rule}
        \mathcal{P}_{\bfX}^{a} (\bfx) =  \mathcal{P}_{\bfX \mid \bfY}^{f}(\bfx^{} \mid \bfy) = \dfrac{\mathcal{P}_{\bfY \mid \bfX}(\bfy \mid \bfx^{}) \mathcal{P}_{\bfX}^{f}(\bfx^{})}{\mathcal{P}_{\bfY}(\bfy)},
    \end{equation}
    where the denominator is a scaling factor and is often not computed/computationally intractable. 
    
    In the remainder of the paper, without loss of generality,  we will ignore the subscripts unless necessary.
    In ensemble-based data assimilation, multiple realizations of a random variable $\bfX$ are represented by the ensemble matrix $\bfx^{f}\in\symbb{R}^{n\times N}$, where $N$ is the number of ensemble members (or particles), and the superscript ``$f$'' denotes the forecast ensemble. 
    Similarly, the analysis ensembles are represented by the matrix $\bfx^{a} \in \symbb{R}^{n \times N} $. 
    Each ensembles in $\bfx^{f}$ and $\bfx^{a}$ are considered independent samples from the distributions $\mathcal{P}^{f} (\bfx)$ and $\mathcal{P}^{a} (\bfx)$ respectively.
    Interchangeably, we will also use $\symrm{\bfE}_{\bfX}$ as the ensemble matrix.

\subsection{Particle Flow Filters}
\label{subsec:particle_flow}
     
    Particle flows evolve particles in state space according to the following stochastic differential equation:
    \begin{equation}
        \label{eq:ito_sde}
        d\, \bfx_\lambda = f(\bfx_\lambda, \lambda)\, d\lambda + g(\bfx_\lambda, \lambda)\,dW_\lambda,         \quad \lambda \ge 0.
    \end{equation}
    where $\lambda$ is synthetic time, $f$ is the drift term, $g$ the diffusion term, and $dW_\lambda$ is a standard Wiener process. The probability density of the particles in state space evolves from $\bfx_0 \sim \mathcal{P}_0(\bfx)$ to $\bfx_\lambda \sim \mathcal{P}_\lambda(\bfx)$ according to the Fokker-Planck equation:  
    \begin{equation}
        \label{eq:Fokker-Planck}
        \dfrac{\partial \mathcal{P}}{\partial \lambda} = - \nabla \cdot (f \mathcal{P}) + \dfrac{1}{2} \nabla \cdot [\bfQ \nabla \mathcal{P}] , 
        \quad 0 \le \lambda \le \Lambda,
        \quad \mathcal{P}(\bfx,\lambda=0) = \mathcal{P}_0(\bfx),
    \end{equation}
    where $\bfQ = g(\bfx_{\lambda}, \lambda) g(\bfx_{\lambda}, \lambda)^{T} $ is the diffusion coefficient , and $\mathcal{P}$ is the probability density.
    The drift and diffusion terms in \cref{eq:ito_sde} are chosen such that the prior distribution at initial time $\mathcal{P}_0(\bfx) = \mathcal{P}^{f}(\bfx)$ evolves to the posterior distribution at the final time $\Lambda$, $\mathcal{P}_\Lambda(\bfx) = \mathcal{P}^{a}(\bfx)$. Specifically, the drift drives the particles from the prior towards the posterior distribution, whereas the diffusion keeps the finite particles spread out and avoids particle collapse \cite{Sandu_2025_VariationalFokkerPlanck}.
    One approach uses $\Lambda = 1$ \cite{reich_gaussian_mixture_ensemble_transform_filter, daum_2017_generalized_gromov_filter, highdimensions_pf_3}, whereas a second approach moves particles from prior to posterior in the pseudo-time limit, i.e. $\Lambda \to \infty $.
    The Gromov flow \cite{daum_2017_generalized_gromov_filter} and Exact flow \cite{daum_exactflow} fall in the first category whereas Stein Variational gradient descent \cite{stein_variational_gradient_descent}, mapping particle filters \cite{pulido_leeuwen_mapping_particle_filter}, Fokker-Planck based filters \cite{reich_fokker_planck_particle_filter, Sandu_2025_VariationalFokkerPlanck} fall into the latter. The implementation moves each particle (prior sample) from time zero to final time via \cref{eq:ito_sde}, where the particles form samples from the posterior distribution.
    
    Several forms of this idea appeared in the literature. Liu et al. found a connection between Stein's discrepancy and the gradient of the Kullback-Leibler (KL) divergence and proposed a general variational inference algorithm \cite{stein_variational_gradient_descent}.
    The idea is to progressively minimize the KL divergence between the current distribution and the posterior/target distribution.
    A closed form solution to the KL divergence gradient was proposed by embedding the particles in Reproducing Kernel Hilbert Space (RKHS), where the kernel has to be chosen meticulously.
    Pulido et al. \cite{pulido_leeuwen_mapping_particle_filter} extended this work to data assimilation setting.
    Embedding in RKHS methodology was used in \cite{highdimensions_pf_3} to propose a Particle Flow Filter (PFF).
    PFF uses a matrix-valued kernel in place of a scalar kernel, which avoids the variance collapse problem.
    Subrahmanya et al. \cite{Sandu_2025_VariationalFokkerPlanck} proposed a particle flow based on Variational Fokker-Planck (VFP) framework that derived an optimal drift based on the selected diffusion term. This framework generalizes the earlier particle flow algorithms \cite{stein_variational_gradient_descent, pulido_leeuwen_mapping_particle_filter, highdimensions_pf_3}. 
    
    Following \cite{gromov_duam_2}, we consider here a particular set of flow filters that work on the time horizon $0 \le \lambda \le 1$ and form homotopies that smoothly transform  prior particles to equally weighted samples from the posterior. The solution of \cref{eq:Fokker-Planck} is
    \begin{equation}        \label{eq:homotopy_cond}
        \mathcal{P}_\lambda (\bfx) = \dfrac{\mathcal{P}_{\bfX}(\bfx)\,\mathcal{P}_{\bfY \mid \bfX}(\bfy \mid \bfx)^{\lambda}}{\mathcal{P}_{\bfY}(\lambda,\bfy)}, \quad \mathcal{P}_{\bfY}(\lambda,\bfy) = \int_{\Re^n} \mathcal{P}_{\bfY \mid \bfX}(\bfy \mid \bfx)^{\lambda}\,\mathcal{P}_{\bfX}(\bfx)\; d\bfx,
    \end{equation}
    such that
    \[
        \mathcal{P}_0(\bfx) = \mathcal{P}_{\bfX}(\bfx), \quad \mathcal{P}_1(\bfx) = \mathcal{P}_{\bfX \mid \bfY}(\bfx \mid \bfy).
    \] 
    The  logarithm of the conditional probability density function \cref{eq:homotopy_cond} is :
    \begin{equation}
        \label{eq:homotopy_log_cond}
        \log \mathcal{P}_\lambda (\bfx) = \log \mathcal{P}_{\bfX}(\bfx) + \lambda\, \log \mathcal{P}_{\bfY \mid \bfX}(\bfy \mid \bfx) - \log \mathcal{P}_{\bfY}(\lambda,\bfy). 
    \end{equation}
    The last term is the normalization constant, which is often intractable to compute and is generally avoided in particle flow algorithms \cite{daum_normalization_const}.
    As $\lambda$ varies from $0$ to $1$ the conditional probability density function in \cref{eq:homotopy_log_cond} gradually incorporates the observation likelihood.
     
    The following particle flow filters \cite{gromov_duam_2} define the drift and diffusion terms in \cref{eq:ito_sde} such that the Fokker-Planck equation \eqref{eq:Fokker-Planck} has a solution that matches the homotopy \eqref{eq:homotopy_cond}.

\subsubsection{Exact Flow}
\label{subsubsec:exact_flow}
    Daum \cite{daum_exactflow} proposed a particle flow filter in which the Bayesian update is implemented in the form of a particle flow instead of point-wise multiplication of prior and likelihood.    
    
    Consider filtering  \cref{eq:likelihood,eq:prior,eq:assimilation} in the Gaussian linear case.
    The noisy observation (or equivalently a sensor data) under a linear assumption is based around the latent truth $\bfx_{}^{tr}$ and is represented as follows:
    \begin{equation}
    \label{eq:exact_observation_linear}
        \bfy_{k} = \bfH_{k} \,\bfx_{k}^{tr} + \eta_{k}, ~~ \text{where} ~~ \eta_{k} \sim \mathcal{N} (\mathbf{0}, \bfR_{k}) \ ,
    \end{equation}
    where $\eta_{k} \sim \mathcal{N} (\mathbf{0}, \bfR_{k})$ and $\bfH_{k}$ is the linearized observation operator.
    At time step $k$, the prior \eqref{eq:prior} and likelihood \eqref{eq:likelihood} are, respectively:
    \[
    \mathcal{P}_{\bfX}(\bfx_k) = \mathcal{N}(\bfx_k \mid \overline{\bfx}_{k}, \bfP_{k}), \quad
    \mathcal{P}_{\bfY | \bfX}(\bfy_k \mid \bfx_k) = \mathcal{N}(\bfy_k \mid \bfH_{k}\,\bfx_{k}, \bfR_{k}).
    \]
    The Exact flow algorithm takes a differential equation \eqref{eq:ito_sde} with a linear drift term and a zero diffusion term:
    \begin{equation}
    \label{eq:exact_drift}
        f(\bfx, \lambda) = \bfA(\lambda)\, \bfx + \bfb(\lambda), \qquad g(\bfx, \lambda) = 0.
    \end{equation}
    The expression for the drift is derived using \cref{eq:homotopy_log_cond} and \cref{eq:Fokker-Planck} and considering zero noise. 
    We list the resulting values of $\bfA$ and $\bfb$ here and refer the readers to the excellent derivation \cite{crouse_flow_filter_bias}:
    \begin{equation}
    \label{eq:general_flow_exact}
        \begin{split}
            \frac{d \bfx_{k,\lambda}^{i}}{d \lambda} &=
            \bfA(\lambda)\, \bfx_{k,\lambda}^{i} + \bfb(\lambda) , \\
            \bfA(\lambda) 
            &=
            - (1 \slash 2) \, \bfP_{k}\, \bfH_{k}^{T}\, \left(\lambda\, \bfH_{k} \,\bfP_{k}\, \bfH_{k}^{T} + \bfR_{k}\right)^{-1}\, \bfH_{k} , \\
            \bfb(\lambda) &= (\bfI_{n} + 2 \lambda \bfA)\, \left[(\bfI_{n} + \lambda\, \bfA)\, \bfP_{k}\,\bfH_{k}^{T}\, \bfR_{k}^{-1}\, \mathbf{\mathscr{Y}}_{k} + \bfA \,\overline{\bfx}_{k} \right],
            \quad \overline{\bfx}_{k} = \frac{1}{N} \sum\limits_{i=1}^{N} \bfx_{k}^{i},
        \end{split}
    \end{equation}
    where $ \overline{\bfx}_{k} $ is the ensemble mean state, $\lambda$ is the synthetic time, $\bfP_{k}$ is the prior ensemble covariance matrix, $\bfR_{k}$ is the observation error covariance, and $\mathscr{Y}_{k}$ is the observation at the current filtering step $k$: 
    \[
        \mathscr{Y}_{k} = \begin{cases}
        \bfy_{k} \quad \cref{eq:exact_observation_linear}, & \text{linear observation operator}, \\
        \bfy_{k} - (\NLObs(\bfx_{k}) - \bfH_{k}\,\bfx_{k}), & \text{nonlinear observation operator}.
        \end{cases} 
     \]
     %
     
    In the case of the linear observation operator, $\mathscr{Y}_{k} = \bfy_{k}$ is given by \cref{eq:exact_observation_linear} . If the observation is nonlinear, the measurement linearization is subtracted and the observation becomes
    $\mathscr{Y}_{k} = \bfy_{k} - (\NLObs(\bfx_{k}) - \bfH_{k}(\bfx_{k}))$.

    The eigenvalues of $\bfA(\lambda)$ are zero or negative, owing to the assumption of positive definiteness of $\bfR$ and $\bfP$ matrices.
    Thus the flow defined by \cref{eq:general_flow_exact} is considered Schur stable \cite{daum_exactflow}.    
     
    The synthetic time interval $[0,\Lambda]$ is divided into $p$ smaller pseudo-steps  of size $\Delta \lambda  = \Lambda/p$. 
    Particle $i$ state at assimilation step $k$ is evolved numerically via standard Euler steps:
    \begin{equation}
        \label{eq:exact_flow_update}
        \bfx_{k, \lambda_{j+1}}^{i} = \bfx_{k, \lambda_{j}}^{i} + \Delta \lambda \,
        \left(\bfA(\lambda_{j})\, \bfx_{k,\lambda}^{i} + \bfb(\lambda_{j})\right).
    \end{equation}
    More sophisticated time stepping methods are left for a future study.

\subsubsection{Gromov Flow}
\label{subsubsec:gromov_flow}
    Daum\cite{daum_generalized_pf_nonlinear_filter}  proposed a stochastic filter which improves the accuracy of the state estimation as well as the uncertainty quantification, and is considered unbiased \cite{crouse_flow_filter_bias}.
    We consider the diffusion in \cref{eq:ito_sde,eq:Fokker-Planck} to be non-zero and follow the derivation given in \cite{crouse_flow_filter_bias}, which gives the following expression for a linear Gaussian assumption: 
    %
    \begin{equation}\label{eq:gromov_flow_drift_noise}
        \begin{split}
            f(\bfx, \, \lambda) &= \left(\bfP_{k}^{-1} + \lambda \, \bfH_{k}^{T} \, \bfR_{k}^{-1} \, \bfH_{k}\right) ^{-1} \ \bfH_{k}^{T} \, \bfR_{k}^{-1} \, \left(\bfy_{k} - \NLObs_{}(\bfx) \right), \\
            \bfQ (\lambda) &= \left(\bfP_{k}^{-1} + \lambda \, \bfH_{k}^{T} \, \bfR_{k}^{-1} \, \bfH_{k} \right) \bfH_{k}^{T} \, \bfR_{k}^{-1} \, \bfH_{k} \, \left(\bfP_{k}^{-1} + \lambda \, \bfH_{k}^{T} \, \bfR_{k}^{-1} \, \bfH_{k}\right), 
        \end{split}
    \end{equation}
    where $\bfP_{k}$ is the prior covariance at the start of the pseudo-time at the current assimilation step $k$, $f(\bfx, \lambda)$ is the drift and $\bfQ(\lambda)$ is the diffusion coefficient as introduced in \cref{eq:ito_sde,eq:Fokker-Planck}. 
    Interested readers can refer to \cite{crouse_flow_filter_bias} for a detailed derivation. 
    We start the flow with an initial condition of $\bfx_{\lambda={0}}^{i} = \bfx_{k}^{i} (1 \leq i \leq N) $. 
    The update for particle $i$ using a Euler-Maruyama method can be formulated as:
    %
    \begin{equation}
    \label{eq:gromov_flow}
        \symbf{x}_{k, \lambda_{j+1}}^{i} = \symbf{x}_{k, \lambda_{j}}^{i} + \Delta \lambda \, f(\symbf{x}_{k, \lambda}^{i}, \lambda_{j}) + \sqrt{\Delta \lambda} \ \symbf{\nu}_{}^{i} (\lambda_{j}), ~~\text{where} ~~ \symbf{\nu}_{}^{i} (\lambda_{j}) \sim \mathcal{N} (\mathbf{0}, \symbf{Q}(\lambda_{j})).
    \end{equation}
    However, inversion of $\bfP_{k}$ can be very expensive in high-dimensional problems. It can be avoided by modifying \cref{eq:gromov_flow_drift_noise} using Sherman-Morrison Woodbury identity as follows: 
    %
    \begin{equation}
    \label{eq:gromov_flow_transformed}
        \begin{split}
            \widetilde{\bfK}_{k}(\lambda) &= \bfP_{k}\,\bfH_{k}^{T}\,\left(\bfH_{k}\,\bfP_{k}\,\bfH_{k}^{T} + \frac{1}{\lambda}\,\bfR_{k}\right)^{-1}, \\
            f(\bfx, \lambda) &= \left(\bfI_{n} - \widetilde{\bfK}_{k}(\lambda)\,\bfH_{k}\right)\, \bfP_{k}\, \bfH_{k}^{T}\, \bfR_{k}^{-1}\, (\bfy_{k} - \NLObs_{}(\bfx)), \\
            \bfQ (\lambda) &= \left(\bfI_{n} - \widetilde{\bfK}_{k}(\lambda)\,\bfH_{k}\right)\,\bfP_{k}\,\bfH_{k}^{T}\,\bfR_{k}^{-1}\,\bfH_{k}\,\left(\bfI_{n} - \widetilde{\bfK}_{k}(\lambda)\,\bfH_{k}\right)\,\bfP_{k},
        \end{split}
    \end{equation}
    %
    where $\widetilde{\bfK}_{k}(\lambda)$ is the modified Kalman gain with a scaled observation error covariance. Intuitively, the filter ``trusts'' the prior/current information more in the initial stages of the particle flow, and incorporates the observation information more as the flow progresses. 
    Finding the square root of $\bfQ (\lambda) \in \symbb{R}^{n \times n} $ can be expensive in high-dimensional setting, so we can approximate the square root as  $(\bfI_{n} - \widetilde{\bfK}_{k}(\lambda) \, \bfH_{k} ) \, \bfP_{k} \, \bfH_{k}^{T} \, \bfR_{k}^{-1/2} \times \eta$, where $\eta \sim \mathcal{N}(\bf0, \bfI_{m})$  as proposed in \cite{crouse_flow_filter_bias}. 
    In order to avoid numerical issues when  $ \lambda = 0 $, we can further manipulate \cref{eq:gromov_flow_transformed} 
    to get the following expression for the Gromov flow \cite{daum_2017_generalized_gromov_filter}: 
    %
    \begin{equation}
    \label{eq:gromov_flow_nodivision_zero}
        \begin{split}
            \widetilde{\bfK}_{k}(\lambda) &= \bfP_{k} \, \bfH_{k}^{T} \, \left(\lambda \bfH_{k} \bfP_{k} \bfH_{k}^{T} + \bfR_{k}\right)^{-1}, \\
            f(\bfx, \lambda) &= \left(\bfI_{n} - \lambda \, \widetilde{\bfK}_{k}(\lambda) \, \bfH_{k} \right) \bfP_{k} \, \bfH_{k}^{T} \, \bfR_{k}^{-1} \, \left( \bfy_{k} - \NLObs_{}(\bfx) \right), \\
            \bfQ (\lambda) &= \left(\bfI_{n} - \lambda \, \widetilde{\bfK}_{k}(\lambda) \, \bfH_{k} \right) \, \bfP_{k} \, \bfH_{k}^{T} \, \bfR_{k}^{-1} \, \bfH_{k} \, \left(\bfI_{n} - \lambda \, \widetilde{\bfK}_{k}(\lambda) \, \bfH_{k} \right) \, \bfP_{k} ~. 
        \end{split}
    \end{equation}
    %
\subsection{Linear Control Variates}\label{subsec:linear_control_variate} 
    Control variate is a variance reduction technique used in Monte Carlo simulations. 
    We will adopt the notation used in \cite{Sandu_2021_MF-EnKF} with some minor modifications.
    Consider a random variable $ \chi$ with support $ \Omega_{\chi} \in \symbb{R}^{n} $ whose mean is represented as $ \bfmu_{\chi} $. 
    We call this a ``principal variate''.
    We have another random variable $ \widehat{v}$ with support $ \Omega_{\widehat{v}} \in \symbb{R}^{r} $, which is a ``control variate'' .
    The control variate is highly correlated with the original random variable $\chi$.  
    We define a new random variable $\mathcal{T} = \chi - \mathcal{S} (\widehat{v} - \symbb{E}[\widehat{v}]) $, where $ \mathcal{S} \in \symbb{R}^{n \times r}$ is a gain matrix. 
    We can call this a ``total variate'' which has the same mean as the principal variate (i.e., it is an unbiased estimator). 
    We seek to find an optimal matrix $\mathcal{S}$ that minimizes the variance of the total variate.
    By simple optimization, we derive the optimal value of $\mathcal{S}$ as: 
    \begin{equation}\label{eq:optimal_S}
        \mathcal{S}_{\rm opt} = \bfSigma_{\chi \widehat{v}} \, \bfSigma_{\widehat{v} \hat{v}}^{-1} \, ,
    \end{equation}
    where we use the short notation $\bfSigma_{AB} = \mathrm{Cov}(A,B)$.
    Putting the optimal $\mathcal{S}$ back into the expression of the total variate and after some algebraic manipulation, the variance becomes:
    \begin{equation}
       \operatorname{Var}(\mathcal{T}) = \bfSigma_{\chi \chi} - \bfSigma_{\chi \widehat{v}} \, \bfSigma_{\widehat{v} \widehat{v}}^{-1} \, \bfSigma_{\widehat{v} \chi},
    \end{equation}
    where we consider the matrix $\bfSigma_{\hat{v} \hat{v}}$ to be full rank.
    This shows a reduction in variance of the total variate.
    However, the variance of the total variate is not reduced if $\chi$ and $\hat{v}$ are not correlated, i.e., $\bfSigma_{\chi \hat{v}} = \mathbf{0}$.
    Next, we consider an ancillary variate ``$v$'' as a proxy for the mean of $\hat{v}$ (i.e., $\mu_{\hat{v}} = \mu_{v}$).
    This variate is considered independent of the primary and control variate.
    Following this assumption and the same algebraic manipulations as above, the optimal value of $\mathcal{S}$ and $\operatorname{Var} (\mathcal{T})$ are:
    \begin{equation}
    \label{eq:s_opt_val_var}
        \mathcal{S}_{\rm opt} = \bfSigma_{\chi \widehat{v}}  (\bfSigma_{\widehat{v} \widehat{v}} + \bfSigma_{v v})^{-1} ~~, ~~
        \operatorname{Var}(\mathcal{T}) = \bfSigma_{\chi \chi} - \bfSigma_{\chi \widehat{v}} (\bfSigma_{\widehat{v} \widehat{v}} + \bfSigma_{v v})^{-1} \bfSigma_{\widehat{v} \chi}
    \end{equation}
    The optimal value is $\mathcal{S}_{\rm opt} = 0.5 ~ \bfSigma_{\chi \widehat{v}} \bfSigma_{\hat{v} \hat{v}}^{-1} $ when both the first and second moment of the ancillary variate matches the control variate. 
    In many applications, the actual distribution of the random variables are hardly available or known. 
    We approximate them using samples to estimate the desired statistics.
    The empirical mean and covariance for a random variable $\chi$ represented by an ensemble $\mathrm{\bfE}_\chi$ of $N$ realizations are:
    \begin{equation}
    \label{eq:ens_representation}
    \begin{split}
        \mathrm{\bfE}_\chi &= \left[ \chi^{1}, \, \chi^{2}, \dots, \, \chi^{N} \right], \quad   \widetilde{\bfmu}_{\chi} = \dfrac{1}{N} \mathrm{\bfE}_\chi \bfone_{N}, \\
        \widetilde{\bfA}_{\chi} &= \dfrac{1}{\sqrt{N-1}} \left(\chi  - \widetilde{\bfmu}_{\chi} \bfone_{N}^{T} \right), \quad 
        \widetilde{\bfSigma}_{\chi \chi} = \widetilde{\bfA}_{\chi} \widetilde{\bfA}_{\chi}^{T},
    \end{split}
    \end{equation}
    where $\mathrm{\bfE}_{\chi} \in \symbb{R}^{n \times N} $ is the ensemble matrix, $\widetilde{\bfmu}_{\chi} \in \symbb{R}^{n}$ is the empirical mean of the ensembles,  $\widetilde{\bfA}_{\chi} \in \symbb{R}^{n \times N}$ denotes the ensemble anomaly, $\widetilde{\bfSigma}_{\chi \chi} \in \symbb{R}^{n \times n}$ denotes empirical covariance and $\bfone_{N}$ is N-dimensional vector of ones.
    The empirical cross covariances can be found out in a similar way which can be used in \cref{eq:s_opt_val_var}.
    The ``inversion'' in the optimal gain in \cref{eq:s_opt_val_var} can be ill-conditioned due to undersampling of the empirical covariance.
    We need to use an optimal gain matrix that is immune to such sampling errors. 
    \begin{remark}
    \label{rem:multilevel_cv}
        This idea can be extended to multiple levels, where we choose the ancillary variate in the current level as our primary in the next level of the hierarchy.
        We keep repeating until we have the desired number of levels.
        Let $\chi$ be denoted as $u_{0}$, $\widehat{v}$ as $\widehat{v}_{1}$ and $v$ as $v_{1}$.
        The total variate at this level is $\mathcal{T}_{1} = u_{0} - \mathcal{S}_{1} (\widehat{v}_{1} - v_{1}) $.
        In the next level, the ancilary variate in the previous level becomes the primary variate. 
        Let us denote $v_{1}$ as $u_{1}$ in the second level. 
        The corresponding control and ancillary variate are $\widehat{v}_{2}$ and $v_{2}$ repectively. 
        At the $\ell$-th level, the variates are denoted as  $u_{\ell-1}$, $\widehat{v}_{\ell}$ and $v_{\ell}$ respectively. 
        The corresponding total variate at level $\ell$ is  $\mathcal{T}_{\ell} = u_{\ell-1} - S_{\ell} (\widehat{v}_{\ell} - v_{\ell} )$.
        The total variate in the first level can be written as:
        \[
        \mathcal{T}_{1} = \chi - \sum\limits_{\ell=1}^{\mathcal{L}} \widetilde{S}_{\ell} \left( \widehat{v}_{\ell} - v_{\ell} \right) \quad \text{with} \quad \widetilde{S}_{\ell} = \prod\limits_{i=1}^{\ell} S_{i}.
        \]
    \end{remark}
    
\subsection{Reduced Order Modeling}
\label{subsec:rom}
    Proper Orthogonal Decomposition (POD) is one of the most successful dimensional reduction techniques used for complex spatio-temporal systems often represented by partial differential equations (PDEs) \cite{pod_Holmes_Lumley_Berkooz_1996}. 
    Discretizing these PDEs in space and time typically yields high-dimensional systems that are expensive to simulate. POD extracts the most informative modes, which are optimal in the mean-square sense and can approximate the high-dimensional system to a desired level of accuracy. 
    The optimal modes, obtained by method of snapshots \cite{method_of_snapshot} and SVD decomposition, are used to project the dynamics onto a reduced space by Galerkin projection.
    The result is a low-dimensional system that serves as a computationally inexpensive surrogate for the underlying PDE. 
    We provide a quick review of the method here; however, we refer the readers to \cite{Brunton_Kutz_2019} for an 
    excellent overview. 
    In the generalized form, let us consider a PDE of the form: 
    %
    \[
    \frac{d}{dt} \bfu(x, t) = \bfL \bfu(x, t) + \bfN(\bfu(x,t)),
    \]
    where $\bfL$ is some linear operator  and $\bfN$ represents the nonlinear terms. 
    
    Suppose we have snapshots of the dynamics at a fixed interval which are represented as $\bfXi = \left[ \bfxi_{1}, \, \bfxi_{2}, ~~ \dots ~~ \bfxi_{s}  \right]$, where $\bfXi \in \symbb{R}^{n \times s}$. 
    Each column $\bfxi_{i}$ is a measurement from a numerical simulation or experiment.
    The number of snapshots is significantly less than the dimensions, i.e. $s \ll n$, resulting in a tall, thin matrix.
    We proceed to take the SVD decomposition of the snapshot matrix $\bfXi = \mathbf{U} \symbf{\Sigma} \mathbf{V}^{T}$, where $ \mathbf{U} \in \symbb{R}^{n \times n}, \mathbf{V} \in \symbb{R}^{s \times s} $ and $\symbf{\Sigma} \in \symbb{R}^{n \times s}$.
    We truncate the singular vectors with the first ``r'' (most) informative modes to approximate $\bfXi$. 
    In short, $\widetilde{\bfXi} = \widetilde{\mathbf{U}} \widetilde{\symbf{\Sigma}} \widetilde{\mathbf{V}}^{T} $ where $ \norm{\bfXi - \widetilde{\bfXi}} < \epsilon $. 
    The optimal orthogonal modes of interest can be derived from the truncated left singular vector $ \mathbf{U} $, where $ \widetilde{\mathbf{U}} = \bfPhi = \left[ \bfphi_{1}, \bfphi_{2}, ~ \dots ~ \bfphi_{r} \right] $, and ``r'' is the number of dominant modes ($ r \ll s $). 
    The states can be approximated using the ansatz $\bfu(x,t) \approx \bfPhi(x) \mathbf{a}(t) $, where $ \mathbf{a(t)} $ are the time dependent coefficients. 
    Substituting it back to the generalized PDE form and taking advantage of orthogonality property of the optimal basis vectors, we get:
    \begin{equation}
        \frac{d}{dt} \mathbf{a}(t) = \bfPhi^{T} \bfL \bfPhi \mathbf{a}(t) + \bfPhi^{T} \bfN(\bfPhi \mathbf{a}(t)), 
    \end{equation}
    where $\bfa(t)$ denotes the states in lower-dimensional space.
    \begin{remark} \label{rem:deim}
        We can resort to more advanced approximate techniques, for example DEIM \cite{deim}, to handle the nonlinear projection terms in PDEs. 
        DEIM is a discrete extension of the \textit{Empirical Interpolation} method proposed  in \cite{empirical_interpolation}.
        DEIM avoids the reconstruction of the reduced dimension to the full space by choosing some ``optimal/magic'' interpolated points for nonlinear function evaluation. 
        This becomes inherently important when the full state dimensions are very high, resulting in an improved computational complexity. 
    \end{remark}

%% file: sections/methodology.tex

\section{Methodology}
\label{sec:methodology}

    \input{concept_mf_flow_filter}

    We focus on constructing a multi-fidelity framework building upon the approach of \cite{Sandu_2021_MF-EnKF}.
    However, this work focuses on two classes of particle flow algorithms, as discussed in \Cref{subsec:particle_flow}.
    A two-level fidelity framework is described in detail in this section  and shown in \Cref{fig:mfpff_schematic}, while an extension to the multi-level setting will be briefly discussed later.
    We assume access to a FOM simulation that propagates particles (or ensembles) in the full-order state space (i.e., $ \symrm{S}_{}^{F} =  \symbb{R}^{n}$) via the FOM dynamics $\NLModel^{F}$. 
    In addition, we assume access to a computationally inexpensive ROM surrogate that approximates the FOM dynamics in the reduced space (i.e., $ \symrm{S}_{}^{R} =  \symbb{R}^{r}$) via the ROM dynamics $\NLModel^{R}$.
    We assume these spaces i.e., $\symrm{S}^{F}$ and $\symrm{S}_{}^{R}$ are endowed with canonical Euclidean inner product.
    We choose and have access to an optimal projection operator $\bfPhi \in \symbb{R}_{}^{n \times r}$ 
    (refer Theorem 3.1 in \cite{Sandu_2021_MF-EnKF}), where $\bfPhi^{\ast} \bfPhi = \bfI_{r} $.
    This optimal operator $\bfPhi$ is problem dependent and will be explained in more details in \cref{sec:numerical_experiments}.
    It is, however, constant throughout the assimilation steps considering we do not change the ROM (refer Theorem 3.1 in \cite{mfenkf_popov} and the assumption therein).
    Suppose $\bfX$ is a random variable that encompasses our knowledge of the states in the FOM space and will be referred to as a principal variate. 
    Similarly, $\bfU$ is a random variable that embodies our knowledge of the states in the ROM space and will be referred to as an ancillary variate.
    The control variate 
    $\widehat{\bfU}$, which is highly correlated with the principal variate $\bfX$, is in the ROM space. 
    The projection from the full space to the reduced space and vise-versa is performed using the operator $\bfPhi$ as:
    \begin{equation}
    \label{eq:projection_fom_rom}
        \bfX = \bfPhi \bfU ,\quad \bfU = \bfPhi^{\ast} \bfX,
    \end{equation}
    where $\bfPhi^{\ast}$ is the adjoint of the mapping operator $\bfPhi$. 

    Let $\NLObs(\cdot) : \symbb{R}^{n} \to \symbb{R}^{m}$ and $\NLObs_{r}(\cdot) : \symbb{R}^{r} \to \symbb{R}^{m}$ denote the potentially nonlinear observation operators mapping the FOM and ROM state spaces, respectively, into the observation space. Following \cite{Sandu_2021_MF-EnKF}, we assume these observation operators, along with their linearized counterparts, satisfy the following consistency condition:
    \begin{equation}
    \label{eq:observation_operator}
        \NLObs_{r} (\bfU) \approx \NLObs (\bfPhi \bfU) , \quad \bfH_{r}(\bfU) \approx \bfH \bfPhi \bfU.
    \end{equation}
    We define a total variate (using the optimal gain matrix) which incorporates the principal, control, and ancillary variate as (refer Remark 5 in \cite{Sandu_2021_MF-EnKF}):
    \begin{equation}
    \label{eq:total_variate}
        \bfZ = \bfX - \frac{\bfPhi}{2} \left( \widehat{\bfU} - \bfU \right),
    \end{equation}
    where $\bfPhi/2$ is the optimal gain matrix.
    Statistics are captured by finite samples of these random variables at the forecast and analysis step, i.e., $\bfX_{k}^{f} \in \symbb{R}^{n \times N}, \widehat{\bfU}_{k}^{f} \in \symbb{R}^{r \times N} $, $\bfU_{k}^{f} \in \symbb{R}^{r \times N_{r}}$ and $\bfX_{k}^{a} \in \symbb{R}^{n \times N}, \widehat{\bfU}_{k}^{a} \in \symbb{R}^{r \times N}, \bfU_{k}^{a} \in \symbb{R}^{r \times N_{r}} $.
    $N$ and $N_{r}$ are the numbers of particles/ensembles corresponding to principal/control and ancillary variate respectively.

\subsection{Multifidelity Forecast Step}
\label{subsec:methodology_forecast}
    To facilitate a strong correlation between the principal and control variate, we project the FOM particles to the ROM space at the assimilation step $k-1$. 
    The FOM and ROM dynamics propagate the analysis ensembles at time $k-1$ to time $k$ as follows:
    \begin{equation}
    \label{eq:forecast_ensembles}
        \begin{split}
        \bfX_{k}^{f[e]} &= \NLModel_{k-1,k}^{F}\left(\bfX_{k-1}^{a[e]}\right), \quad 
        \widehat{\bfU}_{k}^{f[e]} = \NLModel_{k-1,k}^{R} \left(\widehat{\bfU}_{k-1}^{a[e]}\right), \quad 
        \quad 1 \leq e \leq N, \\
        \bfU_{k}^{f[e]} &= \NLModel_{k-1,k}^{R} \left(\bfU_{k-1}^{a[e]}\right) \quad \quad 1\leq e \leq N_{r} .
        \end{split}
    \end{equation}
    We assume that the control variate and the ancillary variate are independent in our setup. 
    The analysis step (explained next) would remove the independence assumption, therefore we discard the control variate after each assimilation step.
    However, the ancillary and principal variate will still have correlation after the analysis step.
    We use the typical Kalman gain independence assumption and consider that $\widehat{\bfU}$ and $\bfX$ remain independent throughout our sequential assimilation process (refer Remark 9 in \cite{Sandu_2021_MF-EnKF}).

\subsection{Multifidelity Analysis Step}
\label{subsec:methodology_analysis}
    The analysis step transforms the samples in the FOM space and the ROM space via a particle flow using the cross covariance information across the principal, control and ancillary variate at the assimilation step $k$.
    Taking the expected value of \cref{eq:total_variate}, the forecast and analysis empirical mean and deviation of the total variate are: 
    \begin{equation}
    \label{eq:mean_total_variate}
        \begin{split}
            \widetilde{\bfmu}_{\bfZ_{k}^{f}} &= \widetilde{\bfmu}_{\bfX_{k}^{f}} - \frac{1}{2} \bfPhi_{k} \left( \widetilde{\bfmu}_{\widehat{\bfU}_{k}^{f}} - \widetilde{\bfmu}_{\bfU_{k}^{f}} \right), \quad
            \widetilde{\bfmu}_{\bfZ_{k}^{a}} = \widetilde{\bfmu}_{\bfX_{k}^{a}} - \frac{1}{2} \bfPhi_{k} \left( \widetilde{\bfmu}_{\widehat{\bfU}_{k}^{a}} - \widetilde{\bfmu}_{\bfU_{k}^{a}} \right), \\
            \bfA_{\bfZ_{k}^{f}} &= \frac{1}{\sqrt{N-1}} \left[ \left( \bfX_{k}^{f} - \widetilde{\bfmu}_{\bfX_{k}^{f}} \right) - \frac{1}{2} \bfPhi_{k} \left( \widehat{\bfU}_{k}^{f} -  \widetilde{\bfmu}_{\widehat{\bfU}_{k}^{f}} \right) + \frac{1}{2} \bfPhi_{k} \left( \bfU_{k}^{f} -  \widetilde{\bfmu}_{\bfU_{k}^{f}}\right) \right], \\
            \bfA_{\bfZ_{k}^{a}} &= \frac{1}{\sqrt{N-1}} \left[ \left( \bfX_{k}^{a} - \widetilde{\bfmu}_{\bfX_{k}^{a}} \right) - \frac{1}{2} \bfPhi_{k} \left( \widehat{\bfU}_{k}^{a} -  \widetilde{\bfmu}_{\widehat{\bfU}_{k}^{a}} \right) + \frac{1}{2} \bfPhi_{k} \left( \bfU_{k}^{a} -  \widetilde{\bfmu}_{\bfU_{k}^{a}}\right) \right].
        \end{split}
    \end{equation}
    The nonlinear ``indirect'' observation operator applied on the total forecast variate and considering the assumption \cref{eq:observation_operator}, we get:
    \begin{equation}\label{eq:non_lin_obs_total_variate}
        \begin{split}
            \NLObs \left(\bfZ_{k}^{f} \right) &= \NLObs \left(\bfX_{k}^{f}\right) - \frac{1}{2} \NLObs \left(\bfPhi_{k} \widehat{\bfU}_{k}^{f}\right) + \frac{1}{2} \NLObs \left(\bfPhi_{k} \bfU_{k}^{f}\right), \\
            \bfmu_{\NLObs \left( \bfZ_{k}^{f} \right)} &= \bfmu_{\NLObs \left( \bfX_{k}^{f} \right)} - \frac{1}{2} \bfmu_{\NLObs \left( \bfPhi_{k} \widehat{\bfU}_{k}^{f} \right)} + \frac{1}{2} \bfmu_{\NLObs \left( \bfPhi_{k} \bfU_{k}^{f} \right)}, \\
            \bfA_{\NLObs \left( \bfZ_{k}^{f} \right)} &= \frac{1}{\sqrt{N-1}} \left[ \left( \NLObs (\bfX_{k}^{f}) -  \bfmu_{\NLObs \left( \bfX_{k}^{f} \right)} \right) - \frac{1}{2} \left( \NLObs \left( \bfPhi_{k} \widehat{\bfU}_{k}^{f} \right) - \bfmu_{\NLObs \left(\bfPhi_{k} \widehat{\bfU}_{k}^{f} \right)} \right)  + \frac{1}{2} \left( \NLObs \left(\bfPhi_{k} \bfU_{k}^{f} \right) - \bfmu_{\NLObs \left(\bfPhi_{k} \bfU_{k}^{f} \right)} \right) \right].
        \end{split}
    \end{equation}
    The cross variance of the total variate can be derived from \cref{eq:total_variate}, by taking relevant expectations and independence assumptions as:
    \begin{equation}
    \label{eq:cross_cov_total_variate}
        \begin{split}
            \widetilde{\bfSigma}_{\bfZ_{k}^{f}, \, \NLObs (\bfZ_{k}^{f})} 
            &=
            \symbb{E} \left[\left(\bfZ_{k}^{f} - \bfmu_{\bfZ_{k}^{f}}^{}\right) \left(\NLObs \left(\bfZ_{k}^{f} \right) - \bfmu_{\NLObs(\bfZ_{k}^{f})}\right)^{T} \right]  \\
            &= \widetilde{\bfSigma}_{\bfX_{k}^{f}, \, \NLObs \left(\bfX_{k}^{f} \right)} + \frac{1}{4} \bfPhi \widetilde{\bfSigma}_{\widehat{\bfU}_{k}^{f}, \, \NLObs \left(\bfPhi \widehat{\bfU}_{k}^{f} \right)} - \frac{1}{2} \widetilde{\bfSigma}_{\bfX_{k}^{f}, \, \NLObs \left(\bfPhi \widehat{\bfU}_{k}^{f} \right)} - \frac{1}{2} \bfPhi \widetilde{\bfSigma}_{\widehat{\bfU}_{k}^{f}, \, \NLObs \left( \bfX_{k}^{f} \right)} \\ & \quad+ \frac{1}{4} \bfPhi \widetilde{\bfSigma}_{\bfU_{k}^{f}, \, \NLObs \left(\bfPhi \bfU_{k}^{f} \right)} \, , \\
            \widetilde{\bfSigma}_{\NLObs (\bfZ_{k}^{f}), \, \NLObs \left( \bfZ_{k}^{f} \right)} 
            &=
            \symbb{E} \left[\left( \NLObs(\bfZ_{k}^{f}) - \bfmu_{\NLObs(\bfZ_{k}^{f})} \right) \left(\NLObs \left(\bfZ_{k}^{f} \right) - \bfmu_{\NLObs(\bfZ_{k}^{f})}\right)^{T} \right] \\
            &= \widetilde{\bfSigma}_{\NLObs \left(\bfX_{k}^{f} \right), \, \NLObs  \left( \bfX_{k}^{f} \right)} + \frac{1}{4} \widetilde{\bfSigma}_{\NLObs \left(\bfPhi \widehat{\bfU}_{k}^{f} \right), \, \NLObs \left( \bfPhi \widehat{\bfU}_{k}^{f} \right)} -\frac{1}{2} \widetilde{\bfSigma}_{\NLObs \left( \bfX_{k}^{f} \right), \, \NLObs \left( \bfPhi \widehat{\bfU}_{k}^{f} \right)} -\frac{1}{2} \widetilde{\bfSigma}_{\NLObs \left( \bfPhi \hat{\bfU}_{k}^{f} \right), \, \NLObs \left( \bfX_{k}^{f} \right)}  \\   & \quad + \frac{1}{4} \widetilde{\bfSigma}_{\NLObs \left( \bfPhi \bfU_{k}^{f} \right), \, \NLObs \left( \bfPhi \bfU_{k}^{f} \right)} \, ,
        \end{split}
    \end{equation}
    where we use \cref{eq:cross_cov_total_variate} to calculate the Kalman gain matrix in the pseudo-step of Gromov and Exact flow as:
    \begin{equation}
    \label{eq:kalman_gain_tot_var}
       \widetilde{\bfK} = \widetilde{\bfSigma}_{\bfZ_{k}^{f}, \, \NLObs \left( \bfZ_{k}^{f} \right)} \left( \widetilde{\bfSigma}_{\NLObs \left( \bfZ_{k}^{f} \right), \, \NLObs \left( \bfZ_{k}^{f} \right)} + \bfR_{k} \right)^{-1} .
    \end{equation}
    The updated Gromov flow using \cref{eq:kalman_gain_tot_var} is:
    \begin{equation}
    \label{eq:gromov_non_linear}
        \begin{split}
            \widetilde{\bfK}_{k} (\lambda) &= \widetilde{\bfSigma}_{\bfZ_{k}^{f}, \,  \NLObs \left( \bfZ_{k}^{f} \right)} \left( \lambda \, \widetilde{\bfSigma}_{\NLObs \left( \bfZ_{k}^{f} \right), \, \NLObs \left( \bfZ_{k}^{f} \right)} + \bfR_{k} \right)^{-1}, \\
            f(\bfx, \lambda) 
            &=
            \widetilde{\bfSigma}_{\bfZ_{k}^{f}, \, \NLObs \left( \bfZ_{k}^{f} \right)} \bfR_{k}^{-1}(\bfy - \NLObs(\bfx)) - \lambda \,  \widetilde{\bfK} (\lambda) \, \widetilde{\bfSigma}_{\NLObs \left( \bfZ_{k}^{f} \right), \, \NLObs \left( \bfZ_{k}^{f} \right)} \bfR_{k}^{-1} (\bfy - \NLObs(\bfx)) \, , \\
            \bfQ^{\frac{1}{2}}(\lambda) 
            &=
            \widetilde{\bfSigma}_{\bfZ_{k}^{f}, \, \NLObs \left( \bfZ_{k}^{f} \right)} \bfR_{k}^{- \frac{1}{2}}  \eta - \lambda \, \widetilde{\bfK} (\lambda) \,  \widetilde{\bfSigma}_{\NLObs \left( \bfZ_{k}^{f} \right), \NLObs \left( \bfZ_{k}^{f} \right)} \bfR_{k}^{-\frac{1}{2}}  \eta \quad \text{where} \quad \eta \sim \mathcal{N} (\bf0, \, \bfI_{m}),
        \end{split}
    \end{equation}
    where $\bfx$ is any generalized state in the FOM space.
    The Gromov flow in the pseudo-time follows \cref{eq:gromov_flow}.
    We use a scaled cross variance $ \left( i.e., \bfPhi^{\ast} \widetilde{\bfSigma}_{\bfZ_{k}^{f}, \, \NLObs \left( \bfZ_{k}^{f} \right)} \right) $ and an updated innovation vector $ \left( \bfy - \NLObs(\bfPhi \bfx) \right) $ in \cref{eq:gromov_non_linear} if $\bfx$ is in the ROM space.

    \noindent The updated Exact flow following \cref{eq:kalman_gain_tot_var} is: 
    %
    \begin{equation}\label{eq:exact_flow_non_linear_1}
        \begin{split}
            \bfA (\lambda) &= -0.5 \, \widetilde{\bfSigma}_{\bfZ_{k}^{f}, \,  \NLObs \left( \bfZ_{k}^{f} \right)} \left( \lambda \widetilde{\bfSigma}_{\NLObs \left( \bfZ_{k}^{f} \right), \, \NLObs \left( \bfZ_{k}^{f} \right)} + \bfR_{k} \right)^{-1} \bfH_{k} \ , \\
            \bfb (\lambda) &= (\bfI_{n} + 2 \lambda \bfA) \left[ (\bfI_{n} + \lambda \bfA) \widetilde{\bfSigma}_{\bfZ_{k}^{f}, \,  \NLObs \left( \bfZ_{k}^{f} \right)} \bfR_{k}^{-1} \symscr{Y}_{k} + \bfA \bar{\bfx}  \right] , 
        \end{split}
    \end{equation}
    where $\symscr{Y}_{k}$ is defined in \cref{subsubsec:exact_flow} and the flow follows \cref{eq:exact_drift}.
    The expression in \cref{eq:gromov_non_linear} can handle both linear and nonlinear observations. 
    The Exact flow in \cref{eq:exact_flow_non_linear_1} is valid for a linear observation, where $\NLObs(\bfZ) = \bfH \bfZ $. 
    The exact flow update in the bi-fidelity simulation following an ensemble approximation with nonlinear observations is:
    \begin{equation}
    \label{eq:exact_flow_non_linear_2}
        \begin{split}
            \bfA(\lambda) \overline{\bfx} 
            &=
            -0.5 \, \widetilde{\bfSigma}_{\bfZ_{k}^{f}, \, \NLObs \left( \bfZ_{k}^{f} \right)} \left( \lambda \widetilde{\bfSigma}_{\NLObs \left( \bfZ_{k}^{f} \right), \NLObs \left( \bfZ_{k}^{f} \right)} + \bfR_{k} \right)^{-1} \NLObs(\overline{\bfx}), \\
            q 
            &=
            \widetilde{\bfSigma}_{\bfZ_{k}^{f}, \, \NLObs \left( \bfZ_{k}^{f} \right)} \bfR_{k}^{-1} \symscr{Y}_{k} + \lambda \left( - 0.5 \, \widetilde{\bfSigma}_{\bfZ_{k}^{f}, \, \NLObs \left( \bfZ_{k}^{f} \right)} \left(\lambda \widetilde{\bfSigma}_{\NLObs \left( \bfZ_{k}^{f} \right), \, \NLObs (\bfZ_{k}^{f})} + \bfR_{k}\right)^{-1} \widetilde{\bfSigma}_{\NLObs \left( \bfZ_{k}^{f} \right), \, \NLObs \left( \bfZ_{k}^{f} \right)} \bfR_{k}^{-1} \symscr{Y}_{k}^{}  \right) \\
            & \quad + \bfA(\lambda) \,  \overline{\bfx}, \\
            \bfb(\lambda) &= q + 2 \lambda \left( -0.5 \,   \widetilde{\bfSigma}_{\bfZ_{k}^{f}, \,  \NLObs \left( \bfZ_{k}^{f} \right)} \left( \lambda \widetilde{\bfSigma}_{\NLObs \left( \bfZ_{k}^{f} \right), \, \NLObs \left( \bfZ_{k}^{f} \right)} + \bfR_{k} \right)^{-1} \NLObs(q) \right), \\
            \frac{d \bfx^{i}}{d \lambda} &= \bfA (\lambda) \bfx^{i} + \bfb (\lambda) , \\
            &= 
            -0.5 \widetilde{\bfSigma}_{\bfZ_{k}^{f}, \,  \NLObs \left( \bfZ_{k}^{f} \right) } \left( \lambda \widetilde{\bfSigma}_{\NLObs \left( \bfZ_{k}^{f} \right), \, \NLObs \left( \bfZ_{k}^{f} \right)} + \bfR_{k} \right)^{-1} \NLObs(\bfx^{i}) + \bfb (\lambda), \quad i =1,\dots, N
        \end{split}
    \end{equation}
    %
    We allow the particle flow algorithm to propagate the principal, control and ancillary variate from the prior distribution to the posterior independently (as denoted by lines $4 - 6$ in \cref{alg:mfpff}) following \cref{eq:exact_drift,eq:general_flow_exact} for the Exact flow and \cref{eq:gromov_flow_nodivision_zero,eq:gromov_flow} for the Gromov flow.
    The update in FOM and ROM space is similar except for the scaled cross covariance $\left(\bfPhi^{T} \widetilde{\bfSigma}_{\bfZ_{k}^{f}, \NLObs \left( \bfZ_{k}^{f} \right)}\right)$ and observation operator ($\NLObs_{r}$) in line $5-6$ of \cref{alg:mfpff}. 
    The cross covariance information, i.e., \cref{eq:cross_cov_total_variate},  captures information across the three variates. 
    The particle flow is followed by a bias correction step as indicated in lines $8-10$ of \Cref{alg:mfpff}.
    The control variate is discarded at this step and we repeat a projection step to build a new control variate which is highly correlated to the forecasted principal variate but remains independent of the ancillary variate. 

    \begin{algorithm}
        \caption{Multi-Fidelity Particle Flow Filter (MFPFF) with two-level fidelity 
        }
        \label{alg:mfpff}
        \begin{algorithmic}[1]
            \Require The forecast ensembles for the principal, ancillary and control variate at current assimilation step $k$ are $\mathrm{\bfE}_{\bfX_{k}^{f}}, \, \mathrm{\bfE}_{\bfU_{k}^{f}}, \, \mathrm{\bfE}_{\widehat{\bfU}_{k}^{f}}$ where $\mathrm{\bfE}_{\bfX_{k}}^{} = \left[  \bfx_{k}^{1}, \, \dots, \, \bfx_{k}^{N}  \right]$
            \Require The projection operator $\bfPhi$
            \Require Nonlinear and linear observation operators $\NLObs$ and $\bfH$ (or $\NLObs_{r}$ and $\bfH_{r} = \bfH \bfPhi$)
            \Require The current observation $\bfy$ and observation error covariance $\bfR$ (considered diagonal here)
            \Require $p$ is the number of pseudo-steps for the flow
            \Ensure Returns the analysis ensembles after one assimilation step
            \State  $\widetilde{\bfmu}_{\bfZ_{k}^{f}} = \widetilde{\bfmu}_{\bfX_{k}^{f}} - 0.5 \bfPhi \left( \widetilde{\bfmu}_{\widehat{\bfU}_{k}^{f}} - \widetilde{\bfmu}_{\bfU_{k}^{f}} \right) $ \quad Mean of total variate
            \State $ \widetilde{\bfSigma}_{\bfZ_{k}^{f}, \, \NLObs \left(\bfZ_{k}^{f} \right)} = \widetilde{\bfSigma}_{\bfX_{k}^{f}, \NLObs \left( \bfX_{k}^{f} \right)} + \ 0.25 \bfPhi \widetilde{\bfSigma}_{\widehat{\bfU}_{k}^{f}, \, \NLObs \left( \bfPhi \widehat{\bfU}_{k}^{f} \right)} - 0.5 \widetilde{\bfSigma}_{\bfX_{k}^{f}, \, \NLObs \left(\bfPhi \widehat{\bfU}_{k}^{f} \right)} - 0.5 \bfPhi \widetilde{\bfSigma}_{\widehat{\bfU}_{k}^{f}, \, \NLObs \left( \bfX_{k}^{f} \right)} + 0.25 \, \bfPhi \, \widetilde{\bfSigma}_{\bfU_{k}^{f}, \, \NLObs \left( \bfPhi \bfU_{k}^{f} \right) }  $
            \State $ \widetilde{\bfSigma}_{\NLObs \left( \bfZ_{k}^{f} \right), \, \NLObs \left( \bfZ_{k}^{f} \right)} = \widetilde{\bfSigma}_{\NLObs \left( \bfX_{k}^{f} \right), \, \NLObs \left( \bfX_{k}^{f} \right)} + 0.25 \widetilde{\bfSigma}_{\NLObs \left( \bfPhi \widehat{\bfU}_{k}^{f} \right), \, \NLObs \left( \bfPhi \widehat{\bfU}_{k}^{f} \right)} -0.5 \widetilde{\bfSigma}_{\NLObs \left( \bfX_{k}^{f} \right), \, \NLObs \left(\bfPhi \widehat{\bfU}_{k}^{f} \right)} -0.5 \widetilde{\bfSigma}_{\NLObs  \left( \bfPhi \widehat{\bfU}_{k}^{f} \right), \, \NLObs \left( \bfX_{k}^{f} \right)} + 0.25 \widetilde{\bfSigma}_{\NLObs \left( \bfPhi \bfU_{k}^{f} \right), \, \NLObs \left( \bfPhi \bfU_{k}^{f} \right)} $
            \State $ \mathrm{\bfE}_{\bfX_{k}^{a}} =$ Gromov/Exact $ \left( \mathrm{\bfE}_{\bfX_{k}^{f}} \, , \widetilde{\bfSigma}_{\bfZ_{k}^{f}, \, \NLObs \left( \bfZ_{k}^{f} \right)} \, , \widetilde{\bfSigma}_{\NLObs \left( \bfZ_{k}^{f} \right), \, \NLObs \left( \bfZ_{k}^{f} \right)} \, , \bfy, \NLObs, \bfH, \bfR, p \right) $ 
            \State $ \mathrm{\bfE}_{\widehat{\bfU}_{k}^{a}} = $ Gromov/Exact $ \left( \mathrm{\bfE}_{\widehat{\bfU}_{k}^{f}} \, , \bfPhi^{T} \widetilde{\bfSigma}_{\bfZ_{k}^{f}, \NLObs \left( \bfZ_{k}^{f} \right)} \, , \widetilde{\bfSigma}_{\NLObs \left( \bfZ_{k}^{f} \right), \, \NLObs\left( \bfZ_{k}^{f} \right)}, \bfy, \NLObs_{r}, \bfH \bfPhi, \bfR, p \right) $
            \State $ \mathrm{\bfE}_{\bfU_{k}^{a}} = $ Gromov/Exact $ \left( \mathrm{\bfE}_{\bfU_{k}^{f}} \, , \bfPhi^{T} \widetilde{\bfSigma}_{\bfZ_{k}^{f}, \, \NLObs \left( \bfZ_{k}^{f} \right)} \, , \widetilde{\bfSigma}_{\NLObs \left( \bfZ_{k}^{f} \right), \, \NLObs \left( \bfZ_{k}^{f} \right)} \, , \bfy, \NLObs_{r}, \bfH \bfPhi, \bfR, p \right) $
            \State $ \widetilde{\bfmu}_{\bfZ_{k}^{a}} = \widetilde{\bfmu}_{\bfX_{k}^{a}} - 0.5 \bfPhi \left( \widetilde{\bfmu}_{\widehat{\bfU}_{k}^{a}} - \widetilde{\bfmu}_{\bfU_{k}^{a}} \right) $
            \State $ \mathrm{\bfE}_{\bfX_{k}^{a}} = \mathrm{\bfE}_{\bfX_{k}^{a}} - \widetilde{\bfmu}_{\bfX_{a}^{k}} + \widetilde{\bfmu}_{\bfZ_{k}^{a}} $
            \State $ \mathrm{\bfE}_{\widehat{\bfU}_{k}^{a}} = \mathrm{\bfE}_{\widehat{\bfU}_{k}^{a}} - \widetilde{\bfmu}_{\widehat{\bfU}_{a}^{k}} + \widetilde{\bfmu}_{\bfZ_{k}^{a}} $ \quad we discard the control variate at this step
            \State $ \mathrm{\bfE}_{U_{k}^{a}} = \mathrm{\bfE}_{U_{k}^{a}} - \widetilde{\bfmu}_{U_{a}^{k}} + \widetilde{\bfmu}_{Z_{k}^{a}} $
        \end{algorithmic}
    \end{algorithm}
    \begin{algorithm}
        \caption{Gromov Flow Filter}
        \label{alg:gromov_flow}
        \begin{algorithmic}[1]
            \Require Nonlinear observation operator $\NLObs$(or $\NLObs_{r}$), linearized observation operator $\bfH$(or $\bfH \bfPhi$), ensembles at the current assimilation step $\bfx_{k}$ and observation error covariance $\bfR$ 
            \Require $p > 0$ pseudo-steps and $  \lambda_{j} = j \times \Delta \lambda $, where $0 \leq j \leq p$ and $\Delta \lambda = \frac{\Lambda}{p} $
            \Require Ensembles $\bfx_{k}^{f} \in \symbb{R}^{n \times N} $  $ \left( \text{or} \ \bfu_{k}^{f} \in \symbb{R}^{r \times N_{r}} \ \text{or} \ \widehat{\bfu}_{k}^{f} \in \symbb{R}_{}^{r \times N} \right)$  
            \Ensure Returns the analysis particles at the end of the flow
            \Procedure{Gromov}{$\mathrm{\bfE}_{\bfX_{k}^{f}} \, , \widetilde{\bfSigma}_{\bfZ_{k}^{f}, \, \NLObs \left( \bfZ_{k}^{f} \right)} \, , \widetilde{\bfSigma}_{\NLObs \left( \bfZ_{k}^{f} \right), \, \NLObs \left( \bfZ_{k}^{f} \right)} \, , \bfy, \NLObs, \bfH, \bfR, p$}
            \For{$j = 0$ to  $p$}
                \State $\lambda_{j}=j \ \times \Delta \lambda $
                \State $\widetilde{\bfK}_{k} (\lambda_{}) = \widetilde{\bfSigma}_{\bfZ_{k}^{f}, \,  \NLObs \left( \bfZ_{k}^{f} \right)} \left( \lambda_{} \, \widetilde{\bfSigma}_{\NLObs \left( \bfZ_{k}^{f} \right), \, \NLObs \left( \bfZ_{k}^{f} \right)} + \bfR_{k} \right)^{-1}$
                \State $f(\bfx, \lambda) 
                =                \widetilde{\bfSigma}_{\bfZ_{k}^{f}, \, \NLObs \left( \bfZ_{k}^{f} \right)} \bfR_{k}^{-1}(\bfy - \NLObs(\bfx)) - \lambda \,  \widetilde{\bfK} (\lambda) \, \widetilde{\bfSigma}_{\NLObs \left( \bfZ_{k}^{f} \right), \, \NLObs \left( \bfZ_{k}^{f} \right)} \bfR_{k}^{-1} (\bfy - \NLObs(\bfx))$
                \State $\bfQ^{\frac{1}{2}}(\lambda) 
                =                \widetilde{\bfSigma}_{\bfZ_{k}^{f}, \, \NLObs \left( \bfZ_{k}^{f} \right)} \bfR_{k}^{- \frac{1}{2}}  \eta - \lambda \, \widetilde{\bfK} (\lambda) \,  \widetilde{\bfSigma}_{\NLObs \left( \bfZ_{k}^{f} \right), \NLObs \left( \bfZ_{k}^{f} \right)} \bfR_{k}^{-\frac{1}{2}}  \eta \quad \text{where} \quad \eta \sim \mathcal{N} (\bf0, \, \bfI_{m})$ 
                \State $\bfx_{k, \lambda_{j+1}}^{i} = \bfx_{k, \lambda_{j}}^{i} + \Delta \lambda \, f(\bfx_{k, \lambda_{j}}^{i}, \lambda_{j}) + \sqrt{\Delta \lambda} \, \bfnu_{}^{i} (\lambda_{j}), ~\text{where} ~~ \bfnu_{}^{i} (\lambda_{j}) \sim \mathcal{N} (\mathbf{0}, \bfQ(\lambda_{j}))$ for particle $i$
            \EndFor
            \EndProcedure
        \end{algorithmic}
    \end{algorithm}
    \begin{algorithm}
        \caption{Exact Flow Filter}
        \label{alg:exact_flow_filter}
        \begin{algorithmic}[1]
            \Require Nonlinear observation operator $\NLObs$ (or $\NLObs_{r}$), linearized observation operator $\bfH$(or $\bfH \bfPhi$), ensembles at the current assimilation step $\bfx_{k}$ and observation error covariance $\bfR$
            \Require $p > 0$ pseudo-steps and $  \lambda_{j} = j \times \Delta \lambda $, where $0 \leq j \leq p$ and $\Delta \lambda = \frac{\Lambda}{p} $
            \Require Ensembles $\bfx_{k}^{f} \in \symbb{R}^{n \times N} $ $ \left( \text{or} \ \bfu_{k}^{f} \in \symbb{R}^{r \times N_{r}} \ \text{or} \ \widehat{\bfu}_{k}^{f} \in \symbb{R}_{}^{r \times N} \right)$  
            \Ensure Returns the analysis particles at the end of the flow
            \Procedure{Exact}{$\mathrm{\bfE}_{\bfX_{k}^{f}} \, , \widetilde{\bfSigma}_{\bfZ_{k}^{f}, \, \NLObs \left( \bfZ_{k}^{f} \right)} \, , \widetilde{\bfSigma}_{\NLObs \left( \bfZ_{k}^{f} \right), \, \NLObs \left( \bfZ_{k}^{f} \right)} \, , \bfy, \NLObs, \bfH, \bfR, p$}
            \For{$j = 0$ to  $p$}
                \State $\lambda_{j}=j \ \times \Delta \lambda $
                \State $\frac{d \bfx_{k, \lambda}^{i} (\lambda)}{d \lambda} = \bfA(\lambda) \bfx_{k, \lambda}^{i} + \bfb(\lambda)$
                \State $\bfA (\lambda) = -0.5 \, \widetilde{\bfSigma}_{\bfZ_{k}^{f}, \,  \NLObs \left( \bfZ_{k}^{f} \right)} \left( \lambda \, \widetilde{\bfSigma}_{\NLObs \left( \bfZ_{k}^{f} \right), \, \NLObs \left( \bfZ_{k}^{f} \right)} + \bfR_{k} \right)^{-1} \bfH_{k}$
                \State $\bfb (\lambda) = (\bfI_{n} + 2 \lambda \bfA) \left[ (\bfI_{n} + \lambda \bfA) \widetilde{\bfSigma}_{\bfZ_{k}^{f}, \,  \NLObs \left( \bfZ_{k}^{f} \right)} \bfR_{k}^{-1} \symscr{Y}_{k} + \bfA \bar{\bfx}  \right]$ where $\Bar{\bfx}_{k} = \frac{1}{N} \bfx_{k}^{f} \bfone_{N}^{}$ and $\symscr{Y}_{k}$ is the observation
                \State $\bfx_{k, \lambda_{j+1}}^{i} = \bfx_{k, \lambda_{j}}^{i} + \Delta \lambda \frac{d \bfx_{k, \lambda}^{i} (\lambda_{j})}{d \lambda}$
            \EndFor
            \EndProcedure
        \end{algorithmic}
    \end{algorithm}

\subsection{Statistical Analysis}
\label{subsec:statistical_analysis}
    We consider a simplified case where observations are linear. 
    We also consider the Gromov flow as a mapping from the prior to the posterior, where we have one pseudo-step, i.e., $p=1$.
    Under such assumptions, the update to the principal variate using the modified Kalman gain (refer \cref{eq:kalman_gain_tot_var}) is:
    \begin{equation}
    \label{eq:gromov_simplified_variate_update}
        \begin{split}
            \bfX_{k}^{a} 
            &=
            \bfX_{k}^{f} + f\left( \bfX_{k}^{f} \right) + \nu_{k}^{} \quad \text{where} ~~ \nu_{k} \sim \mathcal{N}(\bf0, \bfQ) \, ,\\
            \widetilde{\bfK} 
            &=
            \widetilde{\bfSigma}_{\bfZ_{k}^{f}, \, \NLObs \left( \bfZ_{k}^{f} \right)} \left( \widetilde{\bfSigma}_{\NLObs \left( \bfZ_{k}^{f} \right), \, \NLObs \left( \bfZ_{k}^{f} \right)} + \bfR \right)^{-1} , \\ 
            f\left(\bfX_{k}^{f}\right) 
            &=
            \widetilde{\bfSigma}_{\bfZ_{k}^{f}, \, \NLObs \left( \bfZ_{k}^{f} \right)} \bfR^{-1} \left( \bfy_{k} - \bfH \bfX_{k}^{f} \right) - \widetilde{\bfK}_{k} \widetilde{\bfSigma}_{\NLObs \left(\bfZ_{k}^{f}\right), \, \NLObs \left(\bfZ_{k}^{f} \right)} \bfR^{-1} \left( \bfy_{k} - \bfH \bfX_{k}^{f} \right) , \\    
            \bfQ^{\frac{1}{2}} 
            &=
            \widetilde{\bfSigma}_{\bfZ_{k}^{f}, \, \NLObs \left( \bfZ_{k}^{f} \right)} \bfR^{-\frac{1}{2}} - \widetilde{\bfK}_{k}   \widetilde{\bfSigma}_{\NLObs \left( \bfZ_{k}^{f} \right), \, \NLObs \left( \bfZ_{k}^{f} \right)} \bfR^{-\frac{1}{2}} , \\
            \implies \bfX_{k}^{a} 
            &=
            \bfX_{k}^{f} +  \widetilde{\bfSigma}_{\bfZ_{k}^{f}, \NLObs \left( \bfZ_{k}^{f} \right)} \bfR^{-1} \underbrace{\left( \bfy_{k} - \bfH \bfX_{k}^{f} + \bfR^{\frac{1}{2}} \eta \right)}_{\text{perturbed observations}} - \widetilde{\bfK}_{k}   \widetilde{\bfSigma}_{\NLObs \left( \bfZ_{k}^{f} \right), \, \NLObs \left( \bfZ_{k}^{f} \right)} \bfR_{}^{-1} \underbrace{\left( \bfy_{k} - \bfH \bfX_{k}^{f} + \bfR^{\frac{1}{2} } \eta \right) }_{\text{perturbed observations}} , \\
            &=
            \bfX_{k}^{f} + \left( \widetilde{\bfSigma}_{\bfZ_{k}^{f}, \, \NLObs \left( \bfZ_{k}^{f} \right)} - \widetilde{\bfK}_{k}   \widetilde{\bfSigma}_{\NLObs \left( \bfZ_{k}^{f} \right), \, \NLObs \left(\bfZ_{k}^{f} \right)} \right) \bfR^{-1} \left(  \bfy_{k} - \bfH \bfX_{k}^{f} + \bfR^{\frac{1}{2} } \eta \right), ~ \text{where} ~ \eta \sim \mathcal{N}(\bf0, \bfI_{m})
        \end{split}
    \end{equation}
    The empirical covariances can be unraveled as $\widetilde{\bfSigma}_{\bfZ_{k}^{f}, \NLObs (\bfZ_{k}^{f})} = \bfP_{\bfz} \bfH^{T}, \widetilde{\bfSigma}_{\NLObs(\bfZ_{k}^{f}), \NLObs(\bfZ_{k}^{f})} = \bfH_{k} \bfP_{\bfz} \bfH_{k}^{T} $ under our assumptions. 
    The last expression in \cref{eq:gromov_simplified_variate_update}, under the given assumptions, is equivalent to perturbed EnKF update.
    Following this logic, the updates in the control and ancillary variate are perturbed EnKF updates too. 
    We can refer to the Section $4.3$ of \cite{Sandu_2021_MF-EnKF}, and conclude that multi-fidelity particle flow with control variate framework helps in reduction of uncertainty and sampling errors under the given assumptions.

\subsection{Extension to Multiple Fidelity Levels}
\label{subsec:extension_multi_level}
    We can extend this bi-fidelity methodology to the multi-fidelity following \Cref{rem:multilevel_cv}.
    We assume access to $\mathcal{L}$ levels of model fidelity, where the cost and accuracy decrease as we go down the levels. 
    The ancillary variate at level $l-1$ acts as the principal variate at level $l$, where $1 \leq l \leq \mathcal{L} $ .
    We also assume access to optimal projection operator ($\widetilde{\bfPhi}_{l}$) at each level, where $\widetilde{\bfPhi}_{l} = \prod\limits_{i=1}^{l} \bfPhi_{i} $.
    Extending the total variate to $\mathcal{L}$ fidelities, we get:
    \begin{equation}
    \label{eq:multi-fidelity_cv_da}
        \bfZ = \bfX - \sum\limits_{l=1}^{\mathcal{L}} \frac{1}{2^{l}} \widetilde{\bfPhi}_{l} \left( \widehat{\bfU}_{l} - \bfU_{l} \right) .
    \end{equation}
    %

%% file: concept_mf_flow_filter.tex
\begin{figure}[htbp]
    \centering
    \resizebox{1.0\columnwidth}{!}{
    \begin{tikzpicture}

        \tikzset{
            large particle/.style={circle, fill=red!80, minimum size=1cm, inner sep=0pt},
            small particle/.style={circle, fill=blue!80, minimum size=0.25cm, inner sep=0pt},
            level label/.style={font=\footnotesize\bfseries, anchor=east, align=center},
            flow arrow/.style={->, thick, >=stealth, shorten >=2pt, shorten <=2pt},
            arrow label/.style={font=\footnotesize, midway},
            wavy/.style={thick, decoration={snake, amplitude=0.5mm, segment length=3mm}, decorate},
            background box/.style={draw=gray, fill=gray, opacity=0.1, text opacity=0.5, font=\Large\bfseries, rounded corners=5pt, inner sep=15pt}
        }


        \node at (0, 0.8) {$\mathbf{X}_{k-1}^{a}$};

        \node[large particle] (FOM1) at (-1.0, 0) {};
        \node[large particle] (FOM2) at (1, 0) {};
        \node[level label] at (-2.0, 0) {FOM model \\ space};

        \node[small particle] (FIR1) at (-1.0, -2.5) {};
        \node[small particle] (FIR2) at (1.0, -2.5) {};
        \node[level label] at (-2.0, -2.5) {Control Variate \\ (Projected)};

        \draw[flow arrow] (FOM1) -- (FIR1) node[arrow label, left] {};
        \draw[flow arrow] (FOM2) -- (FIR2) node[arrow label, right] {};

        \node at (0, -1.25) {\large $\Phi^{\ast}$};

        \node at (0, -3) {\(\widehat{\mathbf{U}}_{k-1}^{a}\)};

        \foreach \y in {-4.75, -5.5, -6.25} {
            \foreach \x in {-1.0, -0.33, 0.33, 1.0} {
                \node[small particle] at (\x, \y) {};
            }
        }
        
        \node[level label] at (-2.0, -5.125) {ROM model \\ space};

        \node at (0, -7.0) {\(\mathbf{U}_{k-1}^{a}\)};

        \node (T2) at (5, 0.8) {$\mathbf{X}_{k}^{f}$};


        \node[large particle] (FOM3) at (4.0, 0) {};
        \node[large particle] (FOM4) at (6, 0) {};
        \draw[flow arrow] (FOM2) -- (FOM3) node[arrow label, above] {\(\mathbf{M}_{k-1,k}^{F}\)};

        \node[small particle] (FIR3) at (4.0, -2.5) {};
        \node[small particle] (FIR4) at (6.0, -2.5) {};
        \draw[flow arrow] (FIR2) -- (FIR3) node[arrow label, above] {\(\mathbf{M}_{k-1,k}^{R}\)};

        \foreach \y in {-4.75, -5.5, -6.25} {
            \foreach \x in {4.0, 4.67, 5.33, 6.0} {
                \node[small particle] at (\x, \y) {};
            }
        }
        \draw[flow arrow] (1.2, -5.5) -- (3.8, -5.5) node[arrow label, above] {\(\mathbf{M}_{k-1,k}^{R}\)};

        \node at (5, -3) {\(\widehat{\mathbf{U}}_{k}^{f}\)};
        \node (B2) at (5, -7.0) {\(\mathbf{U}_{k}^{f}\)};


        \node (T3) at (10, 0.8) {$\mathbf{X}_{k}^{a}$};

        \node[large particle] (FOM5) at (9, 0) {};
        \node[large particle] (FOM6) at (11, 0) {};

        \draw[wavy, ->, >=stealth] (FOM4) -- (FOM5) node[midway, above=4pt, font=\large] {Flow Filter};

        \node[small particle] (FIR5) at (9.0, -2.5) {};
        \node[small particle] (FIR6) at (11.0, -2.5) {};

        \draw[wavy, ->, >=stealth] (FIR4) -- (FIR5) node[midway, above=4pt, font=\large] {Flow Filter};

        \node at (10, -3) {\(\widehat{\mathbf{U}}_{k}^{a}\)};

        \foreach \y in {-4.75, -5.5, -6.25} {
            \foreach \x in {9.0, 9.67, 10.33, 11.0} {
                \node[small particle] at (\x, \y) {};
            }
        }
        
        \draw[wavy, ->, >=stealth] (6.2, -5.5) -- (8.8, -5.5) node[midway, above=4pt, font=\large] {Flow Filter};

        \node (B3) at (10, -7.0) {\(\mathbf{U}_{k}^{a}\)};

        \begin{pgfonlayer}{background}
            \node[background box, fit=(FIR3) (FOM3) (FOM6) (FIR6) (T3) (B3)] (box) {};
            \node[font=\Large\bfseries, text opacity=0.5, anchor=north, yshift=-5pt] at (box.south) {Analysis using Particle Flow};
        \end{pgfonlayer}

        \node[fill=teal!10,draw,font=\large\bfseries] (Obs) at (7.5, 3.0) {\(\bfY_{k}\): Observation(Noisy and sparse)};
        \draw[flow arrow, very thick] (Obs) -- (box.north);

        \draw[flow arrow] (11.5, 0) -- (13.8, 0) node[arrow label, above] {Bias Correction};
        
        \draw[flow arrow] (11.2, -5.5) -- (13.8, -5.5) node[arrow label, above] {Bias Correction};
        
    \end{tikzpicture}
    }
    \caption{Schematic representation of the model hierarchy in the Multi-Fidelity Flow Filter.}
    \label{fig:mfpff_schematic}
\end{figure}
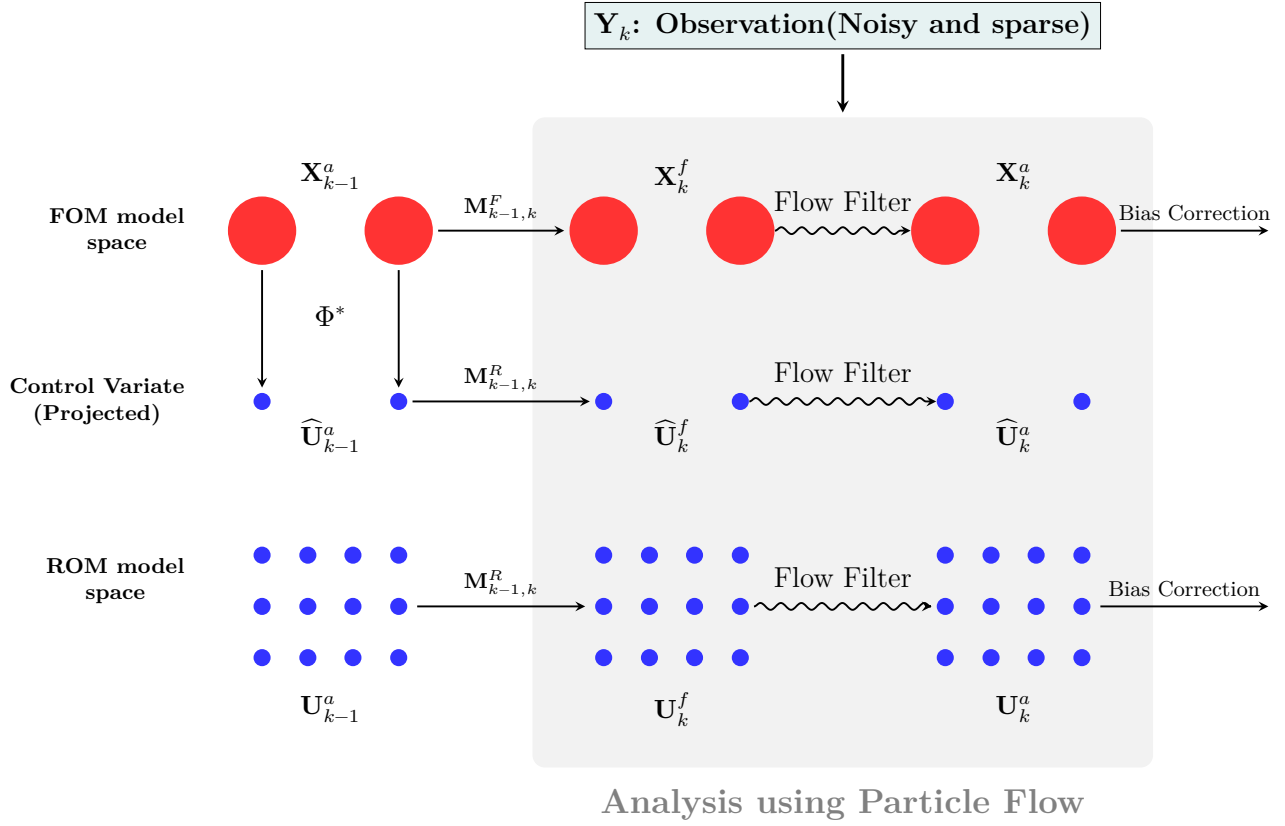

%% file: sections/numerical_experiments.tex
\section{Numerical Experiments}
\label{sec:numerical_experiments}
    We explore the newly proposed bi-fidelity filter and test its efficacy in two standard test problems.
    The first problem is the nonlinear advection dispersion KdV, where the initial condition is chosen such that two traveling solitons interact with each other.
    The barotropic vorticity QG equations \cite{Sandu_2021_MF-EnKF, stabilized_pod_qg_omer_trian} are widely used in the data assimilation and ROM literature, making them a suitable choice for our setup.
    We explore and extend covariance localization for the QG test problem.
    To avoid redundant results, we run each experiment with ten different initializations and report the averaged spatio-temporal Root Mean Square Error (RMSE). 
    We run twin experiments where the spatio-temporal RMSE between a ``truth'' and analysis trajectories for a single simulation is defined as:
    \begin{equation}
    \label{eq:rmse}
        \operatorname{RMSE} (\bfx_{1:T}^{tr}, \bfx_{1:T}^{a}) = \sqrt{ \dfrac{1}{N_{T}} \dfrac{1}{n} \sum\limits_{i=1}^{N_{T}}  \norm{\overline{\bfx}_{i}^{a} - \bfx_{i}^{tr}}_{2}^{2}
        } ~~,
    \end{equation}
    where $N_{T}$ is the total number of assimilation steps, $\bfx_{i}^{tr}$ is the truth at the i-th assimilation step, and $\overline{\bfx}_{i}^{a}$ is the analysis mean at that step.

    The failed experiments are represented by white patches in the heatmaps. 
    Any failed experiment implies that atleast one initialization has failed for a given combination of ensembles and covariance inflation. 
    We use intrusive POD-ROMs for both test problems, which serve as cheap  surrogates in our multi-fidelity setup. 
    We compare FOM and ROM runtimes to ensure a consistent comparison between the bi-fidelity and single-fidelity test cases.
    We consider that the principal variate encompasses the uncertainty in the FOM space, and the control variate and ancillary variate determine the uncertainty in the ROM space.
    To prevent the simulation from rejecting incoming observations, we apply covariance inflation to overestimate the background covariance/uncertainty.
    All experiments were carried out on MATLAB 2024b on a Linux workstation with AMD Ryzen Threadripper (16 physical cores) with $64$ GB of RAM.
    The code can be made available upon request.


\subsection{Korteweg-De Vries (KdV)}
\label{subsec:kdv}
    The one-dimensional KdV equation is a nonlinear hyperbolic PDE that is advection-dominated.
    Its solutions remain smooth for all times and are nondissipative, although the equation exhibits dispersive behavior.

    In this study, the chosen initial condition admits a two-soliton solution in which the solitons interact non-destructively with phase shifts.
    The governing PDE in the continuous form is:
    \begin{equation}
    \label{eq:kdv}
        \begin{aligned}
        \dfrac{\partial u (t,x)}{\partial t} + 2 \alpha u(t,x) \dfrac{\partial u(t,x)}{\partial x} &+ \nu \dfrac{\partial^3 u(t,x)}{\partial x^3}  = 0, \\
        -10  \leq x \leq 10, ~~  u(t=0,x) &= 6 \sech(x),
        \end{aligned}
    \end{equation}
    where $\alpha = 3$ and $ \nu=1 $ with periodic boundary condition. 
    For spatial discretization, we adopt the scheme from \cite{kdv_ascher, kdv_geometric_denys}, which conserves mass, momentum and energy.
    The resulting semi-discrete system is integrated in time using MATLAB's ode45 solver with adaptive step-size control, providing stable and accurate simulations over extended time horizons.
    The number of discrete points in space is $500$.
    This constitutes the FOM in this problem setup.
    Observations are available at a time interval of $0.01$, where we observe at every $20$ discrete points in the grid. 
    The observation error covariance is identity.
    We run the simulations for $350 $ assimilation steps, and we discard the first $50$ steps for spinups. 
    We make sure to use the ROM within the training period, as Galerkin POD ROM can diverge outside the training period \cite{opinf_qian_ionut_karen}.

\begin{figure}[htbp]
        \centering
        \begin{subfigure}[b]{0.45\textwidth}
            \includegraphics[width=\textwidth, alt={Line graph comparing the analytical solution and the 10-mode ROM solution, where the ROM solution exhibits severe oscillations, significant bias, and fails to track the analytial solution.}]{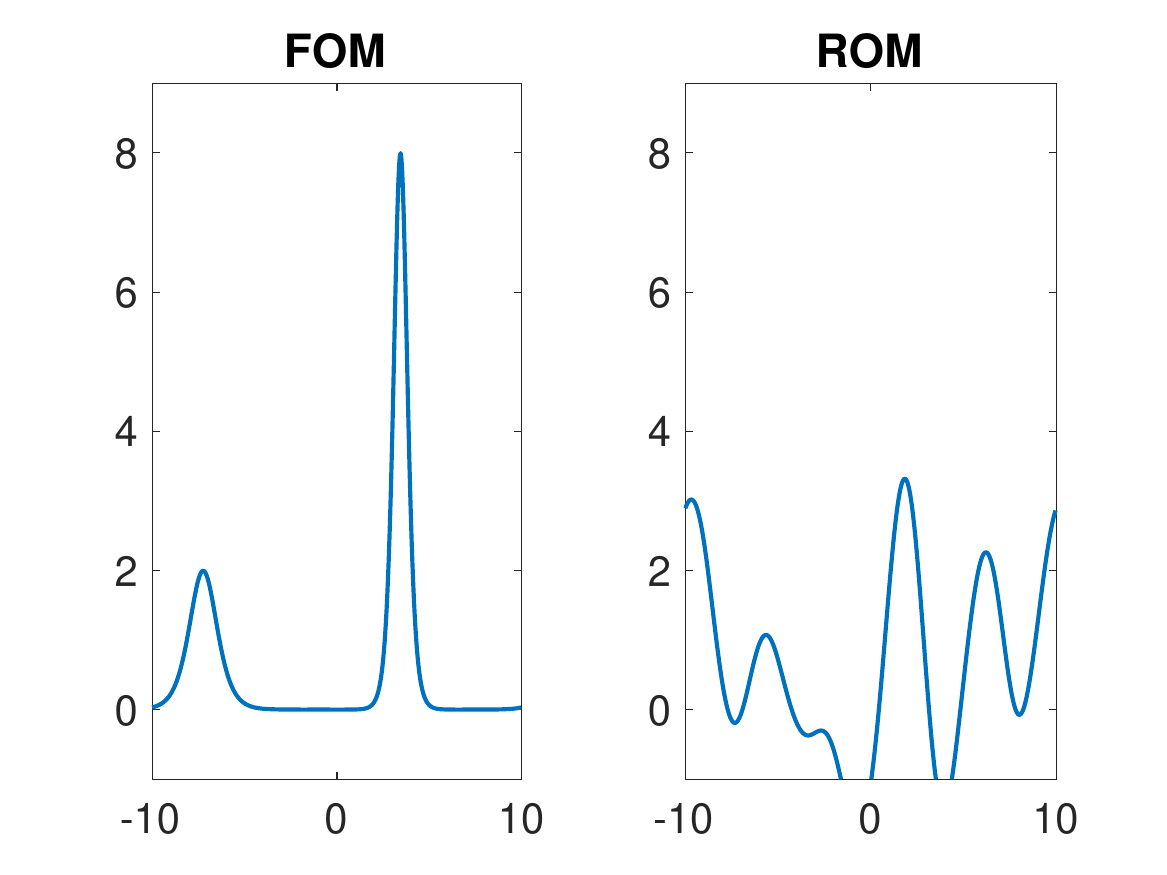}
            \caption{POD modes = 10, Relative energy = 0.7054}
            \label{subfig:kdv_fom_rom_10}
        \end{subfigure}
        \begin{subfigure}[b]{0.45\textwidth}
            \includegraphics[width=\textwidth, alt={Line graph comparing the analytical solution and the 20-mode ROM solution, where the ROM solution exhibits severe oscillations and significant bias.}]{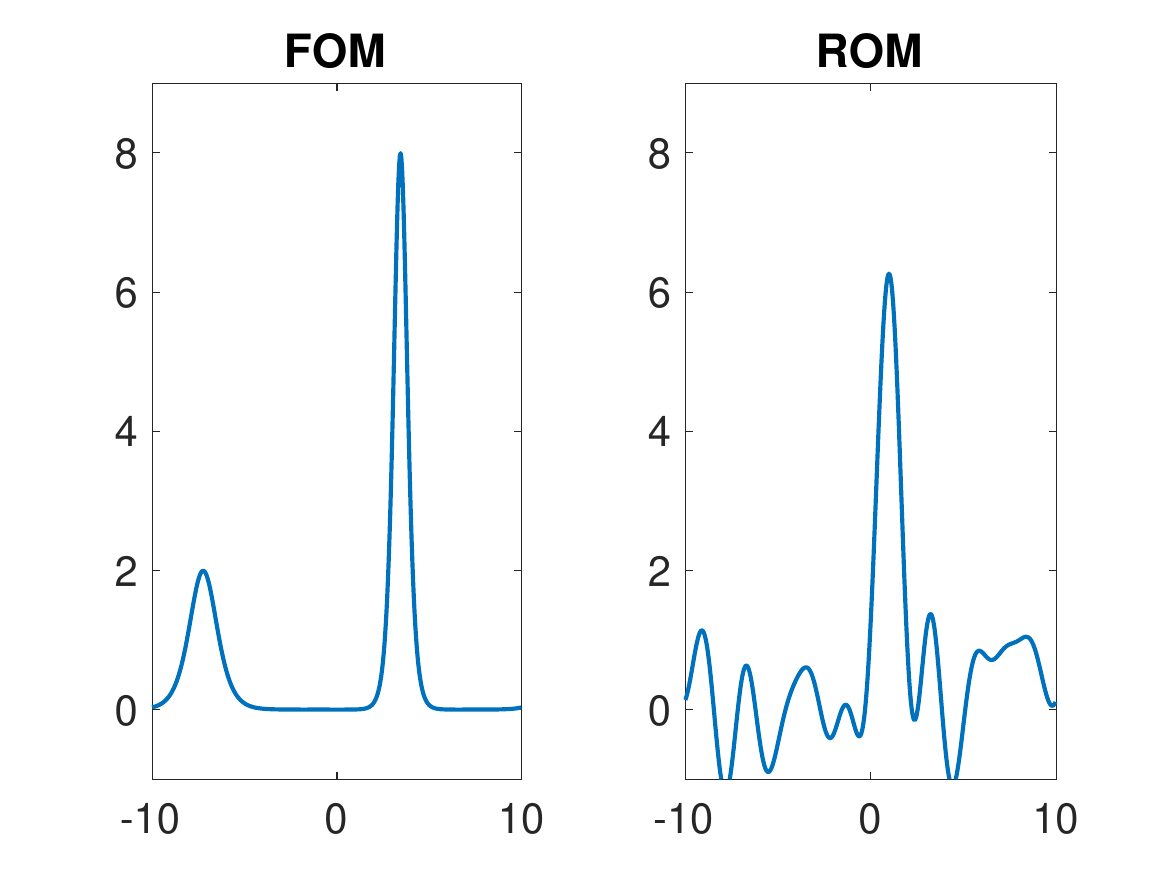}
            \caption{POD modes = 20, Relative energy = 0.9360 }
            \label{subfig:kdv_fom_rom_20}
        \end{subfigure}
        \vskip\baselineskip
        \begin{subfigure}[b]{0.45\textwidth}
            \includegraphics[width=\textwidth, alt={Line graph comparing the analytical solution and the 50-mode ROM solution, where the ROM solution some spurious oscillations and is biased.}]{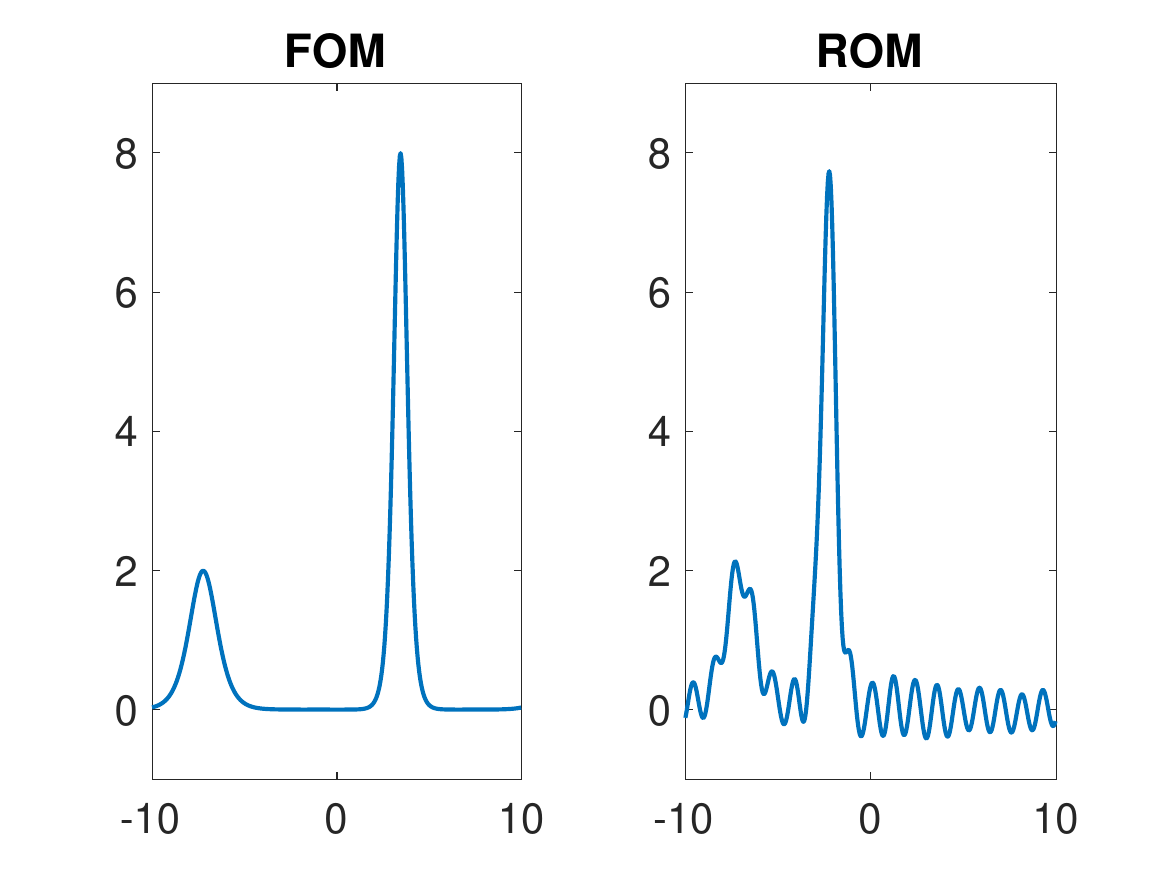}
            \caption{POD modes = 50, Relative energy = 0.9998}
            \label{subfig:kdv_fom_rom_50}
        \end{subfigure}
        \begin{subfigure}[b]{0.45\textwidth}
            \includegraphics[width=\textwidth, alt={Line graph comparing the analytical solution and the 75-mode ROM solution, showing both curves are in agreement}]{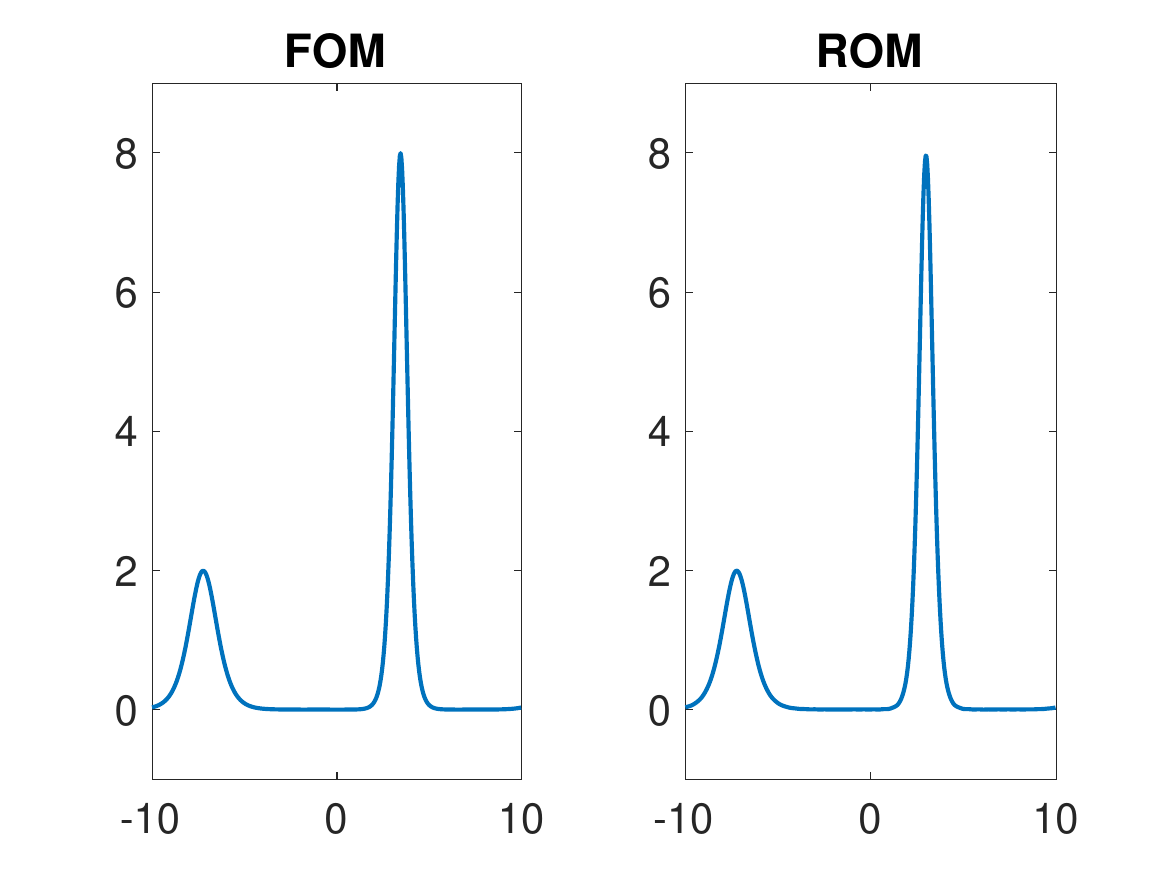}
        \caption{POD modes = 75, Relative energy = 0.9999}
        \label{subfig:kdv_fom_rom_75}
        \end{subfigure}
        \vspace{.3in}
        \caption{Comparison of forward simulation between the ROM and FOM runs for different POD modes.
        The timestep is $0.01$ for a total time of $10$ unit time.
        }
        \label{fig:KdV_fom_rom_comparison}
    \end{figure}

\begin{figure}[htbp]
    \centering
    \begin{subfigure}{0.48\textwidth}
        \centering
        \includegraphics[width=\linewidth, alt={Heatmap depicting analysis RMSE for a single-fidelity Gromov flow for the KdV test problem.}]{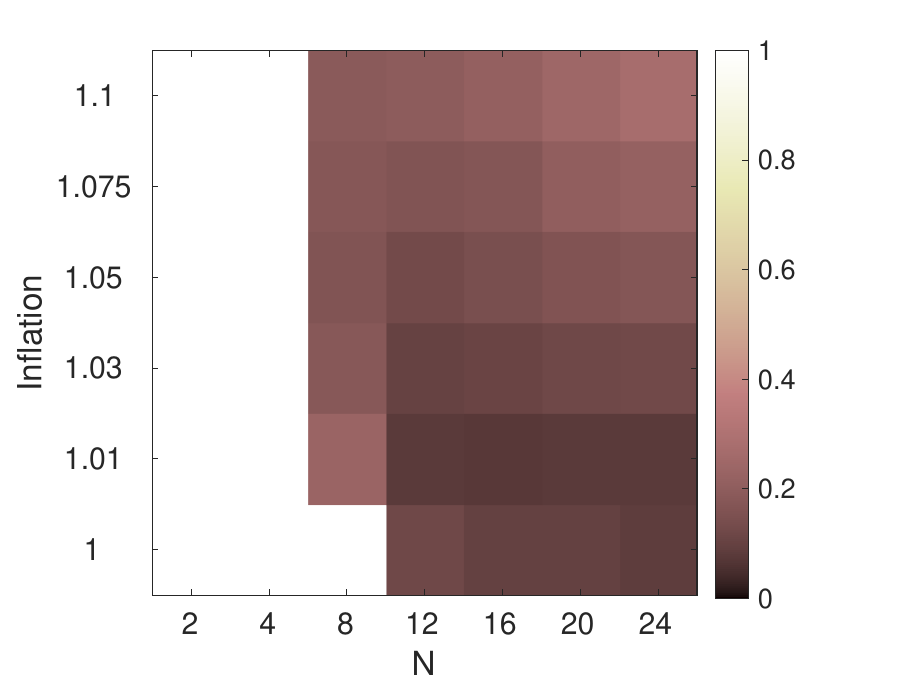}
        \caption{Single-fidelity FOM simulation}
        \label{subfig:kdv_gromov_fom}
    \end{subfigure}
    \hfill
    \begin{subfigure}{0.48\textwidth}
        \centering
        \includegraphics[width=\linewidth, alt={Heatmap depicting analysis RMSE for a bi-fidelity Gromov flow for the KdV test problem.}]{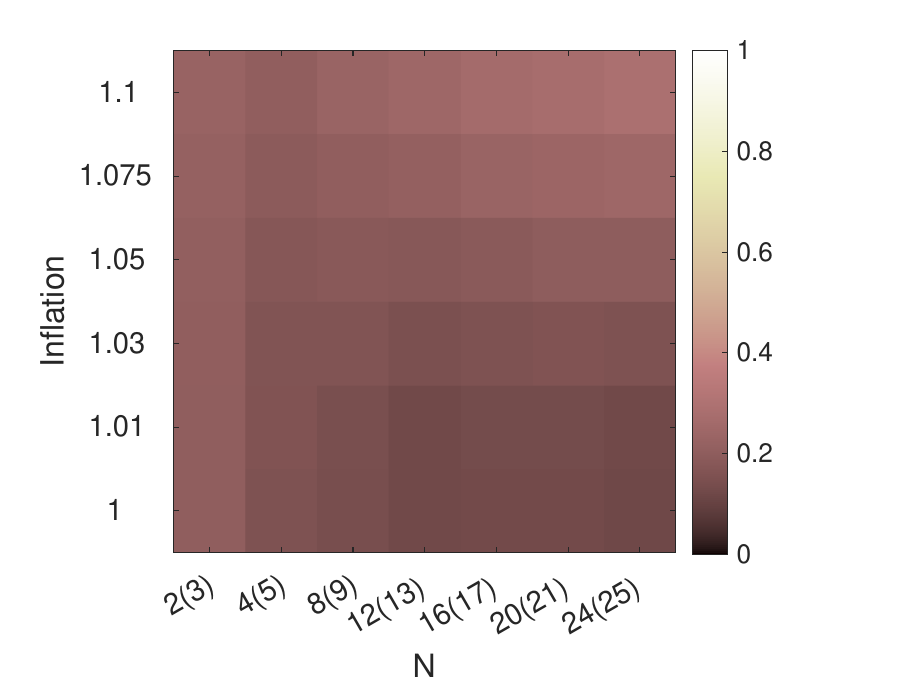}
        \caption{Bi-fidelity FOM and ROM simulation}    \label{subfig:kdv_gromov_multifidelity}
    \end{subfigure}
    \caption{
    Analysis RMSE heatmap comparing single-fidelity and bi-fidelity Gromov flow for the KdV test problem.} 
    \label{fig:kdv_gromov}
    \end{figure}

\begin{figure}[htbp]
    \centering
    \begin{subfigure}{0.48\textwidth}
        \centering
        \includegraphics[width=\linewidth, alt={Heatmap depicting analysis RMSE for a single-fidelity Exact flow for the KdV test problem.}]{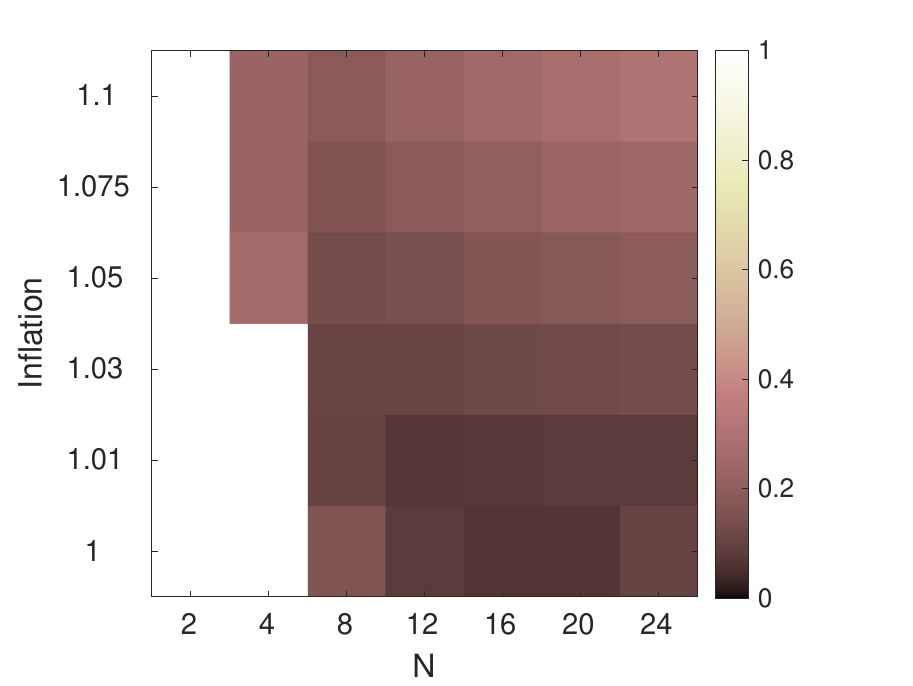}
        \caption{Single Fidelity FOM simulation}
        \label{subfig:kdv_exact_fom}
    \end{subfigure}
    \hfill
    \begin{subfigure}{0.48\textwidth}
        \centering
        \includegraphics[width=\linewidth, alt={Heatmap depicting analysis RMSE for a bi-fidelity Exact flow for the KdV test problem.}]{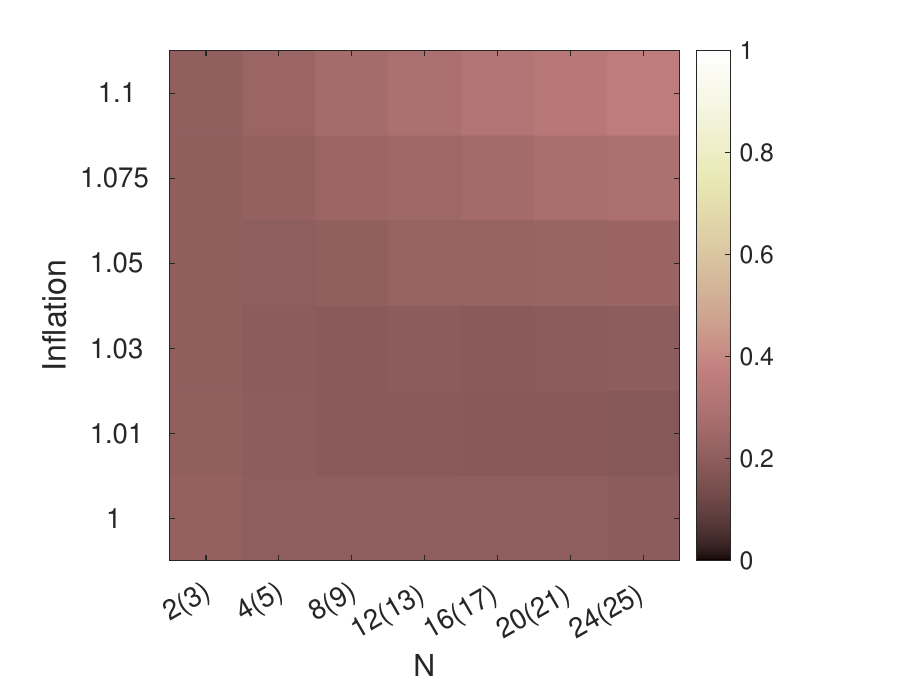}
        \caption{Bi-fidelity FOM and ROM simulation}    \label{subfig:kdv_exact_multifidelity}
    \end{subfigure}
    \caption{Analysis RMSE heatmap comparing single-fidelity and bi-fidelity Exact flow for the KdV test problem.} 
    \label{fig:KdV_exact}
    \end{figure}

    We develop an intrusive Galerkin ROM for the KdV equation (refer \cref{appsec:kdv_rom_development}).
    The quadratic tensor ($Q$ in \cref{eq:kdv_fd_rom_projection}) evaluation involves projecting out to the full order space, which is a disadvantage of POD-Galerkin ROM (refer \cref{rem:deim}). 
    We run and compare ROM runs (with different modes) with the actual FOM simulations as shown in \Cref{fig:KdV_fom_rom_comparison}.
    The comparison in \Cref{fig:KdV_fom_rom_comparison} supports our claim that this problem is challenging for POD–Galerkin ROMs.
    The ROM simulation with $75$ modes closely follows the FOM solution, whereas simulations with fewer modes exhibit significantly different trajectories, characterized by phase shifts, spurious oscillations, and, in the worst case, complete dissolution of the solitons.
    The spurious waves in \cref{subfig:kdv_fom_rom_50} are not resolved because of the absense of diffusion.  
    We select $20$ POD modes (\cref{subfig:kdv_fom_rom_20}) to construct the ROM, which complements the FOM simulation and yields satisfactory results in the bi-fidelity setting.
    We assume that model propagation accounts for most of the computational overhead and compare the runtime of a single FOM propagation with that of a single ROM propagation. 
    We find that a single FOM propagation is approximately $20$ times more expensive than a single ROM propagation.
    In \cref{subfig:kdv_gromov_multifidelity,subfig:kdv_exact_multifidelity}, the number in parenthesis indicate the equivalent computational cost in terms of FOM propagation.
    For example, $2(3)$ indicates two FOMs augmented with $20$ ROMs, where $20$ ROMs are equivalent, in forward propagation cost, to one FOM.   
    The simulation results with bi-fidelity Gromov (\cref{subfig:kdv_gromov_multifidelity}) and Exact flow filter (\cref{subfig:kdv_exact_multifidelity}) show that we have clear distinctive advantage of running a bi-fidelity simulation compared to a single-fidelity FOM run.
    We see that the multi-fidelity setup works with just two FOM ensembles, and is even cheaper if we include the ROM cost.
    This is more advantageous in higher dimensional setting where the cost of FOM propagation can be very high. 
    Higher inflation can introduce spurious modes, which might not get dissipated and can lead to higher error values as indicated in both \cref{fig:kdv_gromov,fig:KdV_exact}.
    %

\subsection{Quasi-Geostrophic Equation (QGE)}\label{subsec:qg}
    The two-dimensional single-layer QGE based on barotropic vorticity, has been studied extensively in data assimilation and ROMs community and we adopt the implementation from the literature \cite{Sandu_2021_MF-EnKF, stabilized_pod_qg_omer_trian, Mou07022020}. 
    In the non-dimensionalized setting, the equations are: 
    %
    \begin{equation}
    \label{eq:qge}
        \begin{split}
            \omega_{t} + J (\psi, \omega) - \mathrm{Ro}_{}^{-1} \psi_{x} = \mathrm{Re}_{} \Delta \omega + \mathrm{Ro}_{}^{-1} F , \\
            J(\psi, \omega) = \psi_{y} \omega_{x} - \psi_{x} \omega_{y}, \quad \Delta \psi = - \omega, 
        \end{split}
    \end{equation}
    where $\omega$ is the vorticity, $\psi$ is the streamfunction, $\mathrm{Re}_{}$ is the Reynolds number and $\mathrm{Ro}_{}$ is the Rossby number. 
    We use a symmetric double gyre for the forcing term $F = \sin (\pi (y-1))$. 
    The computational domain is $\Omega = [0,1] \times [0,2]$ with homogeneous boundary conditions $\omega(x,y) = 0$ (slip boundary condition for the velocity) and $\psi(x,y) = 0$ (imperbeability boundary condition). 
    Following \cite{stabilized_pod_qg_omer_trian}, we consider $\mathrm{Re}_{} = 450$ and $\mathrm{Ro}_{} = 0.0036$ in this study.
    We integrate the PDE given in \cref{eq:qge} starting from the initial condition until it reaches a steady state where wind forcing, dissipation and Jacobian balance each other out \cite{stabilized_pod_qg_omer_trian}.
    The time scale uses 80 units of time to represent 20.12 years. 

\begin{figure}[htbp]
    \centering
    \begin{subfigure}{0.48\textwidth}
        \centering
        \includegraphics[width=\linewidth, alt={Heatmap depicting analysis RMSE for a single-fidelity Gromov flow for the QG test problem.}]{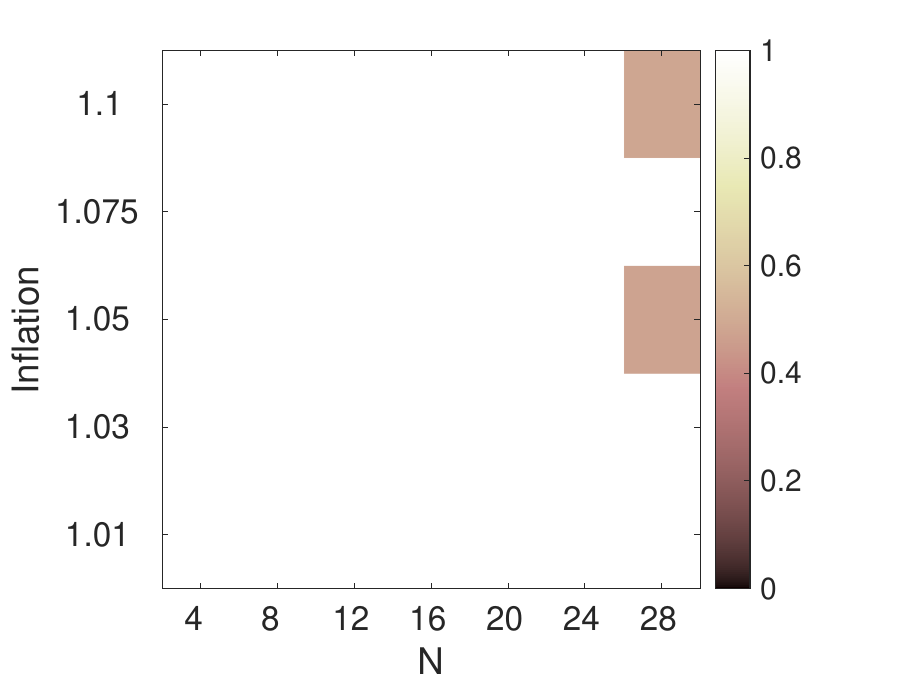}
        \caption{Single Fidelity FOM simulation}
        \label{subfig:qg_gromov_fom_no_local}
    \end{subfigure}
    \hfill
    \begin{subfigure}{0.48\textwidth}
        \centering
        \includegraphics[width=\linewidth, alt={Heatmap depicting analysis RMSE for a bi-fidelity Gromov flow for the QG test problem.}]{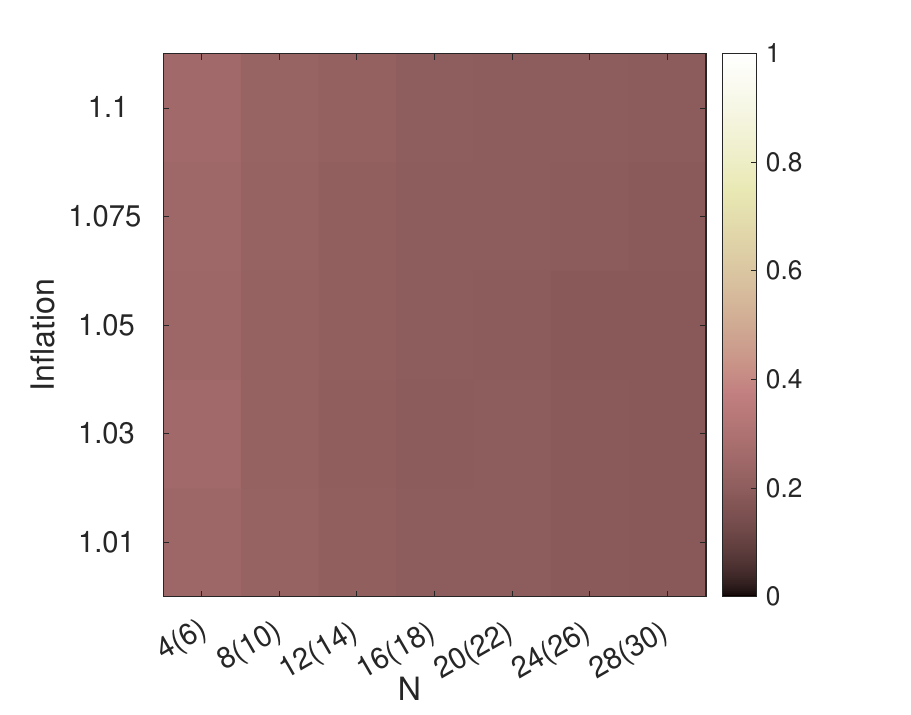}
        \caption{Bi-fidelity FOM and ROM simulation}    \label{subfig:qg_gromov_multifidelity_no local}
    \end{subfigure}
    \caption{
    Analysis RMSE heatmap comparing single-fidelity and bi-fidelity Gromov flow for the QG test problem.}
    \label{fig:qg_gromov}
    \end{figure}

\begin{figure}[htbp]
    \centering
    \begin{subfigure}{0.48\textwidth}
        \centering
        \includegraphics[width=\linewidth, alt={Heatmap depicting analysis RMSE for a single-fidelity Exact flow for the QG test problem.}]{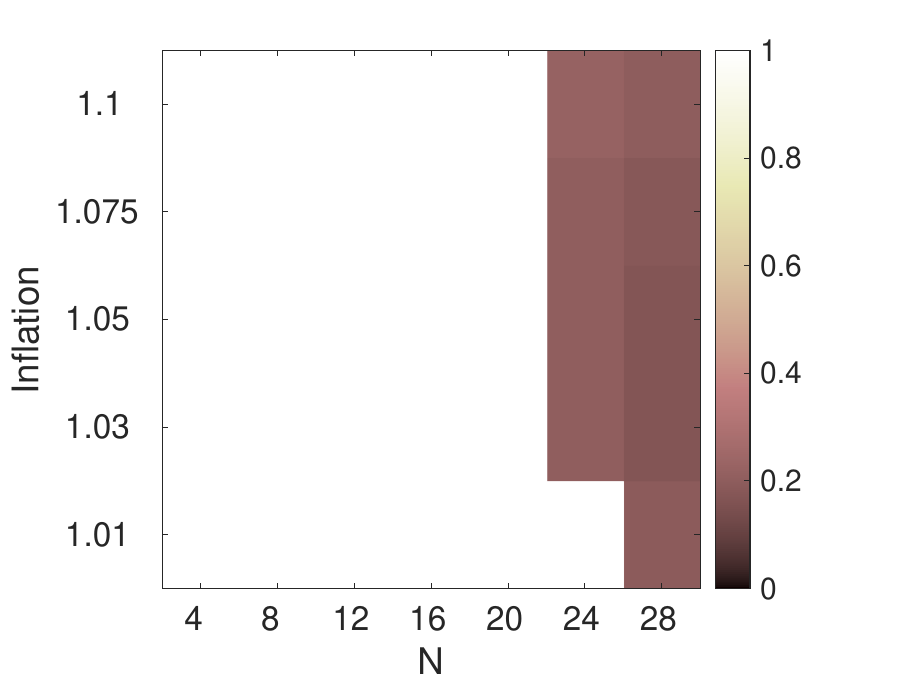}
        \caption{Single Fidelity FOM simulation}
        \label{subfig:qg_exact_fom_no_local}
    \end{subfigure}
    \hfill
    \begin{subfigure}{0.48\textwidth}
        \centering
        \includegraphics[width=\linewidth, alt={Heatmap depicting analysis RMSE for a bi-fidelity Exact flow for the QG test problem.}]{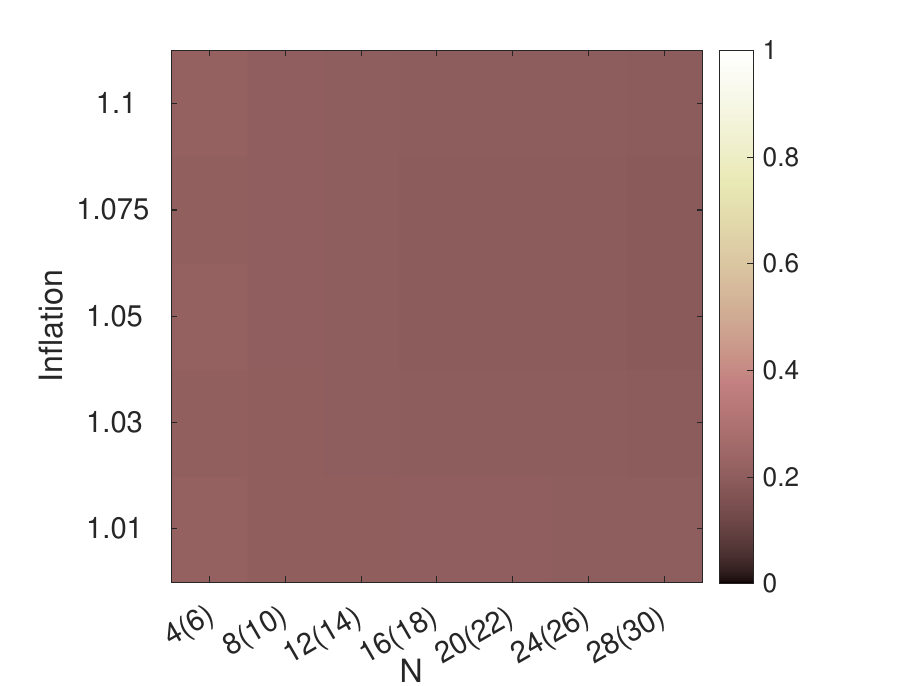}
        \caption{Bi-fidelity FOM and ROM simulation}    \label{subfig:qg_exact_multifidelity_no local}
    \end{subfigure}
    \caption{Analysis RMSE heatmap comparing single-fidelity and bi-fidelity Exact flow for the QG test problem.}
    \label{fig:qg_exact}
    \end{figure}

    The FOM simulation is performed on a spatial grid consisting of $63$ discrete points in the x-direction and $127$ points in the y-direction, resulting in a total of 8001 degrees of freedom. 
    Spatial derivatives are approximated using a second-order central finite difference scheme.
    The resulting semi-discrete system of ODE is integrated in time using the variable step-size Dormand–Prince method. 
    For data assimilation, observations are recorded every $0.0109$ dimensionless time units, corresponding approximately to $24$ hours taken at 150 equally spaced grid points. 
    The QG ROM is constructed using method of snapshots and a projection step with POD following \cite{stabilized_pod_qg_omer_trian,Mou07022020}.
    Further details of the ROM can be found in \cref{appsec:qg_rom} and optimal projection operators are taken directly from the literature \cite{mfenkf_popov}.
    We run a twin experiment where we use a FOM simulation as the ``truth''. 
    Observations are linear, sparse where we observe every alternate $150$ grid point.  
    Observation error covariance is $\bfSigma = \mathbf{I}_{150}$.
    The simulation is run for $350$ assimilation steps where we discard the first $50$ steps for spinups. 
    The numerical testbed \cite{roberts2019odetestproblemsmatlab} is used for our simulation.
    The comparative RMSE heatmap, without localization, for both the filters are shown in \Cref{fig:qg_gromov} and \Cref{fig:qg_exact}.
    We see a significant advantage in our bi-fidelity setup compared to our single-fidelity simulations. 
    We find that a single FOM propagation is approximately $40$ times more expensive than a single ROM propagation. 
    In \cref{subfig:qg_gromov_fom_no_local,subfig:qg_gromov_multifidelity_no local}, the number in parenthesis indicate the equivalent computational cost in terms of FOM propagation.
    The results clearly demonstrate the advantage of using ROMs as surrogates, which gives comparable accuracy at a substantially lower cost.
    Notably, the bi-fidelity setup achieves comparable performance with much smaller ensemble than single-fidelity experiments, even after considering the cost of the augmented ROMs. 
    This experiment shows that complementing a few expensive FOM simulations with a large number of inexpensive, biased ROMs yield a better estimate than running a standalone FOM simulation. 
    %

\subsubsection{Localization}
\label{subsubsec:localization_qg}
\begin{figure}[htbp]
    \centering
    \begin{subfigure}{0.48\textwidth}
        \centering
        \includegraphics[width=\linewidth, alt={Heatmap depicting analysis RMSE for a single-fidelity Gromov flow with localization for the QG test problem.}]{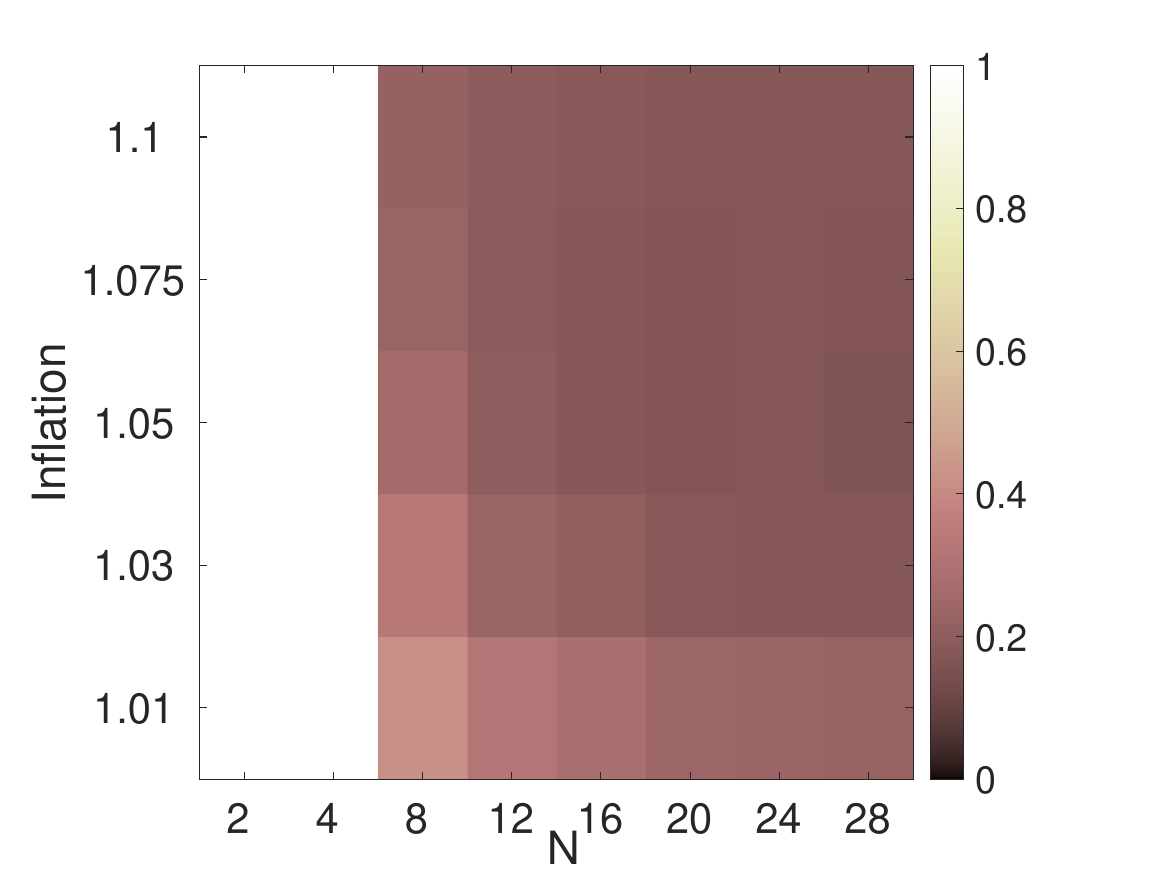}
        \caption{Single Fidelity FOM simulation with localization}
        \label{subfig:qg_gromov_fom_local}
    \end{subfigure}
    \hfill
    \begin{subfigure}{0.48\textwidth}
        \centering
        \includegraphics[width=\linewidth, alt={Heatmap depicting analysis RMSE for a bi-fidelity Gromov flow with localization for the QG test problem.}]{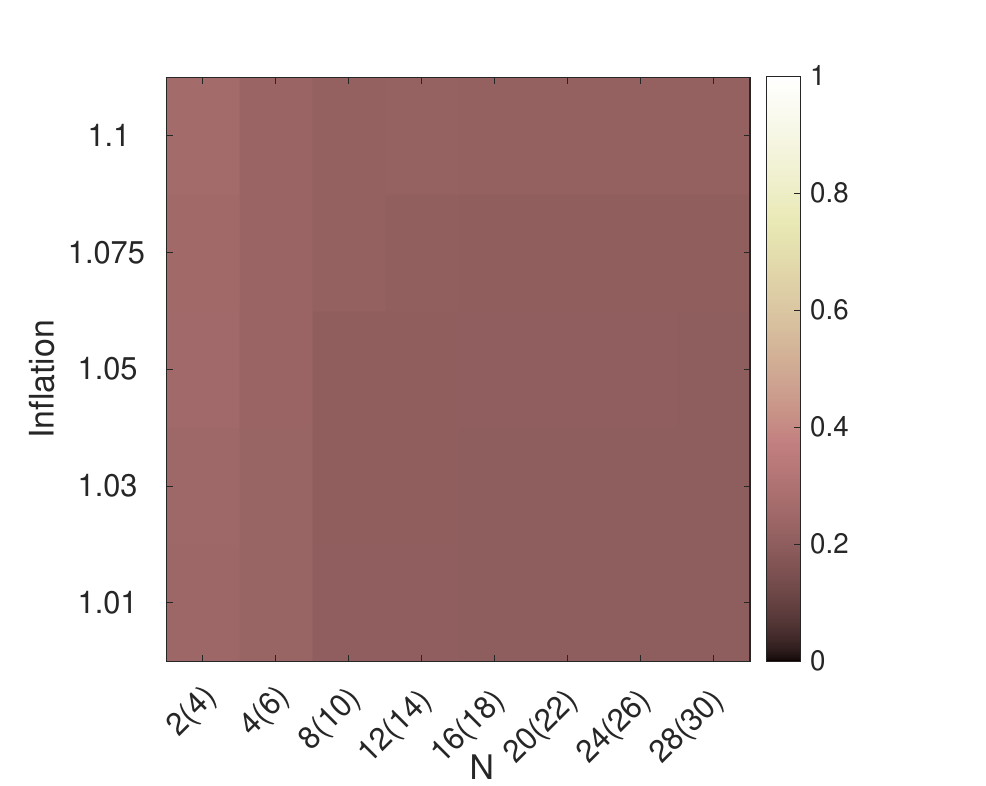}
        \caption{Bi-fidelity FOM and ROM simulation with localization}    \label{subfig:qg_gromov_multifidelity_local}
    \end{subfigure}
    \caption{Analysis RMSE heatmap comparing single-fidelity and bi-fidelity Gromov flow for the QG test problem using localization.} 
    \label{fig:qg_gromov_local}
    \end{figure}

\begin{figure}[htbp]
    \centering
    \begin{subfigure}{0.48\textwidth}
        \centering
        \includegraphics[width=\linewidth, alt={Heatmap depicting analysis RMSE for a single-fidelity Exact flow for the QG test problem with localization.}]{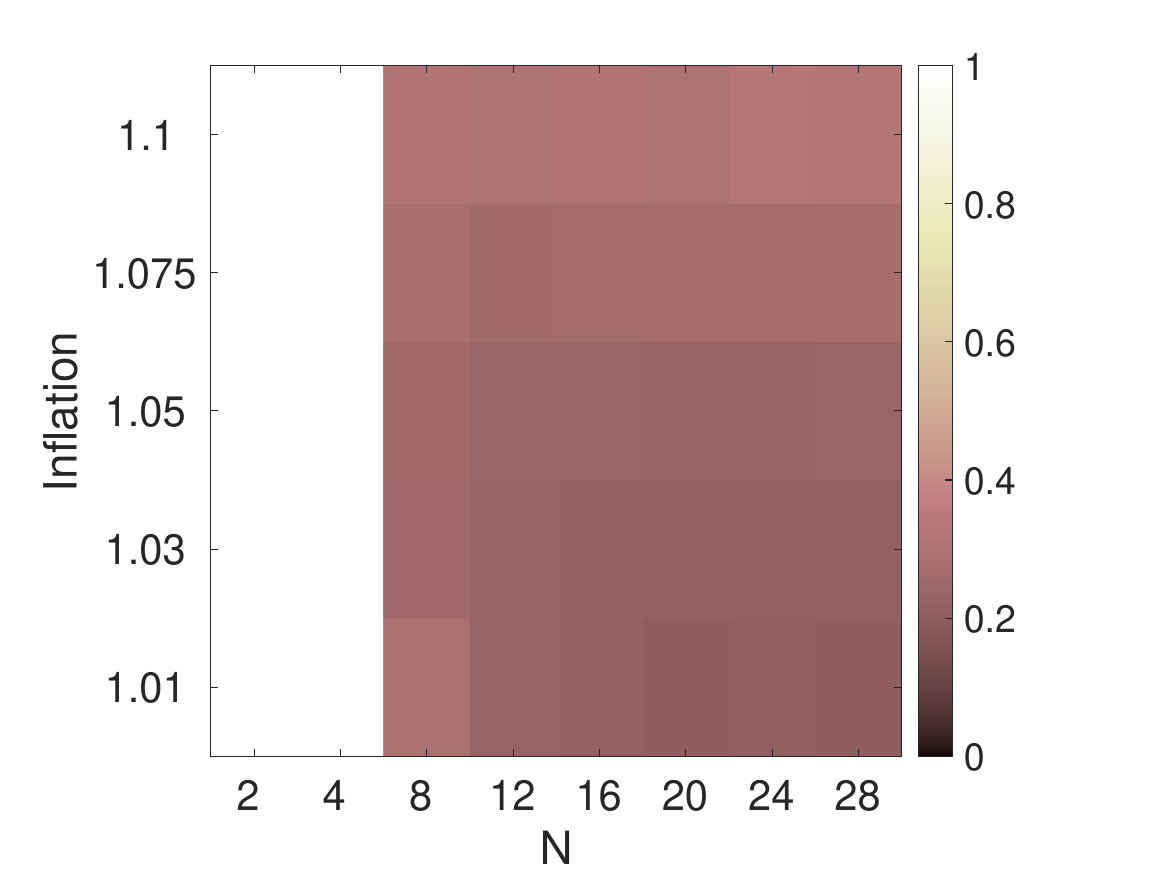}
        \caption{Single Fidelity FOM simulation with localization}
        \label{subfig:qg_exact_fom_local}
    \end{subfigure}
    \hfill
    \begin{subfigure}{0.48\textwidth}
        \centering
        \includegraphics[width=\linewidth, alt={Heatmap depicting analysis RMSE for a bi-fidelity Exact flow with localization for the QG test problem.}]{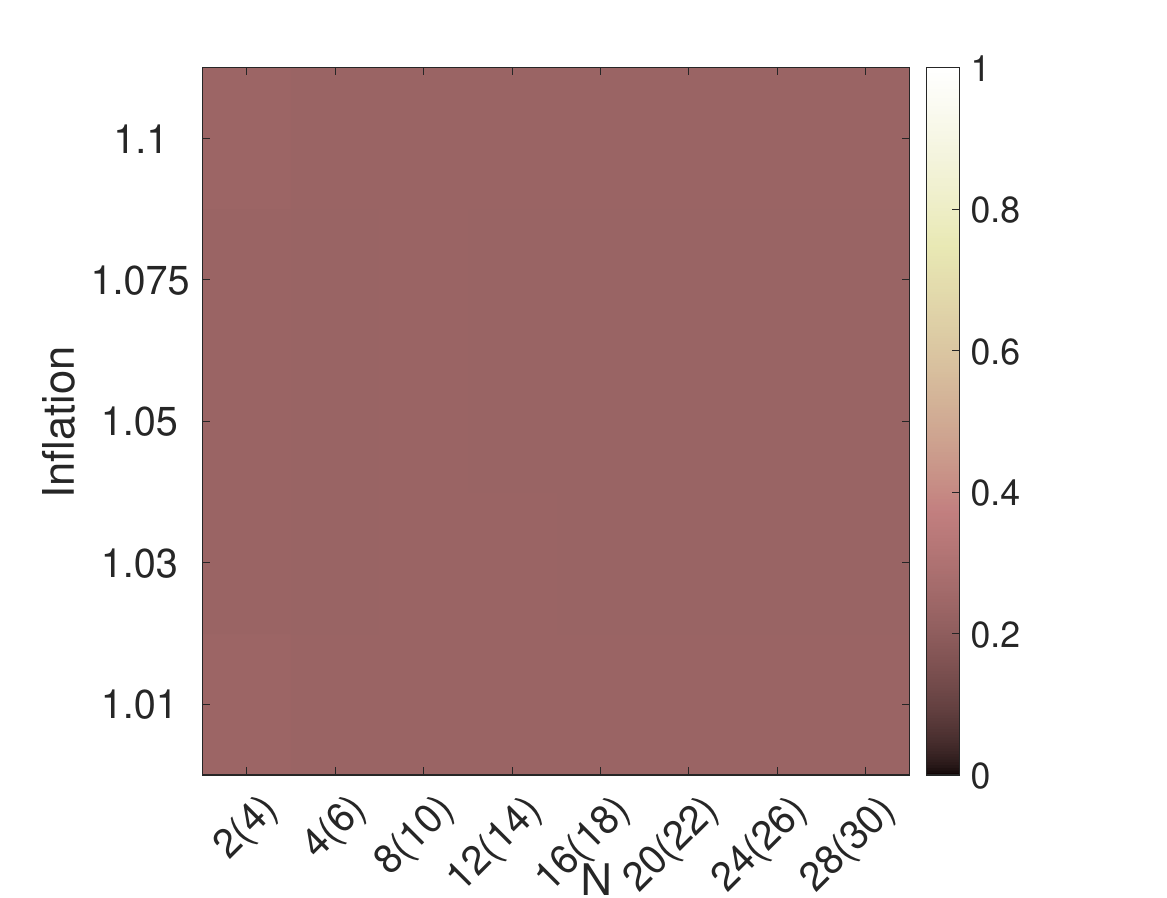}
        \caption{Bi-fidelity FOM and ROM simulation with localization}    \label{subfig:qg_exact_multifidelity_local}
    \end{subfigure}
    \caption{Analysis RMSE heatmap comparing single-fidelity and bi-fidelity Exact flow for the QG test problem with localization.} . 
    \label{fig:qg_exact_local}
    \end{figure}

    We implement localization via covariance tapering techniques \cite{blocalization}. 
    Ensemble based methods often suffer from sampling errors in higher dimensions where the estimated covariance is low rank and has spurious correlation. 
    Localization by covariance tapering aims to remove these spurious correlations and increase the rank of the covariance matrix. 
    We use the piecewise rational function of Gaspari Cohn \cite{gaspari_cohn} that has compact support:
    \begin{equation}
        \mathcal{L}(\upsilon)= 
        \begin{cases}
            - \frac{1}{4} \upsilon^{5} + \frac{1}{2} \upsilon^{4} +\frac{5}{8} \upsilon^{3} - \frac{5}{3} \upsilon^{2} + 1  ,&  0 \leq \upsilon \leq 1,\\
            \frac{1}{12} \upsilon^{5} -\frac{1}{2} \upsilon^{4} + \frac{5}{8} \upsilon^{3} + \frac{5}{3} \upsilon^{2} -5\upsilon +4 -\frac{2}{3 \upsilon}  ,              & 1 \leq \upsilon \leq 2, \\
            0, & \text{otherwise}
        \end{cases}
    \end{equation}
    The covariance of the total variate \cref{eq:cross_cov_total_variate} is modified as:
    \begin{equation}
        \widetilde{\bfSigma}_{\bfZ_{k}^{f} \NLObs (\bfZ_{k}^{f})}^{loc} = \rho_{nm} \odot \widetilde{\bfSigma}_{\bfZ_{k}^{f} \NLObs (\bfZ_{k}^{f})}, \quad \quad
        \widetilde{\bfSigma}_{\NLObs (\bfZ_{k}^{f}), \NLObs (\bfZ_{k}^{f})}^{loc} = \rho_{mm} \odot \widetilde{\bfSigma}_{\NLObs (\bfZ_{k}^{f}), \NLObs (\bfZ_{k}^{f})},
    \end{equation}
    where $\rho_{nm}$ and $\rho_{mm}$ are the tapered matrices in the state-observation and observation-observation space respectively. 
    We use these modified empirical covariances in \cref{eq:gromov_non_linear,eq:exact_flow_non_linear_1}.
    Localization experiments were carried out 
    with localization radius $r=15$ grid points. 
    Results in \Cref{fig:qg_gromov_local} and \Cref{fig:qg_exact_local} show that the proposed filter is compatible with the covariance localization heuristic.
    Interestingly, the bi-fidelity setup works with as few as two ensemble members and remains computationally cheaper even after accounting for the cost of the augmented ROMs. 

\subsubsection{Nonlinear Observations Operators}
\label{subsubsec:non_linear_observation_qg}
\begin{figure}[htbp]
    \centering
    \begin{subfigure}{0.48\textwidth}
        \centering
        \includegraphics[width=\linewidth, alt={Heatmap depicting analysis RMSE for a bi-fidelity Gromov flow with nonlinear observations.}]{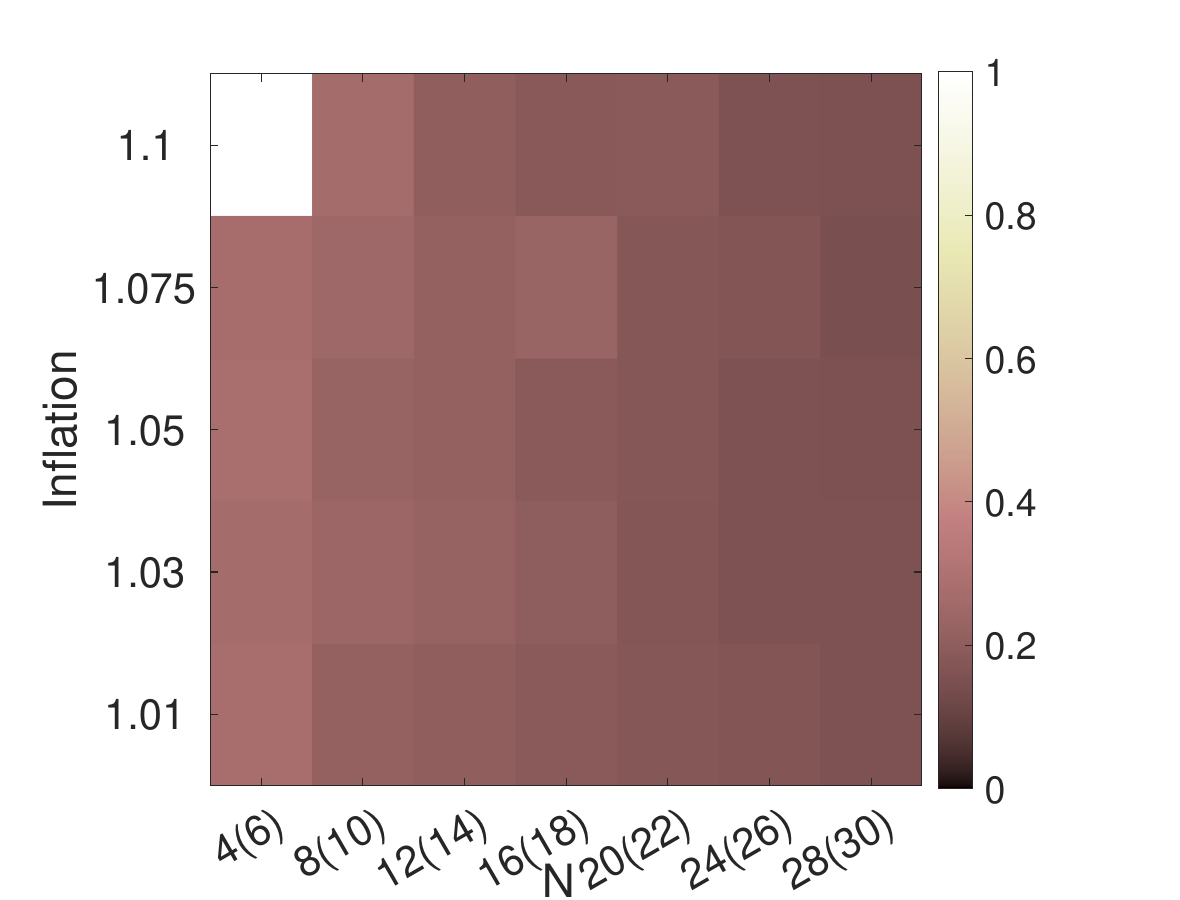}
        \caption{Bi-fidelity Gromov flow with nonlinear observation}
        \label{subfig:mfgromov_non_linear}
    \end{subfigure}
    \hfill
    \begin{subfigure}{0.48\textwidth}
        \centering
        \includegraphics[width=\linewidth, alt={Heatmap depicting analysis RMSE for a bi-fidelity Exact flow with nonlinear observations.}]{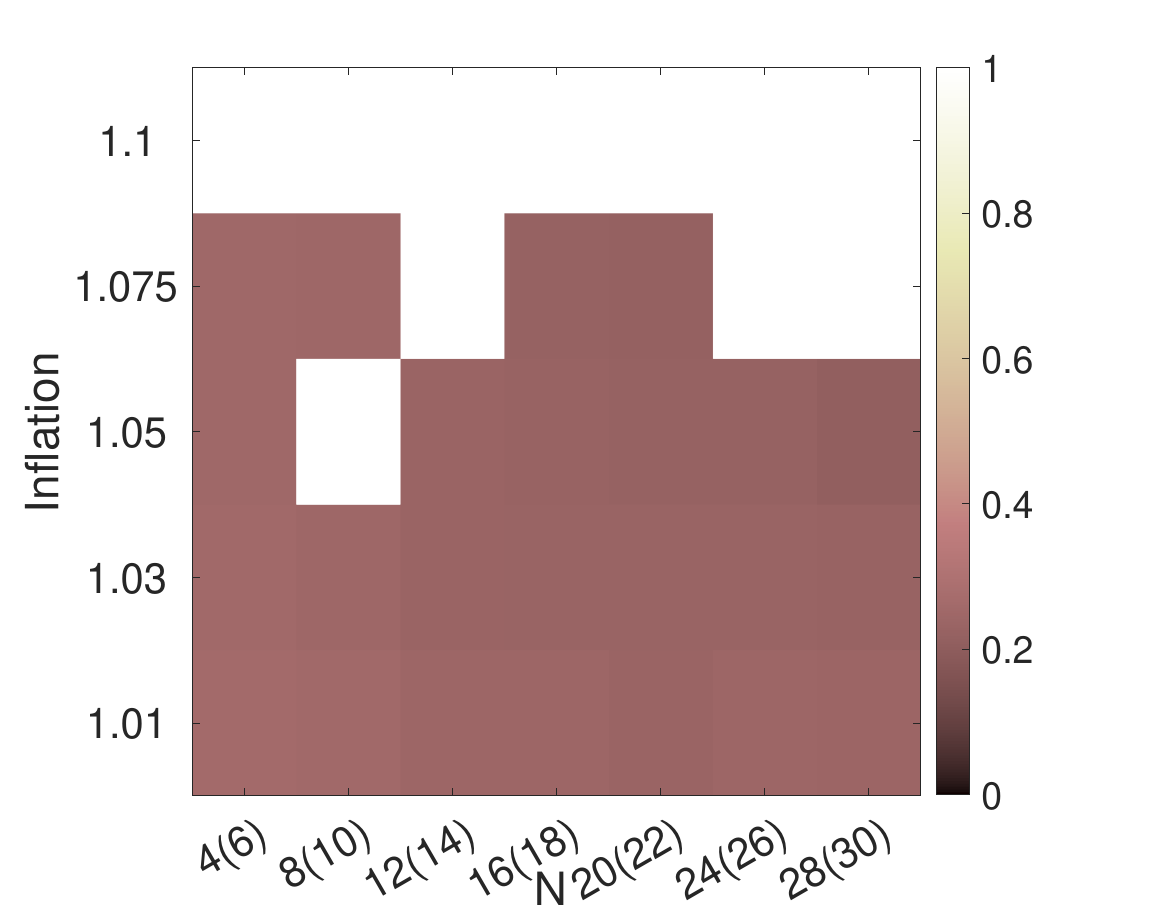}
        \caption{Bi-fidelity Exact flow with nonlinear observation}    \label{subfig:mfexact_non_linear}
    \end{subfigure}
    \caption{Analysis RMSE heatmap comparing bi-fidelity Gromov and Exact flow for the QG test problem using a nonlinear observation operator.}
    \label{fig:qg_non_linear}
    \end{figure}

    In most realistic applications, we get observations from sensor data that are not only sparse but can exhibit a nonlinear relationship between the states and the observables. 
    We consider the velocity magnitude at each of the discrete spatial grid points in our QG experiments as nonlinear observation operator. 
    The velocity magnitude is defined as:
    \begin{equation}
    \label{eq:non_linear_observation_qg}
        \begin{split}
            u &= \frac{\partial \psi (x,y)}{\partial y}, \quad v = - \frac{\partial \psi (x,y)}{\partial x} \\
            \mathcal{H} (x, y) &= \sqrt{u^2 + v^2} = \sqrt{\left(\frac{\partial \psi (x,y)}{\partial y}\right)^2 + \left(-\frac{\partial \psi (x,y)}{\partial x}\right)^2}
        \end{split}
    \end{equation}
    The nonlinear observation is considered sparse and we follow a similar sparsity pattern as in the linear observation case. 
    The observation error covariance is a scaled identity where the scaling factor is $16$. 
    The results are shown in \Cref{fig:qg_non_linear}. The bi-fidelity Gromov flow can handle nonlinear observations quite well as can be seen from the heatmap. 
    However, the Exact flow is unstable at higher inflation and this unsatisfactory performance can be attributed to the assumptions made in \cref{eq:exact_flow_non_linear_2}.
    %

\subsubsection{Cost vs RMSE}
\label{subsubsec:qg_cost}
\begin{figure}[htbp]
    \centering
    \begin{subfigure}{0.48\textwidth}
        \centering
        \includegraphics[width=\linewidth, alt={Line graph showing analysis rmse of single-fidelity vs bi-fidelity simulation for the Gromov flow depicting advantage in RMSE reduction as we augment ROMs with FOMs}]{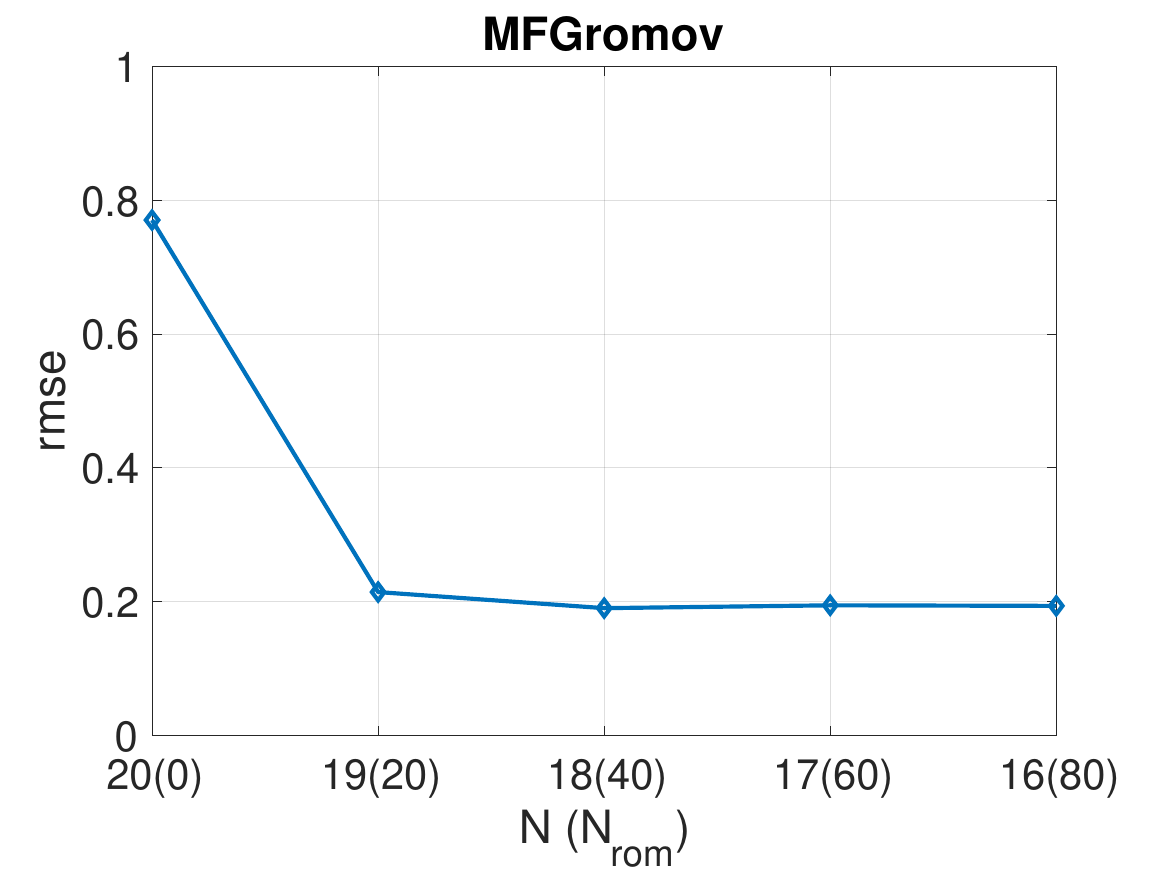}
        \caption{Gromov}
        \label{subfig:mfgromov_cost_analysis}
    \end{subfigure}
    \hfill
    \begin{subfigure}{0.48\textwidth}
        \centering
        \includegraphics[width=\linewidth, alt={Line graph showing analysis rmse of single-fidelity vs bi-fidelity simulation for the Exact flow depicting advantage in RMSE reduction as we augment ROMs with FOMs}]{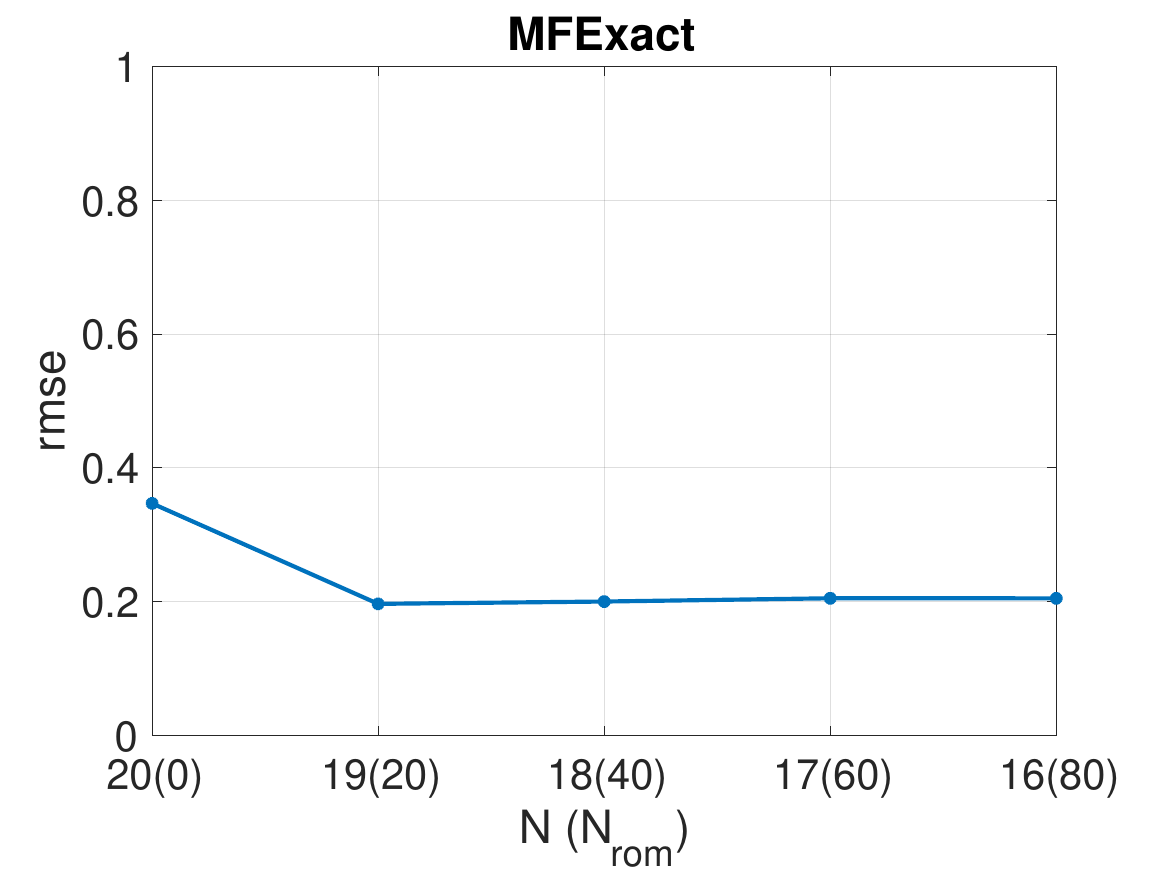}
        \caption{Exact}    \label{subfig:mfexact_cost_analysis}
    \end{subfigure}
    \caption{RMSE of single-fidelity vs multi-fidelity Gromov and Exact flow.
    The number in parentheses is the number of ROM simulations. } 
    \label{fig:qg_cost_analysis}
    \end{figure}

    The RMSE plots in \Cref{fig:qg_cost_analysis} measure the gain in RMSE at a comparable computational cost.
    To simplify, we currently ignore assimilation costs and focus solely on model propagation costs.
    In real-world applications, multiple FOM runs often mask the computational cost of assimilation.
    In our experiment, we ran a single-fidelity test with 
    FOM ensembles and recorded the RMSE over multiple initializations.
    Next, we replaced one FOM with a computationally equivalent number of ROM models and repeated the process.
    We observed a slight advantage in the Exact flow and a significant advantage in the Gromov flow experiments in terms of RMSE.

%% file: sections/conclusion.tex
\section{Conclusions}
\label{sec:conclusion}
    Particle flow algorithms have proven useful in high dimensional problems by alleviating degeneracy and weight collapse, features that plague standard particle filters based on importance sampling. 
    We derive a multi-fidelity data assimilation algorithm with particle flow based on log homotopy and control variate formulation. 
    We specifically focus on the deterministic Exact and stochastic Gromov flow.
    Our numerical experiments show a clear advantage by complementing a small number of accurate and expensive high-fidelity model runs with a large number of biased but cheaper lower-fidelity model runs. 
    This is achieved at a comparable or even cheaper computational cost.
    We see a very similar advantage when we employ localization and nonlinear observations in our QG test problem.
    Different variates/random variables can exist in different spaces in our method, and we use an optimal projection/interpolation operator to map from one space to the other.
    This methodology is easily extensible to multiple levels each with a varying level of accuracy, where the uncertainty in coarser level acts as a control variate for the uncertainty in the finer level. 
    Models at different levels of hierarchy can be defined on distinct spaces, with potentially different dimensions and inner products.    

    There are multiple future directions to further expand this work.
    One possible extension would be to replace intrusive Galerkin ROMs with data-driven or Machine Learning based ROMs, a case which arises in real life application where we do not have access to the underlying governing PDEs.
    Another extension can be to extend this to particle filters based on multi-level importance sampling \cite{karen_wilcox_multi_level_importance_sampling}. 
    Extending multi-fidelity Exact flow to nonlinear measurement in a more mathematically accurate way would be an interesting challenge.
    More stable time integration schemes have recently been proposed in \cite{crouse_flow_filter_bias}.
    Investigating their incorporation into the proposed multi-fidelity framework to assess potential gains in stability and computational efficiency can be another interesting direction.
    This work will be extended and employed in state and parameter estimation in large realistic models in a digital twin setup \cite{monica_math_dt_1}.


    \vspace{1cm}
    \noindent \textbf{Acknowledgment:}
    This work was supported by NSF Computational Mathematics program through award DMS-2436357, and by the Computational Science Laboratory at Virginia Tech.
    Generative AI was used in producing parts of the tikz code for \Cref{fig:mfpff_schematic}, and the authors have checked for the correctness and take full responsibility for it.

%% file: sections/appendix.tex

\section{ROM for KdV}\label{appsec:kdv_rom_development}
    \setcounter{figure}{0} 
    The two solitons solution advects and interacts non-destructively. 
    The presence of traveling solitons makes this problem challenging for POD ROMs, since accurately capturing the dynamics requires a large number of pod modes.
    We propagate the semi-discretized model for 1000 time steps with a step size $0.01$, and stack them in the snapshot matrix $\bfXi$.
    The singular values of the snapshot matrix ($\bfXi$) decays slowly in this particular setup as shown in \Cref{fig:kdv_svd_singular_values}.
    \begin{figure}[htbp]
        \centering    \includegraphics[width=0.5\linewidth, alt={Line grtaph showing the singular values of the snapshot matrix of KdV forward simulation.}]{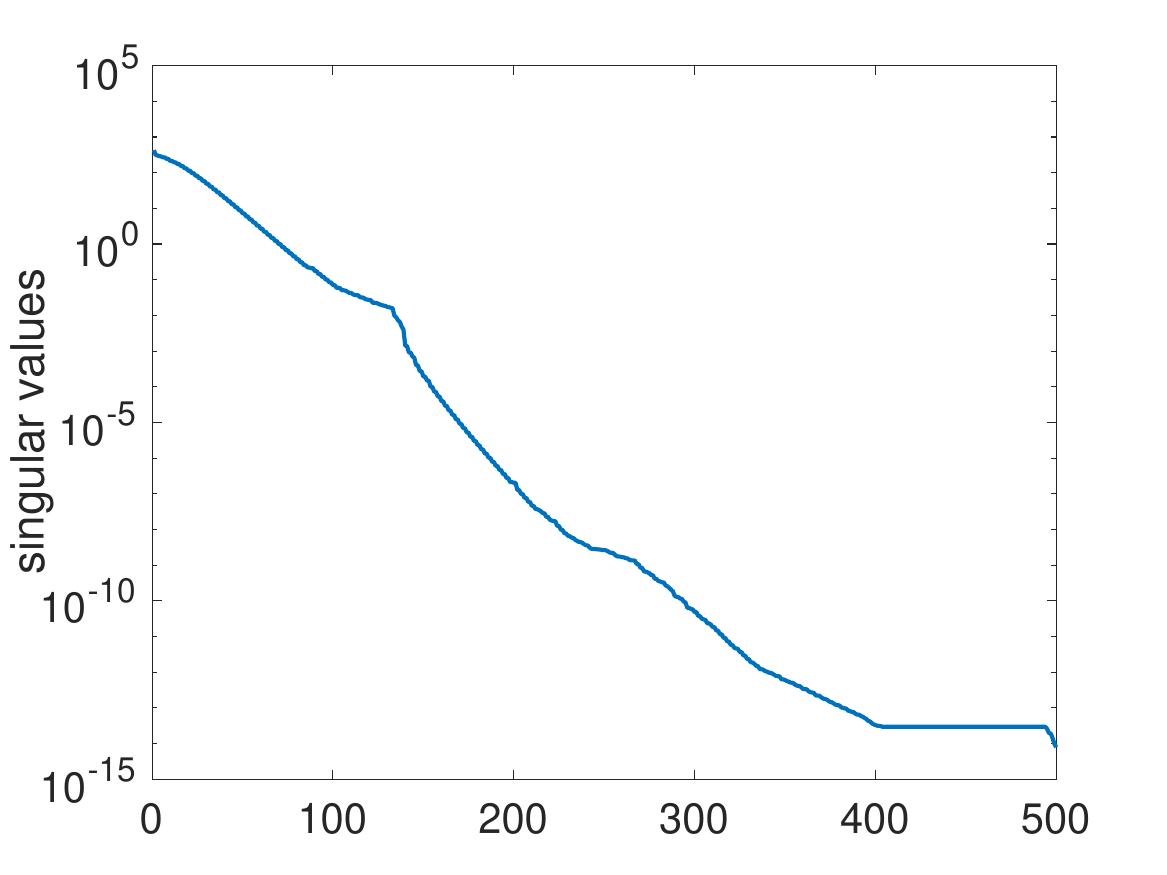}
        \caption{Singular values of the snapshot matrix for the KdV forward simulation.}
        \label{fig:kdv_svd_singular_values}
    \end{figure}
    We use the ansatz that $ u(x,t)  \approx \sum\limits_{j=1}^{r}  a_{j}(t) \phi_{j}(x) $ following \Cref{subsec:rom}, and replace them in \cref{eq:kdv} to get the following:
    \begin{equation}
    \label{eq:kdv_approx_u}
    \begin{split}
        \dfrac{\partial}{\partial t} \sum\limits_{j=1}^{r} a_{j}(t) \phi_{j}(x) + 2 \alpha \left( \sum\limits_{j=1}^{r} a_{j}(t) \phi_{j}(x) \right) \dfrac{\partial}{\partial x} \left( \sum\limits_{k=1}^{r} a_{k}(t) \phi_{k}(x) \right) + \nu \dfrac{\partial^3}{\partial x^{3}} \left( \sum\limits_{j=1}^{r} a_{j}(t) \phi_{j}(x) \right) &= 0,  \\
        \sum\limits_{j=1}^{r} \dot{a}_{j}^{} (t) \phi_{j}(x) + 2 \alpha \left(  \sum\limits_{j=1}^{r} a_{j}(t) \phi_{j}(x) \right) \left( \sum\limits_{k=1}^{r} a_{k}^{} (t) \dfrac{d \phi_{k}(x)}{dx}  \right) + \nu \left( \sum\limits_{j=1}^{r} a_{j}(t) \dfrac{d^{3} \phi_{j}(x)}{dx^{3}}  \right) &= 0 .
    \end{split}    
    \end{equation}
    Multiplying \cref{eq:kdv_approx_u} by $\phi_{i}$, taking inner products  and using the orthogonality property of the basis function, we get:
    \begin{equation}
    \label{eq:kdv_rom_projection}
        \begin{split}
            \sum\limits_{j=1}^{r} & \dot{a}_{j} \inner{\phi_{i}}{\phi_{j}} + 2 \alpha \sum\limits_{j=1}^{r} \sum\limits_{k=1}^{r} a_{j}(t) a_{k}(t) \inner{\phi_{i}}{\phi_{j} \dfrac{d \phi_{k}(x)}{dx} }  + \nu \sum\limits_{j=1}^{k} a_{j} (t) \inner{\phi_{i}}{\dfrac{d^{3} \phi_{j}(x)}{dx^{3}}} =0, \\
            & \dot{a}_{i} + 2 \alpha \sum\limits_{j=1}^{r} \sum\limits_{k=1}^{r} Q_{i,j,k} a_{j}(t) a_{k}(t) + \nu \sum\limits_{j=1}^{r} L_{i,j} a_{j}(t)  = 0, \quad 1 \leq i \leq r,
        \end{split}
    \end{equation}
    where $Q_{i,j,k}$ is the quadratic tensor and $L_{i,j}$ is the linear operator. 
    Using finite difference to discretize \cref{eq:kdv} in space, we get the following equation:
    \begin{equation}
    \label{eq:kdv_fd_space}
        \dot{\bfu} + 2 \alpha \bfu \odot \left(\bfD_{1} \bfu\right) + \nu \bfD_{3} \bfu = \bfzero,
    \end{equation}
    where $\bfu \in \symbb{R}^{n}$ are the nodal values, 
    $\bfD_{1} \in \symbb{R}^{n \times n} $ and $ \bfD_{3} \in \symbb{R}^{n \times n} $ are the finite-difference approximations of the first and third order spatial derivative operators respectively, and $ \odot $ denotes the Hadamard product.
    Using the state approximation  $ \left(u(x,t)  \approx \sum\limits_{j=1}^{r}  a_{j}(t) \phi_{j}(x) \right) $mentioned earlier, 
    we get:
    \begin{equation}
    \label{eq:kdv_fd_rom_projection}
        \begin{split}
            \bfPhi \dot{\bfa} + 2 \alpha \left( \bfPhi \bfa \right) \odot \left( D_{1} \bfPhi \bfa \right) + \nu \bfD_{3} \bfPhi \bfa = \bfzero \\ 
            \bfPhi^{T} \bfPhi \dot{\bfa} + 2 \, \alpha \, \bfPhi^{T} \left[ \left( \bfPhi \bfa \right) \odot \left( D_{1} \bfPhi \bfa \right) \right] + \nu \left( \bfPhi^{T} D_{3} \bfPhi \right) \bfa = \bfzero \\
            \implies \dfrac{d \bfa_{i}}{dt} = - \sum\limits_{j}^{r} L_{i,j} \, \bfa_{j} - 6 \sum\limits_{j}^{r} \sum\limits_{k}^{r} Q_{i,j,k} \, \bfa_{j} \, \bfa_{k}
    , \quad 1 \leq i \leq r   \end{split}
    \end{equation}
    where $L_{i,j} = \bfPhi_{i}^{T} D_{3} \bfPhi_{j} $ and $Q_{i,j,k} = \bfPhi_{i}^{T} \left[ \bfPhi_{j} \odot (D_{1} \bfPhi_{k}) \right] $.
    The $\bfPhi$ is our optimal operator in this test problem (refer \cref{subsec:kdv}).

\section{ROM for QG}\label{appsec:qg_rom}
    \setcounter{figure}{0} 
    We start with building the vorticity basis function with $2000$ snapshots at a time interval of $30 \times 0.0109$ unit time, which is equivalent to nearly a month. 
    Following \cite{stabilized_pod_qg_omer_trian}, we define a correlation matrix and evaluate it using a quadrature rule:
    \begin{equation}
    \label{eq:qg_correlation}
        \mathcal{C}_{i}^{j} = \inner{\omega_{i}}{\omega_{j}}, \quad i,j = 1,2,\dots, s,
    \end{equation}
    where $s$ is the number of snapshots derived from numerical simulation. 
    Solving the eigenvalue problem, i.e. $\mathcal{C} \Gamma = \Gamma \Sigma\ $, provides us the optimal and most informative POD basis functions \cite{rom_book_1, Holmes_Lumley_Berkooz_1996_book_roms, rom_pod_literature_1}. 
    The diagonal matrix $\Sigma$ contains the eigen values in the descending order, i.e. $ \Sigma  = [ \varsigma_{1}, \varsigma_{2}, \dots, \varsigma_{n}], $ where $ \varsigma_{1} \geq \varsigma_{2} \geq \varsigma_{3} \geq \dots \geq \varsigma_{n} $.

\begin{figure}[htbp]
        \centering    \includegraphics[width=0.55\linewidth, alt={Line graph depicting the eigenvalues of the correlation  matrix}]{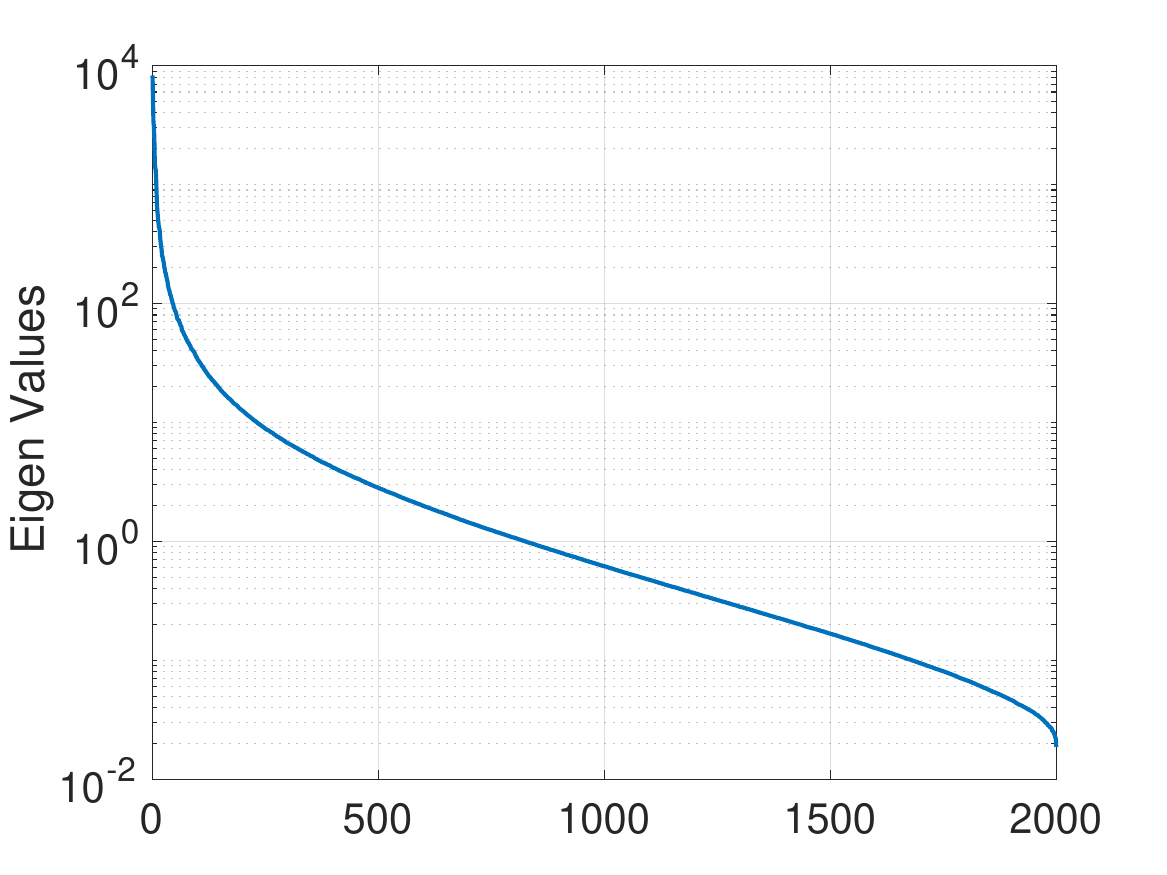}
        \caption{Eigenvalues of the correlation matrix $\mathcal{C}$}
        \label{fig:qg_correlation_eigen_values}
    \end{figure}

    \Cref{fig:qg_correlation_eigen_values} shows the eigenvalues of correlation matrix.
    The matrix $ \Gamma = [\zeta_{1}, \zeta_{2}, \dots, \zeta_{n}] $ are the corresponding eigenvectors. 
    We truncate and keep the first $r$ most informative modes (where $r \ll n $) to create the basis for the vorticity. 
    The ROM streamfunction basis can be derived using the relationship $ \Delta \phi_{i} = - \zeta_{i} $, and the corresponding approximations are:
    \begin{equation}
    \label{eq:vorticity_steamfunction_rom_approx}
        \widetilde{\omega} (t,x,y) = \sum\limits_{i=1}^{r} a_{i} (t) \zeta_{i} (x,y) ~, \quad \quad \widetilde{\psi}_{} (t,x,y) =  \sum\limits_{i=1}^{r} a_{i}(t) \phi_{i} (x,y),  
    \end{equation}
    where we use the same time coefficients $a_{i} (t)$ for both approximations.
    Plugging \cref{eq:vorticity_steamfunction_rom_approx} into \cref{eq:qge}, multiplying by the vorticity basis vector $\zeta_{i}$, and integrating, we get \cite{Sandu_2021_MF-EnKF,stabilized_pod_qg_omer_trian}:
    \begin{multline}
    \label{eq:qg_galerkin_projection_1}
    a'_{i} (t) = - \left\{ \sum\limits_{j =1}^{r} \sum\limits_{k=1}^{r} a_{j} (t) \inner{\zeta_{i}}{\frac{d \phi_{j}}{dy}  \frac{d \zeta_{k}}{dx}} a_{k}(t) -  \sum\limits_{j=1}^{r} \sum\limits_{k=1}^{r} a_{j} (t) \inner{\zeta_{i}}{\frac{d \phi_{j}}{dx}  \frac{d \zeta_{k}}{dy}} a_{k}(t)  \right\} \\
    + \mathrm{Ro}_{}^{-1} \sum\limits_{j=1}^{r} a_{j}(t) \inner{\zeta_{i}}{\frac{d \phi_{j}}{dx}}
    - \mathrm{Re}_{}^{-1} \sum\limits_{j=1}^{r} a_{j}(t) \inner{ \nabla \zeta_{i}}{\nabla \zeta_{j} } + \mathrm{Ro}_{}^{-1} \inner{F}{\zeta_{i}}, \quad 1 \leq i \leq r.
    \end{multline}
    %
    The final condensed form of \cref{eq:qg_galerkin_projection_1} can be written as:
    \begin{equation}
    \label{eq:qg_galerkin_projection_2}
        a' (t) = b + \mathcal{A} a + a^{T} \mathcal{B} a, 
    \end{equation}
    where $b_{i} = \mathrm{Ro}_{}^{-1} \inner{F}{\zeta_{i}}$, $\mathcal{A}_{i,j} = \mathrm{Ro}_{}^{-1} \inner{\zeta_{i}}{\frac{d \phi_{j}}{dx}} - \mathrm{Re}_{}^{-1} \inner{ \nabla \zeta_{i}}{\nabla \zeta_{j} }, \mathcal{B}_{i,j,k} = \inner{\zeta_{i}}{\frac{d \phi_{j}}{dy}  \frac{d \zeta_{k}}{dx}} - \inner{\zeta_{i}}{\frac{d \phi_{j}}{dx}  \frac{d \zeta_{k}}{dy}} $ .

%% file: bib/reference.bib
@book{reich_2015_dabook, 
place={Cambridge}, 
title={Probabilistic Forecasting and Bayesian Data Assimilation}, 
publisher={Cambridge University Press}, 
author={Reich, Sebastian and Cotter, Colin},
year={2015}}

@book{markasch_2016_dabook,
author = {Asch, Mark and Bocquet, Marc and Nodet, Maëlle},
title = {Data Assimilation},
publisher = {Society for Industrial and Applied Mathematics},
year = {2016},
doi = {10.1137/1.9781611974546},
address = {Philadelphia, PA},
edition   = {},
URL = {https://epubs.siam.org/doi/abs/10.1137/1.9781611974546},
eprint = {https://epubs.siam.org/doi/pdf/10.1137/1.9781611974546}
}

@article{vanleeuwen_2019_pf_highdimension,
author = {van Leeuwen, Peter Jan and Künsch, Hans R. and Nerger, Lars and Potthast, Roland and Reich, Sebastian},
title = {Particle filters for high-dimensional geoscience applications: A review},
journal = {Quarterly Journal of the Royal Meteorological Society},
volume = {145},
number = {723},
pages = {2335-2365},
doi = {https://doi.org/10.1002/qj.3551},
url = {https://rmets.onlinelibrary.wiley.com/doi/abs/10.1002/qj.3551},
eprint = {https://rmets.onlinelibrary.wiley.com/doi/pdf/10.1002/qj.3551},
year = {2019}
}

@inproceedings{daum_2017_generalized_gromov_filter,
author = {Fred Daum and Jim Huang and Arjang Noushin},
title = {{Generalized Gromov method for stochastic particle flow filters}},
volume = {10200},
booktitle = {Signal Processing, Sensor/Information Fusion, and Target Recognition XXVI},
editor = {Ivan Kadar},
organization = {International Society for Optics and Photonics},
publisher = {SPIE},
pages = {102000I},
year = {2017},
doi = {10.1117/12.2248723},
URL = {https://doi.org/10.1117/12.2248723}
}

@INPROCEEDINGS{gromov_pal_1,
  author={Pal, Soumyasundar and Coates, Mark},
  booktitle={2019 IEEE 8th International Workshop on Computational Advances in Multi-Sensor Adaptive Processing (CAMSAP)}, 
  title={Particle Flow Particle Filter using Gromov's method}, 
  year={2019},
  volume={},
  number={},
  pages={634-638},
  doi={10.1109/CAMSAP45676.2019.9022494}}

@INPROCEEDINGS{gromov_duam_2,
  author={Daum, Fred and Huang, Jim and Noushin, Arjang},
  booktitle={2018 21st International Conference on Information Fusion (FUSION)}, 
  title={New Theory and Numerical Results for Gromov's Method for Stochastic Particle Flow Filters}, 
  year={2018},
  volume={},
  number={},
  pages={108-115},
  doi={10.23919/ICIF.2018.8455287}}

@inproceedings{daum_log_homotopy,
author = {Fred Daum and Jim Huang},
title = {{Particle flow for nonlinear filters with log-homotopy}},
volume = {6969},
booktitle = {Signal and Data Processing of Small Targets 2008},
editor = {Oliver E. Drummond},
organization = {International Society for Optics and Photonics},
publisher = {SPIE},
pages = {696918},
year = {2008},
doi = {10.1117/12.764909},
URL = {https://doi.org/10.1117/12.764909}
}

@inproceedings{daum_normalization_const,
author = {Fred Daum and Jim Huang},
title = {{How to avoid normalization of particle flow for nonlinear filters, Bayesian decisions, and transport}},
volume = {9092},
booktitle = {Signal and Data Processing of Small Targets 2014},
editor = {Oliver E. Drummond},
organization = {International Society for Optics and Photonics},
publisher = {SPIE},
pages = {90920B},
year = {2014},
doi = {10.1117/12.2044122},
URL = {https://doi.org/10.1117/12.2044122}
}

@article{mfenkf_popov,
author = {Popov, Andrey A. and Mou, Changhong and Sandu, Adrian and Iliescu, Traian},
title = {A Multifidelity Ensemble Kalman Filter with Reduced Order Control Variates},
journal = {SIAM Journal on Scientific Computing},
volume = {43},
number = {2},
pages = {A1134-A1162},
year = {2021},
doi = {10.1137/20M1349965},
URL = { 
       https://doi.org/10.1137/20M1349965
},
eprint = {    
    https://doi.org/10.1137/20M1349965   
}

}

@article{Mou07022020,
author = {Changhong Mou and Honghu Liu and David R. Wells and Traian Iliescu},
title = {Data-driven correction reduced order models for the quasi-geostrophic equations: a numerical investigation},
journal = {International Journal of Computational Fluid Dynamics},
volume = {34},
number = {2},
pages = {147--159},
year = {2020},
publisher = {IAHR Website},
doi = {10.1080/10618562.2020.1723556},
URL = {             https://doi.org/10.1080/10618562.2020.1723556},
eprint = {   
        https://doi.org/10.1080/10618562.2020.1723556}
}

@article{stabilized_pod_qg_omer_trian,
author = {San, Omer and Iliescu, Traian},
year = {2014},
month = {01},
pages = {},
title = {A stabilized proper orthogonal decomposition reduced-order model for large scale quasigeostrophic ocean circulation},
volume = {41},
journal = {Advances in Computational Mathematics},
doi = {10.1007/s10444-015-9417-0}
}

@INPROCEEDINGS{daum_exactflow,
  author={Daum, Fred and Huang, Jim},
  booktitle={2010 Conference Record of the Forty Fourth Asilomar Conference on Signals, Systems and Computers}, 
  title={Exact particle flow for nonlinear filters: Seventeen dubious solutions to a first order linear underdetermined PDE}, 
  year={2010},
  volume={},
  number={},
  pages={64-71},
  doi={10.1109/ACSSC.2010.5757468}}

@inproceedings{daum_generalized_pf_nonlinear_filter,
author = {Fred Daum and Jim Huang},
title = {{Generalized particle flow for nonlinear filters}},
volume = {7698},
booktitle = {Signal and Data Processing of Small Targets 2010},
editor = {Oliver E. Drummond},
organization = {International Society for Optics and Photonics},
publisher = {SPIE},
pages = {76980I},
year = {2010},
doi = {10.1117/12.839421},
URL = {https://doi.org/10.1117/12.839421}
}

@article{kovachki_kalman_inversion,
doi = {10.1088/1361-6420/ab1c3a},
url = {https://doi.org/10.1088/1361-6420/ab1c3a},
year = {2019},
month = {aug},
publisher = {IOP Publishing},
volume = {35},
number = {9},
pages = {095005},
author = {Kovachki, Nikola B and Stuart, Andrew M},
title = {Ensemble Kalman inversion: a derivative-free technique for machine learning tasks},
journal = {Inverse Problems}
}

@article {leeuwen_pf_paper_1,
      author = "Peter Jan van Leeuwen",
      title = "Particle Filtering in Geophysical Systems",
      journal = "Monthly Weather Review",
      year = "2009",
      publisher = "American Meteorological Society",
      address = "Boston MA, USA",
      volume = "137",
      number = "12",
      doi = "10.1175/2009MWR2835.1",
      pages=      "4089 - 4114",
      url = "https://journals.ametsoc.org/view/journals/mwre/137/12/2009mwr2835.1.xml"
}

@Inbook{pf_vanLeeuwen2015,
author="van Leeuwen, Peter Jan",
title="Nonlinear Data Assimilation for high-dimensional systems",
bookTitle="Nonlinear Data Assimilation",
year="2015",
publisher="Springer International Publishing",
address="Cham",
pages="1--73",
isbn="978-3-319-18347-3",
doi="10.1007/978-3-319-18347-3_1",
url="https://doi.org/10.1007/978-3-319-18347-3_1"
}

@article{pf_vanLeeuwen2010,
author = {van Leeuwen, P. J.},
title = {Nonlinear data assimilation in geosciences: an extremely efficient particle filter},
journal = {Quarterly Journal of the Royal Meteorological Society},
volume = {136},
number = {653},
pages = {1991-1999},
doi = {https://doi.org/10.1002/qj.699},
url = {https://rmets.onlinelibrary.wiley.com/doi/abs/10.1002/qj.699},
eprint = {https://rmets.onlinelibrary.wiley.com/doi/pdf/10.1002/qj.699},
year = {2010}
}

@article {Synder_ObstaclestoHighDimensionalParticleFiltering,
      author = "Chris Snyder and Thomas Bengtsson and Peter Bickel and Jeff Anderson",
      title = "Obstacles to High-Dimensional Particle Filtering",
      journal = "Monthly Weather Review",
      year = "2008",
      publisher = "American Meteorological Society",
      address = "Boston MA, USA",
      volume = "136",
      number = "12",
      doi = "10.1175/2008MWR2529.1",
      pages=      "4629 - 4640",
      url = "https://journals.ametsoc.org/view/journals/mwre/136/12/2008mwr2529.1.xml"
}

@INPROCEEDINGS{daum_degeneracy_1,
       author = {{Daum}, Fred and {Huang}, Jim},
        title = "{Particle degeneracy: root cause and solution}",
    booktitle = {Signal Processing, Sensor Fusion, and Target Recognition XX},
         year = 2011,
       editor = {{Kadar}, Ivan},
       series = {Society of Photo-Optical Instrumentation Engineers (SPIE) Conference Series},
       volume = {8050},
        month = jun,
          eid = {80500W},
        pages = {80500W},
          doi = {10.1117/12.877167},
       adsurl = {https://ui.adsabs.harvard.edu/abs/2011SPIE.8050E..0WD}
}

@INPROCEEDINGS{highdimensions_pf_1,
  author={Djurić, Petar M. and Bugallo, Mónica F.},
  booktitle={2013 5th IEEE International Workshop on Computational Advances in Multi-Sensor Adaptive Processing (CAMSAP)}, 
  title={Particle filtering for high-dimensional systems}, 
  year={2013},
  volume={},
  number={},
  pages={352-355},
  doi={10.1109/CAMSAP.2013.6714080}}

@inproceedings{highdimensions_pf_2,
  title={Particle filters, the “optimal” proposal and high-dimensional systems},
  author={Snyder, Chris},
  booktitle={Proceedings of the ECMWF Seminar on Data Assimilation for atmosphere and ocean},
  pages={1--10},
  year={2011}
}

@article{highdimensions_pf_3,
author = {Hu, Chih-Chi and van Leeuwen, Peter Jan},
title = {A particle flow filter for high-dimensional system applications},
journal = {Quarterly Journal of the Royal Meteorological Society},
volume = {147},
number = {737},
pages = {2352-2374},
doi = {https://doi.org/10.1002/qj.4028},
url = {https://rmets.onlinelibrary.wiley.com/doi/abs/10.1002/qj.4028},
eprint = {https://rmets.onlinelibrary.wiley.com/doi/pdf/10.1002/qj.4028},
year = {2021}
}

@article{etpf_reich_2013,
author = {Reich, Sebastian},
title = {A Nonparametric Ensemble Transform Method for Bayesian Inference},
journal = {SIAM Journal on Scientific Computing},
volume = {35},
number = {4},
pages = {A2013-A2024},
year = {2013},
doi = {10.1137/130907367},
URL = { 
        https://doi.org/10.1137/130907367
},
eprint = { 
        https://doi.org/10.1137/130907367
}
}

@article{giles2015multilevel,
  title={Multilevel monte carlo methods},
  author={Giles, Michael B},
  journal={Acta numerica},
  volume={24},
  pages={259--328},
  year={2015},
  publisher={Cambridge University Press}
}

@article{mlenkf,
author = {Hoel, H\r{a}kon and Law, Kody J. H. and Tempone, Raul},
title = {Multilevel ensemble Kalman filtering},
journal = {SIAM Journal on Numerical Analysis},
volume = {54},
number = {3},
pages = {1813-1839},
year = {2016},
doi = {10.1137/15M100955X},

URL = {     
        https://doi.org/10.1137/15M100955X
},
eprint = {     
        https://doi.org/10.1137/15M100955X
}
}

@article{mlenkf_2,
author = {Chernov, Alexey and Hoel, Håkon and Law, Kody and Nobile, Fabio and Tempone, Raul},
year = {2016},
month = {08},
pages = {},
title = {Multilevel ensemble Kalman filtering for spatially extended models},
journal={arXiv preprint arXiv:1608.08558},
doi = {10.48550/arXiv.1608.08558}
}

@article{mletpf,
author = {Gregory, A. and Cotter, C. J. and Reich, S.},
title = {Multilevel Ensemble Transform Particle Filtering},
journal = {SIAM Journal on Scientific Computing},
volume = {38},
number = {3},
pages = {A1317-A1338},
year = {2016},
doi = {10.1137/15M1038232},
URL = {     
        https://doi.org/10.1137/15M1038232
},
eprint = {     
        https://doi.org/10.1137/15M1038232
}
}

@article{rabier_2000_ecmwf_4dvar,
  title={The ECMWF operational implementation of four-dimensional variational assimilation. I: Experimental results with simplified physics},
  author={Rabier, Florence and J{\"a}rvinen, Heikki and Klinker, Erland and Mahfouf, J-F and Simmons, Adrian},
  journal={Quarterly Journal of the Royal Meteorological Society},
  volume={126},
  number={564},
  pages={1143--1170},
  year={2000},
  publisher={Wiley Online Library}
}

@article{wang_2010_economical_4dvar,
  title={An economical approach to four-dimensional variational data assimilation},
  author={Wang, Bin and Liu, Juanjuan and Wang, Shudong and Cheng, Wei and Juan, Liu and Liu, Chengsi and Xiao, Qingnong and Kuo, Ying-Hwa},
  journal={Advances in Atmospheric Sciences},
  volume={27},
  number={4},
  pages={715--727},
  year={2010},
  publisher={Springer}
}

@article{4dvar_rom_1,
author = {Cao, Yanhua and Zhu, Jiang and Navon, I. M. and Luo, Zhendong},
title = {A reduced-order approach to four-dimensional variational data assimilation using proper orthogonal decomposition},
journal = {International Journal for Numerical Methods in Fluids},
volume = {53},
number = {10},
pages = {1571-1583},
doi = {https://doi.org/10.1002/fld.1365},
url = {https://onlinelibrary.wiley.com/doi/abs/10.1002/fld.1365},
eprint = {https://onlinelibrary.wiley.com/doi/pdf/10.1002/fld.1365},
year = {2007}
}

@article{4dvar_rom_2,
author = {Daescu, D. N. and Navon, I. M.},
title = {Efficiency of a POD-based reduced second-order adjoint model in 4D-Var data assimilation},
journal = {International Journal for Numerical Methods in Fluids},
volume = {53},
number = {6},
pages = {985-1004},
doi = {https://doi.org/10.1002/fld.1316},
url = {https://onlinelibrary.wiley.com/doi/abs/10.1002/fld.1316},
eprint = {https://onlinelibrary.wiley.com/doi/pdf/10.1002/fld.1316},
year = {2007}
}

@article {liu_2009_4DEnVar,
      author = "Chengsi Liu and Qingnong Xiao and Bin Wang",
      title = "An Ensemble-Based Four-Dimensional Variational Data Assimilation Scheme. Part II: Observing System Simulation Experiments with Advanced Research WRF (ARW)",
      journal = "Monthly Weather Review",
      year = "2009",
      publisher = "American Meteorological Society",
      address = "Boston MA, USA",
      volume = "137",
      number = "5",
      doi = "10.1175/2008MWR2699.1",
      pages=      "1687 - 1704",
      url = "https://journals.ametsoc.org/view/journals/mwre/137/5/2008mwr2699.1.xml"
}

@Article{smoothers_1,
AUTHOR = {Bocquet, M. and Sakov, P.},
TITLE = {Joint state and parameter estimation with an iterative ensemble Kalman smoother},
JOURNAL = {Nonlinear Processes in Geophysics},
VOLUME = {20},
YEAR = {2013},
NUMBER = {5},
PAGES = {803--818},
URL = {https://npg.copernicus.org/articles/20/803/2013/},
DOI = {10.5194/npg-20-803-2013}
}

@article {smoothers_2,
      author = "Geir Evensen and Peter Jan van Leeuwen",
      title = "An Ensemble Kalman Smoother for Nonlinear Dynamics",
      journal = "Monthly Weather Review",
      year = "2000",
      publisher = "American Meteorological Society",
      address = "Boston MA, USA",
      volume = "128",
      number = "6",
      doi = "10.1175/1520-0493(2000)128<1852:AEKSFN>2.0.CO;2",
      pages=      "1852 - 1867",
      url = "https://journals.ametsoc.org/view/journals/mwre/128/6/1520-0493_2000_128_1852_aeksfn_2.0.co_2.xml"
}

@book{evensen2022_book,
  title={Data assimilation fundamentals: A unified formulation of the state and parameter estimation problem},
  author={Evensen, Geir and Vossepoel, Femke C and Van Leeuwen, Peter Jan},
  year={2022},
  publisher={Springer}
}

@article{evensen_1994_enkf,
author = {Evensen, Geir},
title = {Sequential data assimilation with a nonlinear quasi-geostrophic model using Monte Carlo methods to forecast error statistics},
journal = {Journal of Geophysical Research: Oceans},
volume = {99},
number = {C5},
pages = {10143-10162},
doi = {https://doi.org/10.1029/94JC00572},
url = {https://agupubs.onlinelibrary.wiley.com/doi/abs/10.1029/94JC00572},
eprint = {https://agupubs.onlinelibrary.wiley.com/doi/pdf/10.1029/94JC00572},
year = {1994}
}

@article{evensen_2003_enkf,
  title={The ensemble Kalman filter: Theoretical formulation and practical implementation},
  author={Evensen, Geir},
  journal={Ocean dynamics},
  volume={53},
  pages={343--367},
  year={2003},
  publisher={Springer}
}

@article{attia_smoothers_mf,
author = {Attia, Ahmed and Ştefănescu, Răzvan and Sandu, Adrian},
title = {The reduced-order hybrid Monte Carlo sampling smoother},
journal = {International Journal for Numerical Methods in Fluids},
volume = {83},
number = {1},
pages = {28-51},
doi = {https://doi.org/10.1002/fld.4255},
url = {https://onlinelibrary.wiley.com/doi/abs/10.1002/fld.4255},
eprint = {https://onlinelibrary.wiley.com/doi/pdf/10.1002/fld.4255},
year = {2017}
}

@book{rom_book_1,
  title={Reduced basis methods for partial differential equations: an introduction},
  author={Quarteroni, Alfio and Manzoni, Andrea and Negri, Federico},
  volume={92},
  year={2015},
  publisher={Springer}
}

@article{rom_non_linear_nguyen,
author = {Nguyen, N. C. and Peraire, J.},
title = {An efficient reduced-order modeling approach for non-linear parametrized partial differential equations},
journal = {International Journal for Numerical Methods in Engineering},
volume = {76},
number = {1},
pages = {27-55},
doi = {https://doi.org/10.1002/nme.2309},
url = {https://onlinelibrary.wiley.com/doi/abs/10.1002/nme.2309},
eprint = {https://onlinelibrary.wiley.com/doi/pdf/10.1002/nme.2309},
year = {2008}
}

@book{Brunton_Kutz_2019,
place={Cambridge},
title={Data-Driven Science and Engineering: Machine Learning, Dynamical Systems, and Control},
publisher={Cambridge University Press},
author={Brunton, Steven L. and Kutz, J. Nathan},
year={2019}}

@article{deim,
author = {Chaturantabut, Saifon and Sorensen, Danny C.},
title = {Nonlinear Model Reduction via Discrete Empirical Interpolation},
journal = {SIAM Journal on Scientific Computing},
volume = {32},
number = {5},
pages = {2737-2764},
year = {2010},
doi = {10.1137/090766498},
URL = { https://doi.org/10.1137/090766498
},
eprint = {     
https://doi.org/10.1137/090766498
}
}

@article{empirical_interpolation,
title = {An ‘empirical interpolation’ method: application to efficient reduced-basis discretization of partial differential equations},
journal = {Comptes Rendus Mathematique},
volume = {339},
number = {9},
pages = {667-672},
year = {2004},
issn = {1631-073X},
doi = {https://doi.org/10.1016/j.crma.2004.08.006},
url = {https://www.sciencedirect.com/science/article/pii/S1631073X04004248},
author = {Maxime Barrault and Yvon Maday and Ngoc Cuong Nguyen and Anthony T. Patera}
}

@inbook{rom_inverse_problems_1,
author = {Frangos, M. and Marzouk, Y. and Willcox, K. and van Bloemen Waanders, B.},
publisher = {John Wiley \& Sons, Ltd},
isbn = {9780470685853},
title = {Surrogate and Reduced-Order Modeling: A Comparison of Approaches for Large-Scale Statistical Inverse Problems},
booktitle = {Large‐Scale Inverse Problems and Quantification of Uncertainty},
chapter = {7},
pages = {123-149},
doi = {https://doi.org/10.1002/9780470685853.ch7},
url = {https://onlinelibrary.wiley.com/doi/abs/10.1002/9780470685853.ch7},
eprint = {https://onlinelibrary.wiley.com/doi/pdf/10.1002/9780470685853.ch7},
year = {2010}
}

@article{rom_inverse_problems_2,
author = {Cui, Tiangang and Marzouk, Youssef M. and Willcox, Karen E.},
title = {Data-driven model reduction for the Bayesian solution of inverse problems},
journal = {International Journal for Numerical Methods in Engineering},
volume = {102},
number = {5},
pages = {966-990},
doi = {https://doi.org/10.1002/nme.4748},
url = {https://onlinelibrary.wiley.com/doi/abs/10.1002/nme.4748},
eprint = {https://onlinelibrary.wiley.com/doi/pdf/10.1002/nme.4748},
year = {2015}
}

@article{rom_inverse_problems_3,
author = {Galbally, D. and Fidkowski, K. and Willcox, K. and Ghattas, O.},
title = {Non-linear model reduction for uncertainty quantification in large-scale inverse problems},
journal = {International Journal for Numerical Methods in Engineering},
volume = {81},
number = {12},
pages = {1581-1608},
doi = {https://doi.org/10.1002/nme.2746},
url = {https://onlinelibrary.wiley.com/doi/abs/10.1002/nme.2746},
eprint = {https://onlinelibrary.wiley.com/doi/pdf/10.1002/nme.2746},
year = {2010}
}

@article{multivariate_control_variate,
 ISSN = {0030364X, 15265463},
 URL = {http://www.jstor.org/stable/170564},
 author = {Reuven Y. Rubinstein and Ruth Marcus},
 journal = {Operations Research},
 number = {3},
 pages = {661--677},
 publisher = {INFORMS},
 title = {Efficiency of Multivariate Control Variates in Monte Carlo Simulation},
 urldate = {2025-12-26},
 volume = {33},
 year = {1985}
}

@article{zanetti_andrey_survey,
  title = {A Survey of Nonlinear Estimation Filters},
  author={Zanetti, Renato and Popov, Andrey A and Michaelson, Kristen and Giraldo-Grueso, Felipe and Durant, Dalton and Servadio, Simone and Hanebeck, Uwe D},
  journal={Journal of Advances in Information Fusion},
  volume={20},
  number={2},
  pages={97},
  year={2025}
}

@techreport{crouse_flow_filter_bias,
    author = {David Frederic Crouse and Codie Tyler Lewis},
    title = {Consideration of Particle Flow Filter Implementations and Biases},
    institution = {Naval Research Laboratory},
    year = {2020}
}

@book{pod_Holmes_Lumley_Berkooz_1996,
place={Cambridge},
series={Cambridge Monographs on Mechanics},
title={Turbulence, Coherent Structures, Dynamical Systems and Symmetry},
publisher={Cambridge University Press},
author={Holmes, Philip and Lumley, John L. and Berkooz, Gal},
year={1996},
collection={Cambridge Monographs on Mechanics}}

@article{method_of_snapshot,
author = {Sirovich, Lawrence},
year = {1987},
month = {10},
pages = {},
title = {Turbulence and the dynamics of coherent structures. I - Coherent structures. II - Symmetries and transformations. III - Dynamics and scaling},
volume = {45},
journal = {Quarterly of Applied Mathematics - QUART APPL MATH},
doi = {10.1090/qam/910463}
}

@article{kdv_geometric_denys,
  title={Geometric numerical schemes for the KdV equation},
  author={Denys Dutykh and Marx Chhay and Francesco G. Fedele},
  journal={Computational Mathematics and Mathematical Physics},
  year={2012},
  volume={53},
  pages={221 - 236},
  url={https://api.semanticscholar.org/CorpusID:53485055}
}

@article{kdv_ascher,
author = {Ascher, U. M. and McLachlan, R. I.},
title = {On Symplectic and Multisymplectic Schemes for the KdV Equation},
year = {2005},
issue_date = {Oct 2005},
publisher = {Plenum Press},
address = {USA},
volume = {25},
number = {1},
issn = {0885-7474},
url = {https://doi.org/10.1007/s10915-004-4634-6},
doi = {10.1007/s10915-004-4634-6},
journal = {J. Sci. Comput.},
month = oct,
pages = {83–104},
numpages = {22}
}

@book{Holmes_Lumley_Berkooz_1996_book_roms,
place={Cambridge},
series={Cambridge Monographs on Mechanics},
title={Turbulence, Coherent Structures, Dynamical Systems and Symmetry},
publisher={Cambridge University Press},
author={Holmes, Philip and Lumley, John L. and Berkooz, Gal},
year={1996},
collection={Cambridge Monographs on Mechanics}
}

@article{rom_pod_literature_1,
author = {Ravindran, S. S.},
title = {A reduced-order approach for optimal control of fluids using proper orthogonal decomposition},
journal = {International Journal for Numerical Methods in Fluids},
volume = {34},
number = {5},
pages = {425-448},
doi = {https://doi.org/10.1002/1097-0363(20001115)34:5<425::AID-FLD67>3.0.CO;2-W},
url = {https://onlinelibrary.wiley.com/doi/abs/10.1002/1097-0363%2820001115%2934%3A5%3C425%3A%3AAID-FLD67%3E3.0.CO%3B2-W},
eprint = {https://onlinelibrary.wiley.com/doi/pdf/10.1002/1097-0363%2820001115%2934%3A5%3C425%3A%3AAID-FLD67%3E3.0.CO%3B2-W},
year = {2000}
}

@article{gaspari_cohn,
author = {Gaspari, Gregory and Cohn, Stephen E.},
title = {Construction of correlation functions in two and three dimensions},
journal = {Quarterly Journal of the Royal Meteorological Society},
volume = {125},
number = {554},
pages = {723-757},
doi = {https://doi.org/10.1002/qj.49712555417},
url = {https://rmets.onlinelibrary.wiley.com/doi/abs/10.1002/qj.49712555417},
eprint = {https://rmets.onlinelibrary.wiley.com/doi/pdf/10.1002/qj.49712555417},
year = {1999}
}

@misc{roberts2019odetestproblemsmatlab,
      title={ODE Test Problems: a MATLAB suite of initial value problems}, 
      author={Steven Roberts and Andrey A. Popov and Adrian Sandu},
      year={2019},
      eprint={1901.04098},
      archivePrefix={arXiv},
      primaryClass={math.NA},
      url={https://arxiv.org/abs/1901.04098}, 
}

@article {blocalization,
      author = "Lili Lei and Jeffrey S. Whitaker",
      title = "Model Space Localization Is Not Always Better Than Observation Space Localization for Assimilation of Satellite Radiances",
      journal = "Monthly Weather Review",
      year = "2015",
      publisher = "American Meteorological Society",
      address = "Boston MA, USA",
      volume = "143",
      number = "10",
      doi = "10.1175/MWR-D-14-00413.1",
      pages=      "3948 - 3955",
      url = "https://journals.ametsoc.org/view/journals/mwre/143/10/mwr-d-14-00413.1.xml"
}

@article{karen_wilcox_multi_level_importance_sampling,
title = {Multifidelity importance sampling},
journal = {Computer Methods in Applied Mechanics and Engineering},
volume = {300},
pages = {490-509},
year = {2016},
issn = {0045-7825},
doi = {https://doi.org/10.1016/j.cma.2015.12.002},
url = {https://www.sciencedirect.com/science/article/pii/S004578251500393X},
author = {Benjamin Peherstorfer and Tiangang Cui and Youssef Marzouk and Karen Willcox}
}

@article{popov2026learning,
  title={Learning to Trust AI and Data-driven models in Data Assimilation through a Multifidelity Ensemble Gaussian Mixture Filter Framework},
  author={Popov, Andrey A},
  journal={arXiv preprint arXiv:2604.23060},
  year={2026}
}

@article{reich_gaussian_mixture_ensemble_transform_filter,
author = {Reich, Sebastian},
title = {A Gaussian-mixture ensemble transform filter},
journal = {Quarterly Journal of the Royal Meteorological Society},
volume = {138},
number = {662},
pages = {222-233},
doi = {https://doi.org/10.1002/qj.898},
url = {https://rmets.onlinelibrary.wiley.com/doi/abs/10.1002/qj.898},
eprint = {https://rmets.onlinelibrary.wiley.com/doi/pdf/10.1002/qj.898},
year = {2012}
}

@inproceedings{stein_variational_gradient_descent,
 author = {Liu, Qiang and Wang, Dilin},
 booktitle = {Advances in Neural Information Processing Systems},
 editor = {D. Lee and M. Sugiyama and U. Luxburg and I. Guyon and R. Garnett},
 pages = {},
 publisher = {Curran Associates, Inc.},
 title = {Stein Variational Gradient Descent: A General Purpose Bayesian Inference Algorithm},
 url = {https://proceedings.neurips.cc/paper_files/paper/2016/file/b3ba8f1bee1238a2f37603d90b58898d-Paper.pdf},
 volume = {29},
 year = {2016}
}

@article{pulido_leeuwen_mapping_particle_filter,
title = {Sequential Monte Carlo with kernel embedded mappings: The mapping particle filter},
journal = {Journal of Computational Physics},
volume = {396},
pages = {400-415},
year = {2019},
issn = {0021-9991},
doi = {https://doi.org/10.1016/j.jcp.2019.06.060},
url = {https://www.sciencedirect.com/science/article/pii/S0021999119304681},
author = {Manuel Pulido and Peter Jan {van Leeuwen}}
}

@article{reich_fokker_planck_particle_filter,
author = {Reich, Sebastian and Weissmann, Simon},
title = {Fokker--Planck Particle Systems for Bayesian Inference: Computational Approaches},
journal = {SIAM/ASA Journal on Uncertainty Quantification},
volume = {9},
number = {2},
pages = {446-482},
year = {2021},
doi = {10.1137/19M1303162},

URL = { 
https://doi.org/10.1137/19M1303162
},
eprint = {         https://doi.org/10.1137/19M1303162
}
}

@article{opinf_qian_ionut_karen,
author = {Qian, Elizabeth and Farca\c{s}, Ionu\c{t}-Gabriel and Willcox, Karen},
title = {Reduced Operator Inference for Nonlinear Partial Differential Equations},
journal = {SIAM Journal on Scientific Computing},
volume = {44},
number = {4},
pages = {A1934-A1959},
year = {2022},
doi = {10.1137/21M1393972},
URL = { 
        https://doi.org/10.1137/21M1393972    
},
eprint = {  
    https://doi.org/10.1137/21M1393972
}
}

@conference{monica_math_dt_1,
author = {Monica Shanmugam and Abhinab Bhattacharjee and K. Todd Lowe and Christopher J. Roy and Adrian Sandu and Rui Qiao and Ionut-Gabriel Farcas},
title = {Building a Digital Twin Framework for Propulsion Engines With Particle Damage},
booktitle = {AIAA AVIATION 2026 Forum},
chapter = {},
pages = {},
year = {2026},
doi = {10.2514/6.2026-4021},
URL = {https://arc.aiaa.org/doi/abs/10.2514/6.2026-4021},
eprint = {https://arc.aiaa.org/doi/pdf/10.2514/6.2026-4021}
}

@article{park_mou_liu_sandu_illescu_2026_two_level_ROM,
author = {Park, Dylan and Mou, Changhong and Liu, Honghu and Sandu, Adrian and Iliescu, Traian},
title = {A Two-Level Galerkin Reduced Order Model for Nonlinear Problems},
journal = {Numerical Methods for Partial Differential Equations},
volume = {42},
number = {6},
pages = {e70148},
doi = {https://doi.org/10.1002/num.70148},
url = {https://onlinelibrary.wiley.com/doi/abs/10.1002/num.70148},
eprint = {https://onlinelibrary.wiley.com/doi/pdf/10.1002/num.70148},
year = {2026}
}


%% file: bib/sandu.bib
@article{Sandu_2026_model-forest-EnKF,
	archiveprefix = {arXiv},
	author = {Andrey A Popov and Adrian Sandu},
	doi = {10.1137/25M1792943},
	eprint = {https://arxiv.org/abs/2210.11971},
	journal = {SIAM/ASA Journal on Uncertainty Quantification},
	number = {3},
	primaryclass = {cs.CE},
	subjects = {data-assimilation},
	title = {The Model Forest Ensemble Kalman Filter},
	url = {https://doi.org/10.1137/25M1792943},
	volume = {14},
	year = {2026}}

@article{Sandu_2025_VariationalFokkerPlanck,
	author = {Amit N. Subrahmanya and Andrey A. Popov and Adrian Sandu},
	doi = {https://doi.org/10.1016/j.jcp.2024.113681},
	issn = {0021-9991},
	journal = {Journal of Computational Physics},
	pages = {113681},
	title = {Ensemble variational Fokker-Planck methods for data assimilation},
	url = {https://www.sciencedirect.com/science/article/pii/S002199912400929X},
	volume = {523},
	year = {2025}}

@article{Sandu_2021_MF-EnKF,
	author = {Popov, Andrey A. and Mou, Changhong and Sandu, Adrian and Iliescu, Traian},
	doi = {10.1137/20M1349965},
	eprint = {https://doi.org/10.1137/20M1349965},
	journal = {SIAM Journal on Scientific Computing},
	number = {2},
	pages = {A1134-A1162},
	subjects = {inverse-problems, surrogate-models, statistical-computing},
	title = {A multifidelity ensemble {Kalman} filter with reduced order control variates},
	url = {https://doi.org/10.1137/20M1349965},
	volume = {43},
	year = {2021}}

@article{Sandu_2016_reducedSampling4DVar,
	author = {A. Attia and R. Stefanescu and A. Sandu},
	doi = {10.1002/fld.4255},
	journal = {International Journal of Numerical Methods in Fluids},
	number = {1},
	pages = {28--51},
	subjects = {inverse-problems,surrogate-models,statistical-computing},
	title = {The reduced-order hybrid {Monte-Carlo} sampling smoother},
	url = {http://dx.doi.org/10.1002/fld.4255},
	volume = {83},
	year = {2016}}

@article{Sandu_2015_POD-inverse-problems,
	author = {R. Stefanescu and A. Sandu and I.M. Navon},
	doi = {10.1016/j.jcp.2015.04.030},
	journal = {Journal of Computational Physics},
	pages = {569--595},
	subjects = {inverse-problems,surrogate-models},
	title = {{POD/DEIM} strategies for reduced data assimilation systems},
	url = {http://dx.doi.org/10.1016/j.jcp.2015.04.030},
	volume = {295},
	year = {2015}}
